\documentclass{article}

\usepackage[
backend=bibtex,
url=false,isbn=false,doi=false,
date=year,
hyperref=auto,
style=numeric-comp,
natbib=true,
sorting=nty,
firstinits=true,
maxnames=2, maxbibnames=10, sorting=none,
]{biblatex}

\usepackage{hyperref}
\hypersetup{
    bookmarks=true,         
    unicode=false,          
    pdftoolbar=true,        
    pdfmenubar=true,        
    pdffitwindow=false,     
    pdfstartview={FitH},    
    pdfauthor={Mohammad Zalbagi Darestani},     
    pdfsubject={Subject},   
    pdfcreator={Mohammad Zalbagi Darestani},   
    pdfproducer={Producer}, 
    pdfnewwindow=true,      
    colorlinks=true,       
    linkcolor=red,          
    citecolor=green,        
    filecolor=magenta,      
    urlcolor=cyan           
}

\bibliography{Machine_Learning.bib}

\def\equationautorefname~#1\null{(#1)\null}

\newcommand\eqAutoref[1]{{\hypersetup{linkcolor=red}\autoref{#1}}}

\usepackage{subfiles}
\usepackage{amsmath,amssymb,amsfonts}
\usepackage{algorithmic}
\usepackage{textcomp}

\usepackage[utf8]{inputenc} 
\usepackage[T1]{fontenc}    
\usepackage{hyperref}       
\usepackage{url}            
\usepackage{booktabs}       
\usepackage{amsfonts}       
\usepackage{nicefrac}       
\usepackage{microtype}      

\usepackage[ruled,vlined]{algorithm2e}
\usepackage{rays_defs_18}

\usepackage{graphicx} 
\usepackage{caption}
\usepackage{subcaption}
\usepackage{amssymb}
\usepackage{bbm}
\usepackage[normalem]{ulem}
\usepackage{float}

\usepackage{multicol}
\usepackage{amsmath}
\usepackage{blkarray}
\usepackage{tikz,pgfplots}
\usetikzlibrary{pgfplots.statistics}
\usepackage{pgfplotstable}
\usepgfplotslibrary{colormaps}
\usetikzlibrary{quotes,angles}
\usepackage{textpos}
\usepackage{pgfplotstable}
\usepackage{array}
\usepackage{authblk}
\usepackage{multirow}
\usepgfplotslibrary{fillbetween}

\usepackage{threeparttable}
\usepackage{booktabs}
\usepackage{colortbl} 
\usepackage{xcolor} 
\usepackage{xfrac}
\usepackage{caption} 

\DeclareMathAlphabet{\mathcal}{OMS}{cmsy}{m}{n}
\SetMathAlphabet{\mathcal}{bold}{OMS}{cmsy}{b}{n}

\usetikzlibrary{matrix}

\pgfplotsset{compat=1.14}
\usepgfplotslibrary{groupplots}
\usepgfplotslibrary{dateplot}

\usetikzlibrary{3d,decorations.text,shapes.arrows,positioning,fit,backgrounds}
\tikzset{pics/fake box/.style args={
#1 with dimensions #2 and #3 and #4}{
code={
\draw[ultra thin,fill=#1]  (0,0,0) coordinate(-front-bottom-left) to
++ (0,#3,0) coordinate(-front-top-right) --++
(#2,0,0) coordinate(-front-top-right) --++ (0,-#3,0) 
coordinate(-front-bottom-right) -- cycle;
\draw[ultra thin,fill=#1] (0,#3,0)  --++ 
 (0,0,#4) coordinate(-back-top-left) --++ (#2,0,0) 
 coordinate(-back-top-right) --++ (0,0,-#4)  -- cycle;
\draw[ultra thin,fill=#1!80!black] (#2,0,0) --++ (0,0,#4) coordinate(-back-bottom-right)
--++ (0,#3,0) --++ (0,0,-#4) -- cycle;
}
}}
\tikzset{pics/empty fake box/.style args={
#1 with dimensions #2 and #3 and #4}{
code={
\draw[ultra thin,fill=#1]  (0,0,0) coordinate(-front-bottom-left) to
++ (0,#3,0) coordinate(-front-top-right) --++
(#2,0,0) coordinate(-front-top-right) --++ (0,-#3,0) 
coordinate(-front-bottom-right) -- cycle;
\draw[ultra thin,fill=#1] (0,#3,0)  --++ 
 (0,0,#4) coordinate(-back-top-left) --++ (#2,0,0) 
 coordinate(-back-top-right) --++ (0,0,-#4)  -- cycle;
\draw[ultra thin,fill=#1!80!black] (#2,0,0) --++ (0,0,#4) coordinate(-back-bottom-right)
--++ (0,#3,0) --++ (0,0,-#4) -- cycle;
}
}}

\pgfplotsset{
    every first x axis line/.style={},
    every first y axis line/.style={},
    every first z axis line/.style={},
    every second x axis line/.style={},
    every second y axis line/.style={},
    every second z axis line/.style={},
    first x axis line style/.style={/pgfplots/every first x axis line/.append style={#1}},
    first y axis line style/.style={/pgfplots/every first y axis line/.append style={#1}},
    first z axis line style/.style={/pgfplots/every first z axis line/.append style={#1}},
    second x axis line style/.style={/pgfplots/every second x axis line/.append style={#1}},
    second y axis line style/.style={/pgfplots/every second y axis line/.append style={#1}},
    second z axis line style/.style={/pgfplots/every second z axis line/.append style={#1}}
}

\makeatletter
\def\pgfplots@drawaxis@outerlines@separate@onorientedsurf#1#2{%
    \if2\csname pgfplots@#1axislinesnum\endcsname
    \else
    \scope[/pgfplots/every outer #1 axis line,
        #1discont,decoration={pre length=\csname #1disstart\endcsname, post length=\csname #1disend\endcsname}]
        \pgfplots@ifaxisline@B@onorientedsurf@should@be@drawn{0}{%
            \draw [/pgfplots/every first #1 axis line] decorate {
                \pgfextra
                \pgfplotspointonorientedsurfaceabsetupfor{#2}{#1}{\pgfplotspointonorientedsurfaceN}%
                \pgfplots@drawgridlines@onorientedsurf@fromto{\csname pgfplots@#2min\endcsname}%
                \endpgfextra 
                };
        }{}%
        \pgfplots@ifaxisline@B@onorientedsurf@should@be@drawn{1}{%
            \draw [/pgfplots/every second #1 axis line] decorate {
                \pgfextra
                \pgfplotspointonorientedsurfaceabsetupfor{#2}{#1}{\pgfplotspointonorientedsurfaceN}%
                \pgfplots@drawgridlines@onorientedsurf@fromto{\csname pgfplots@#2max\endcsname}%
                \endpgfextra 
                };
        }{}%
    \endscope
    \fi
}%
\makeatother

\pgfplotsset{select coords between index/.style 2 args={
    x filter/.code={
        \ifnum\coordindex<#1\fi
        \ifnum\coordindex>#2\fi
    }
}}

\pgfplotsset{
    compat=1.14,
    colormap={no data}{
        color=(white)
        color=(red)
    },
    colormap/bluered,
    colormap={parula}{
        rgb255=(53,42,135)
        rgb255=(15,92,221)
        rgb255=(18,125,216)
        rgb255=(7,156,207)
        rgb255=(21,177,180)
        rgb255=(89,189,140)
        rgb255=(165,190,107)
        rgb255=(225,185,82)
        rgb255=(252,206,46)
        rgb255=(249,251,14)
    },
}

\providecommand{\abs}[1]{\lvert#1\rvert}
\usepackage{adjustbox}
\usepackage{booktabs}
\usepackage{relsize,stackengine}

\usepackage{ragged2e,tabularx}
\newcolumntype{L}[1]{>{\raggedright\let\newline\\\arraybackslash\hspace{0pt}}p{#1}}
\definecolor{steelblue}{rgb}{0.27, 0.51, 0.71}
\definecolor{amber}{rgb}{1.0, 0.49, 0.0}
\definecolor{bluedf}{rgb}{0.19, 0.55, 0.91}
\definecolor{caribbeangreen}{rgb}{0.0, 0.8, 0.6}
\definecolor{pastelgreen}{rgb}{0.47, 0.87, 0.47}
\definecolor{darkpastelgreen}{rgb}{0.01, 0.75, 0.24}
\definecolor{aliceblue}{rgb}{0.94, 0.97, 1.0}
\definecolor{carminered}{rgb}{1.0, 0.0, 0.22}
\definecolor{dodgerblue}{rgb}{0.12, 0.56, 1.0}
\definecolor{psychedelicpurple}{rgb}{0.87, 0.0, 1.0}
\definecolor{neonfuchsia}{rgb}{1.0, 0.25, 0.39}
\definecolor{mauve}{rgb}{0.88, 0.69, 1.0}
\definecolor{manatee}{rgb}{0.59, 0.6, 0.67}
\definecolor{brightpink}{rgb}{1.0, 0.0, 0.5}
\definecolor{brilliantrose}{rgb}{1.0, 0.33, 0.64}
\definecolor{bubblegum}{rgb}{0.99, 0.76, 0.8}
\definecolor{electriclavender}{rgb}{0.96, 0.73, 1.0}
\definecolor{buff}{rgb}{0.94, 0.86, 0.51}
\definecolor{carminepink}{rgb}{0.92, 0.3, 0.26}
\definecolor{grannysmithapple}{rgb}{0.66, 0.89, 0.63}
\definecolor{darkspringgreen}{rgb}{0.09, 0.45, 0.27}
\definecolor{ash}{rgb}{0.7, 0.75, 0.71}
\definecolor{nblue}{rgb}{0,0.263,0.576}
\definecolor{mblue}{rgb}{0.075,0.541,0.855}
\long\def\comment#1{}

\begin{document}

\begin{center}

{\bf{\LARGE{
Accelerated MRI with Un-trained Neural Networks
}}}

\vspace*{.2in}

{\large{
\begin{tabular}{cccc}
Mohammad Zalbagi Darestani$^{\ast}$
and 
Reinhard Heckel$^{\dagger,\ast}$
\end{tabular}
}}

\vspace*{.05in}

\begin{tabular}{c}
$^\ast$Dept. of Electrical and Computer Engineering, Rice University\\
$^\dagger$Dept. of Electrical and Computer Engineering, Technical University of Munich
\end{tabular}

\vspace*{.1in}

\today

\vspace*{.1in}

\end{center}

\begin{abstract}
Convolutional Neural Networks (CNNs) are highly effective for image reconstruction problems. Typically, CNNs are trained on large amounts of training images. Recently, however, un-trained CNNs such as the Deep Image Prior and Deep Decoder have achieved excellent performance for image reconstruction problems such as denoising and inpainting, \emph{without using any training data}. Motivated by this development, we address the reconstruction problem arising in accelerated MRI with un-trained neural networks. 
We propose a highly optimized un-trained recovery approach based on a variation of the Deep Decoder and show that it significantly outperforms other un-trained methods, in particular sparsity-based classical compressed sensing methods and naive applications of un-trained neural networks.
We also compare performance (both in terms of reconstruction accuracy and computational cost) in an ideal setup for trained methods, specifically on the fastMRI dataset, where the training and test data come from the same distribution. We find that our un-trained algorithm achieves similar performance to a baseline trained neural network, but a state-of-the-art trained network outperforms the un-trained one. Finally, we perform a comparison on a non-ideal setup where the train and test distributions are slightly different, and find that our un-trained method achieves similar performance to a state-of-the-art accelerated MRI reconstruction method.
\end{abstract}

\section{Introduction}
\label{sec:intro}
CNNs are highly successful tools for image reconstruction tasks.
In recent years, a large number of works have shown that they can outperform traditional image processing methods for tasks such as image denoising, compressive sensing, and image compression~\citep{bora_CompressedSensingUsing_2017a,theis2017lossy, toderici2015variable, agustsson2017soft, burger2012image}. 
Almost exclusively, CNNs are trained on large sets of images, and their success is typically attributed to their ability to learn from those training images.

However, starting with the Deep Image Prior (DIP)~\citep{ulyanov2018deep}, a number of works have demonstrated that the architecture of a CNN can act as a sufficiently strong prior to enable image reconstruction, even without any training data. Un-trained networks perform well for denoising~\citep{ulyanov2018deep,heckel2019deep}, compressive sensing~\citep{van2018compressed,arora2020untrained}, 
phase retrieval~\citep{jagatap2019algorithmic,bostan2020deep,wang2020phase}, and even for reconstructing videos~\citep{jin2019time,hyder2020generative}. 
Moreover, they provably succeed in denoising and reconstructing smooth signals from few measurements~\cite{heckel2020denoising,heckel2020compressive}. They have, however, mainly been studied in relatively controlled setups (such as denoising with Gaussian noise), and their performance in practical medical imaging problems is relatively unexplored.

In this work, we address the problem of accelerating Magnetic Resonance Imaging (MRI) with un-trained neural networks, to understand if un-trained networks can give state-of-the-art performance in practice, and to study if they can compete with trained neural networks for image reconstructions problems.

MRI is an important medical imaging technique and is extremely popular because it is non-invasive. 
However, performing an MRI scan is slow due to physical limitations of the scanning process. 
These limitations have led to a line of research known as compressed sensing which focuses on accelerating MRI~\citep{lustig2008compressed} by reconstructing an image from a few measurements, which in turn results in a faster scan time.


Traditional compressed sensing methods are based on $\ell_1$-norm minimization or Total-Variation (TV) norm minimization~\citep{block2007undersampled} and are un-trained, i.e., they do not rely on any training data.
Those methods are implemented in modern MRI scanners, but are outperformed by an emerging class of deep-learning-based reconstruction techniques, as demonstrated for example by the fastMRI challenge~\citep{knoll2020advancing,muckley2020state}, a competition for accelerated MRI reconstruction. 

Off-the-shelf un-trained neural networks have been applied to accelerated MRI reconstruction
before~\cite{arora2020untrained,heckel2019regularizing,jin2019time}, and have shown promising improvements over traditional un-trained methods for example images. 
At the same time, the two works~\cite{arora2020untrained,heckel2019regularizing} reported visibly worse images than those produced by the state-of-the-art trained neural networks. 
The goal of this paper is to develop an un-trained neural network based reconstruction method specifically tailored to multi-coil MRI and understand its imaging performance relative to competing un-trained and trained approaches. 
Our contributions are:
\begin{itemize}
    \item We improve image recovery performance with un-trained neural networks through i) architectural improvements, ii) introducing a data consistency step, iii) including sensitivity maps in the reconstruction process, and iv) combining reconstructions of different runs.
    Taken together, those steps give a significant accuracy improvement over naive application of un-trained networks.
    \item As an architectural improvement, we propose a variation of the Deep Image Prior (DIP)~\citep{ulyanov2018deep} and Deep Decoder~\citep{heckel2019deep} architectures, called ConvDecoder. We show that this architecture performs best for knee MRI images and similar to Deep Decoder and DIP for brain MRI images, which have less detail. ConvDecoder is a simple convolutional generator comprised of up-sampling, convolution, batch normalization, and ReLU blocks in each layer.
    \item 
    We show that un-trained neural networks significantly outperform traditional un-trained methods based on sparse and low-rank models. We consider TV-minimization as a baseline~\citep{block2007undersampled} and ENLIVE~\citep{holme2019enlive} as a representative un-trained method based on low-rank solutions~\citep{haldar2013low,shin2014calibrationless,jin2016general}. Traditional sparse and low-rank approaches are a natural comparison, because they perform well, are used in practice, and do not rely on training data, just like our un-trained method.
    \item Un-trained neural networks are relatively slow because they require a neural network to be fitted to observations via an iterative procedure. We propose an initialization technique that accelerates the reconstruction by a factor of 10.
    %
    \item We compare un-trained methods to trained methods on a dataset where trained methods shine: the fastMRI dataset~\citep{zbontar2018fastmri}, a dataset for deep learning based accelerated MRI, where the training and test distribution are the same. We find that, perhaps surprisingly, even in this setup, un-trained methods have on-par performance with the U-net~\citep{ronneberger2015u}, a baseline trained method,
    and are only 1.78 dB worse (see Table~\ref{tab:do-better}) than the state-of-the art trained method, the VarNet~\citep{sriram2020end}. 
    This suggests that there is less benefit in learning-based approaches for imaging, at least in the context of MRI, than currently thought.
    \item Finally, we compare un-trained methods to trained methods in a setup where train and test distributions are slightly different: we train U-net and VarNet and tune un-trained methods on the fastMRI knee dataset, and test on the fastMRI brain dataset. All methods perform significantly worse under this distribution shift. 
    In this regard, we propose two approaches to mitigate performance degradation due to distribution shifts for un-trained methods: (i) performing meta learning via a few examples from the test domain, and (ii) we propose an auto-tuning technique for un-trained networks with which our un-trained method achieves performance on par to the state-of-the-art VarNet under the distribution shift. 
    Distribution shifts are common in practice, and a key advantage of un-trained neural networks (with auto-tuning) over trained ones is robustness to out-of-distribution examples.
\end{itemize}
 
As a baseline for trained methods, we consider U-net based reconstruction because of its popularity and ease of use. As the current state-of-the art method tested on the fastMRI dataset, we consider the end-to-end Variational Network (VarNet) introduced in~\citep{sriram2020end}.
We emphasize that \emph{our method operates without any training data} and only leverages few data points for hyper-parameter tuning, while the U-net and VarNet are trained. 
The U-net is a baseline in the fastMRI challenge~\citep{knoll2020advancing,muckley2020state}. Other interesting approaches that perform close to VarNet are the Pyramid Convolutional Recurrent Neural Network (PCRNN)~\citep{wang2019pyramid}, $\Sigma$-net~\citep{schlemper2019sigma}, XPDNet~\citep{ramzi2020xpdnet}, Cascade net~\citep{schlemper2017deep}, and invertible Recurrent Inference Machines (i-RIM)~\citep{putzky2019invert}. 


\section{Problem statement: Accelerated multi-coil MRI}\label{sec:prob_statement}


We consider accelerated multi-coil MRI. 
Our goal is to recover an image $\vx \in \mathbb{C}^N$ from a set of measurements. The measurements are obtained as
\[
\vy_i = \mM \mF \mS_i \vx + \text{noise},
\quad i=1,\ldots, n_c,
\]
where $\mS_i$ is a complex-valued position-dependent sensitivity map, that is applied through entry-wise multiplication to the image $\vx$, $\mF$ implements the 2D discrete Fourier transform, $n_c$ is the number of magnetic coils, and $\mM$ is a mask that implements under-sampling. 
The measurements $\vy_i$ are called $k$-space measurements. 

First, suppose that we are only given one measurement, this is called single-coil imaging.
Also, suppose that the sensitivity map is equal to identity, and that the mask also corresponds to the identity, i.e., we are given a single measurement $\vy = \mF \vx + \text{noise}$. In this case, we can estimate the image up to the uncertainty of the additive noise as $\hat \vx = \inv{\mF} \vy$. 

In accelerated imaging, the $k$-space is under-sampled which is modeled through multiplication with the mask $\mM$ which simply sets some of the frequencies of the $k$-space measurement to zero. Under-sampling by a factor of $K$ results in a scan speed-up by the same factor. Reconstruction from the under-sampled measurements amounts to estimating an image from few measurements which is known as compressed sensing.

The practically most relevant problem is multi-coil reconstruction. In multi-coil imaging, each of the coils results in a different measurement. The sensitivity maps which determine those measurements are typically not given, but can be estimated from the under-sampled measurements. 
In this paper we consider the problem of reconstructing an image from under-sampled multi-coil measurements.

We work with the recently-released fastMRI dataset~\citep{zbontar2018fastmri}. 
The fastMRI dataset consists of a training and validation set each consisting of full $k$-space measurements of knees taken with $n_c = 15$ coils, and of brains taken with a varying number of coils.
The dataset also contains ``reference'' images which are obtained by reconstructing the coil images from each coil measurement as $\hat \vx_i = \inv{\mF} \vy_i$ and then combining the coil images via the root-sum-of-squares (RSS) algorithm to a final single image:
\begin{align}
    \hat{\vx} = \sqrt{\sum_{i=1}^{n_c} \abs{\hat{\vx}_i}^2}.
    \label{eq:rss}
\end{align}
Here, $\abs{\cdot}$ and $\sqrt{\cdot}$ denote element-wise absolute value and squared root operations. 
Since different coil sensitivities overlap only little, the RSS algorithm works well for combining the images.

We consider accelerated imaging by obtaining measurements with a mask. We utilize a standard 1D variable-density mask (i.e., random or equi-spaced vertical lines across the $k$-space), because those masks are challenging but practically most relevant, and are the default in the fastMRI challenge. 
For evaluation, we compare to the ``reference'' images reconstructed from the full $k$-space.

\section{Image recovery with un-trained neural networks}\label{sec:method}

We recover images from measurements by using un-trained networks as image priors as follows. 
We view un-trained neural networks as convolutional image priors mapping a parameter space to an image space, i.e., $G\colon \reals^p \to \reals^{c \times w \times h}$, where $c$ is the number of channels of the output image (for example $c=1$ for single grayscale image), and $w$ and $h$ are the width and height of the image. 

Deep-learning-based image models are typically trained; specifically, they are parameterized functions mapping an input to an output, with trainable (weight) parameters. For such trained image priors, the input parameterizes the image, the output is an image, and the weights are trained on a distribution of images. In contrast, an un-trained neural network is an image model where the input to the network is relatively irrelevant and fixed, the weights are the parameters of the model, and the output of the model is again an image. 

As mentioned before, un-trained neural networks were first proposed for image restoration problems by \citet{ulyanov2018deep}.
\citet{ulyanov2018deep} proposed to use a simple U-net architecture consisting of an encoder, decoder, and skip connections for image reconstruction by fitting this architecture to a measurement. 
It has relatively quickly been realized that the encoder and the skip connections are irrelevant for performance and the decoder can even be further simplified~\cite{heckel2019deep}. We discuss architectures in Section~\ref{sec:convdecoder}. 

\subsection{Single-coil reconstruction}\label{sec:single-coil}

To explain how images are recovered  with an  un-trained  neural network,  we start with single-coil reconstruction.
Let $\vy \in \mathbb{C}^M$ be the under-sampled $k$-space measurement and let $\mM$ and $\mF$ be the mask and Fourier transform defined in Section \ref{sec:prob_statement} that maps an image to a measurement. 
Given an un-trained neural network $G \colon \reals^p \to \reals^{c \times w \times h}$ with parameter vector $\mC \in \mathbb{R}^p$, we estimate an image based on the measurements $\vy$ by first minimizing the mean-squared loss function
\begin{align}
    \mathcal{L}(\mC) = \; \frac{1}{2} 
    \norm[2]{\vy - \mM \mF G(\mC)}^2,
    \label{eq:optim2}
\end{align}
with an iterative first order method to obtain the estimate $\hat \mC$, and second, estimating the image as $\hat \vx = G(\hat \mC)$. 
The image consists of a real and imaginary part, each described by one channel, therefore $c=2$. 
As will become clear later, the choice of optimization algorithm is critical for performance: The network together with the optimization acts as a prior.


\subsection{Multi-coil accelerated MRI reconstruction}\label{sec:multi-coil}

In multi-coil imaging, multiple magnetic coils take different $k$-space measurements of the same image, and thus there are a variety of ways to use un-trained methods for image reconstruction. We found the following two to work best in different problem setups (i.e., one works best for brain images, the other for knee images); the first reconstructs an image with estimated sensitivity maps, and the second works without sensitivity map estimates.

\paragraph*{Step 1-A: Reconstruction without coil sensitivity maps} 
The perhaps most straightforward way to recover the image is to treat each measurement/coil independently, reconstruct as described in the single-coil reconstruction section above, and then combine the images using the RSS algorithm in \eqAutoref{eq:rss}. 

A better performing and computationally more efficient approach is to impose the same prior to all images pertaining to the coils. This approach was first used by~\citet{arora2020untrained}.
Here, the first two output channels of the un-trained network generate the real and imaginary parts of the first image, the second two channels generate the the real and imaginary parts of the second image and so on.
The final single image is then obtained with the RSS algorithm in \eqAutoref{eq:rss}. 
The loss function pertaining to this method is
\begin{equation}
	\mathcal{L}(\mC) = \frac{1}{2}
	\sum_{i=1}^{n_c} 
	\norm[2]{ \vy_i - \mM \mF G_i(\mC) }^2,
	\label{eq:loss}
\end{equation}
where $G(\mC)$ is a stack of reconstructed images for all coils, in that we use one generator to generate multiple images at once. Therefore, $G_i(\mC)$ (the $i-$th element of the output stack) is the reconstructed image associated with the measurements from the $i$-th coil $\vy_i$.
We found that using a single image prior for all images as proposed here gives slightly better reconstruction quality relative to reconstructing image by image and is significantly more efficient (approximately $n_c$-times faster).

\paragraph*{Step 1-B: Reconstruction with sensitivity map estimates} Instead of reconstructing all coil images separately, this method reconstructs the final output image directly, and takes the sensitivity maps $\hat \mS$ estimated from the under-sampled data by applying ESPIRiT \citep{Uecker2014} into account. Specifically, the loss function is

\begin{equation}
	\mathcal{L}(\mC) = \frac{1}{2}
	\sum_{i=1}^{n_c} 
	\norm[2]{ \vy_i - \mM \mF \hat \mS_i G(\mC) }^2.
	\label{eq:optim4}
\end{equation}

Note that the convolutional generator $G(\mC)$ has only two output channels, corresponding to the final image (since we reconstruct real and imaginary parts of the image in two separate channels).

\paragraph*{Step 2: Enforcing data consistency}
After fitting the network by minimizing the respective loss function, we enforce data constancy. Specifically, recall that  the measurement consists of under-sampled frequencies of the original image located at $x$ axis coordinates $\mathcal{S}_x=\{i_1,\ldots,i_m\}$. 
After reconstructing the image, in its corresponding Fourier domain representation, we replace frequency components located at $\mathcal{S}_x$ with the ones from the under-sampled measurement.


\subsection{Network architectures}\label{sec:convdecoder}

In this section, we discuss the network architectures we consider in this paper. 
All of the architectures we considered for MRI reconstruction are convolutional image-generating networks that map an input volume to an output. We choose a fixed input (specifically, we choose it randomly at initialization) and optimize over the weights of the network. 
The DIP is the first and most popular architecture~\citep{ulyanov2018deep}, and consists of an encoder, decoder and skip connections. 

The Deep Decoder~\citep{heckel2019deep} is a simple decoder architecture, even simpler than the decoder part of the DIP, only consisting of convolutional operations with fixed convolutional filters followed by linear combinations (i.e., 1x1 convolutions). 

In this paper, we found a variation of the Deep Decoder architecture, that we call ConvDecoder to perform best in most instances; and similar to the original Deep Decoder in other instances. 
We also tried a variety of other architectures, including combinations of ConvDecoders that reconstruct an image at different resolutions, but found the simple ConvDecoder to perform best (more details on the architectures we analyzed can be found in the supplementary material).


Both the Deep Decoder and the ConvDecoder are convolutional neural networks mapping a parameter space to images, i.e., $G\colon \reals^p \to \reals^{c \times w \times h}$, where $c$ is the number of output channels, and $w$ and $h$ are the width and height of the image in each channel. 
In each layer, except the last one, the network is composed of the following components: up-sampling, a convolutional operation, ReLU activation function, and finally a Batch Normalization (BN) block~\citep{ioffe2015batch}. 
BN normalizes each channel of its input volume independent of other channels. 
The Deep Decoder uses bi-linear upsampling and 1x1 convolutions, while the ConvDecoder uses Nearest-Neighbor up-sampling and a 3x3 convolutional layer. Figure~\ref{fig:architecture} depicts the network. 


\begin{figure}[t!]
    \centering

\begin{tikzpicture}[x={(1,0)},y={(0,1)},z={({cos(60)},{sin(60)})},
font=\sffamily\small,scale=2]
\node [single arrow, draw=blue!30, fill=blue!10, minimum height=3em, minimum width=1.5em,yscale=0.55,outer sep=0pt,label={[label distance=3mm]180:\large $\vz$}] at (0.2,0.4,-0.5) {\vphantom{x}};
\node[text width=1cm] at (-0.3,0.15,-0.5) {$(fixed)$};
\foreach \X/\Y/\Z [count=\U] in {0.3/0.17/electriclavender,0.4/0.17/amber,0.5/0.17/steelblue,0.6/0.17/aliceblue}
\draw pic (box1-\U) at (\X,-\Y,-\Y/2) 
{fake box=\Z!80! with dimensions {0.2} and {4*\Y} and 2*\Y};
\node [single arrow, draw=blue!30, fill=blue!10, minimum height=3em, minimum width=1.5em,yscale=0.6,outer sep=0pt] at (1.17,0.4,-0.5) {\vphantom{x}};
\foreach \X/\Y/\Z [count=\U] in {1.3/0.3/electriclavender,1.4/0.3/amber,1.5/0.3/steelblue,1.6/0.3/aliceblue}
\draw pic (box1-\U) at (\X,-\Y,-\Y/2) 
{fake box=\Z!80! with dimensions {0.2} and {4*\Y} and 2*\Y};
\node [single arrow, draw=blue!30, fill=blue!10, minimum height=3em, minimum width=1.5em,yscale=0.6,outer sep=0pt] at (2.2,0.4,-0.5) {\vphantom{x}};
\node[draw=none] (ellipsis1) at (2.7,0.4,-0.5) {$\cdots$};
\node [single arrow, draw=blue!30, fill=blue!10, minimum height=3em, minimum width=1.5em,yscale=0.6,outer sep=0pt] at (3.1,0.4,-0.5) {\vphantom{x}};
\foreach \X/\Y/\Z [count=\U] in {3.3/0.5/electriclavender,3.4/0.5/amber,3.5/0.5/steelblue,3.6/0.5/aliceblue}
\draw pic (box1-\U) at (\X,-\Y,-\Y/2) 
{fake box=\Z!80! with dimensions {0.2} and {4*\Y} and 2*\Y};
\node [single arrow, draw=blue!30, fill=blue!10, minimum height=3em, minimum width=1.5em,yscale=0.6,outer sep=0pt] at (4.2,0.4,-0.5) {\vphantom{x}};
\foreach \X/\Y/\Z [count=\U] in {4.4/0.5/amber,4.5/0.5/steelblue,4.6/0.5/aliceblue,4.7/0.5/grannysmithapple}
\draw pic (box1-\U) at (\X,-\Y,-\Y/2) 
{fake box=\Z!80! with dimensions {0.2} and {4*\Y} and 2*\Y};
\node (rect) at (-1.2,-1) [draw,minimum width=0.1cm,minimum height=0.1cm, fill=electriclavender!80] {};
\node[text width=2cm, anchor=west, right, font=\rmfamily] at (-1.15,-1) {\normalfont{up-sampling}};
\node (rect) at (0.05,-1) [draw,minimum width=0.1cm,minimum height=0.1cm, fill=amber!80] {};
\node[text width=6cm, anchor=west, right, font=\rmfamily] at (0.1,-1) {convolution};
\node (rect) at (1.25,-1) [draw,minimum width=0.1cm,minimum height=0.1cm, fill=steelblue!80] {};
\node[text width=6cm, anchor=west, right, font=\rmfamily] at (1.3,-1) {ReLU};
\node (rect) at (2.05,-1) [draw,minimum width=0.1cm,minimum height=0.1cm, fill=aliceblue!80] {};
\node[text width=6cm, anchor=west, right, font=\rmfamily] at (2.1,-1) {batch normalization};
\node (rect) at (3.85,-1) [draw,minimum width=0.1cm,minimum height=0.1cm, fill=grannysmithapple!80] {};
\node[text width=6cm, anchor=west, right, font=\rmfamily] at (3.9,-1) {channels\_to\_image};

\end{tikzpicture}
\caption{ConvDecoder architecture. It is comprised of up-sampling, convolutional, ReLU, batch normalization, and linear combination layers.
}
\label{fig:architecture}
\end{figure}

The parameters of the convolutional layers (and BN) are optimized when fitting the network to the given under-sampled measurement.
The final layer excludes the up-sampling layer and simply combines the images via a 1x1 convolutional layer that performs linear combinations of the channels.

Convolutional and up-sampling blocks are essential.
The former is responsible for capturing local information among the pixels and refines that information from one layer to another. The up-sampling block induces a notion of resolution into each layer as elaborated in supplementary material. 
Note that an ($n+1$)-layer ConvDecoder outputs an image $\hat{\vx} \in \mathbb{R}^{c_{n+1} \times w_{n+1} \times h_{n+1}}$ from a fixed input $\vz \in \mathbb{R}^{c_0 \times w_0 \times h_0}$, which we drawn from a Gaussian distribution and then fix. The default architecture we consider 
has 8 layers (including the last layer) and 256 channels per layer.


\section{Reconstruction accuracy of un-trained neural networks for MRI}\label{sec:performance}

In this section, we study the performance of un-trained neural networks for multi-coil 4x accelerated knee MRI reconstruction. We also provide results for 8x accelerated knee measurements and 4x accelerated brain measurements in the supplementary material.
We focus on multi-coil reconstruction, because multi-coil is clinically more relevant than single-coil reconstruction. 
Here, we focus on a setup that is ideal for trained methods: specifically, we train on the fastMRI training set, and evaluate on the fastMRI validation set, and both of those sets come from the same distribution. In practice, the train an test set often come from slightly different distributions, hence we study this aspect in Section~\ref{sec:distribution_shift}. 

After discussing how to evaluate performance (Section \ref{sec:evaluation_dilemma}), we provide the evaluation results for 4x accelerated knee measurements (Section \ref{sec:knee4x}).
Our main findings are: 
(i) perhaps surprisingly, our un-trained networks perform as well as a standard baseline \emph{trained} method, the U-net, and only slightly worse than the state-of-the-art network VarNet---but without any training data, and (ii) our un-trained network significantly outperforms other un-trained methods, in particular total-variation minimization (TV), a standard sparsity based baseline method, and ENLIVE, a calibration-less low-rank based method.

We performed a grid search to find the best parameters for each method; for the ConvDecoder, that resulted in an 8-layer network with 256 channels in each layer. 
More details on the architecture setup and optimization parameters can be found in the supplementary material. We also include a discussion in the supplementary material on how the results depend on the initialization and the choice of hyper-parameters.


\subsection{Considerations when evaluating reconstruction performance}\label{sec:evaluation_dilemma}

It is surprisingly non-trivial to compare different reconstruction methods for MRI. 
We have faced the following challenges and we have made the following choices for measuring the reconstruction performance:

\paragraph*{Image comparison metrics} Popular image metrics between a ground-truth image and a reconstructed image, like peak-signal-to-noise ratio (PSNR), are often unsuitable to capture reconstruction performance: 
\citet{mason2019comparison} investigated a number of metrics based on the diagnostic quality of MR images according to the feedback given by a group of five radiologists. 
Their study shows that Visual Information Fidelity (VIF)~\citep{sheikh2006image} is often a better metric for assessing diagnostic quality than widely-used metrics such as PSNR and Structural Similarity Index (SSIM)~\citep{wang2004image}. 
Interestingly, the study ranked PSNR and SSIM---the perhaps most widely-used metrics for assessing image quality--among the most inept metrics in this study. 
But even a higher VIF score sometimes corresponds to a visibly worse performance; to evaluate algorithms, we therefore measure performance in VIF, PSNR, SSIM, and Multi-Scale SSIM (MS-SSIM)~\citep{wang2003multiscale}, and also show example images for visual comparison. 

\paragraph*{Normalization} It is often necessary to normalize images when comparing them to a ground truth image. 
\citet{marcin2020does} have shown that image normalization techniques have a significant impact on various texture features being extracted from a medical image. 
We chose mean-std normalization (applied to the ground-truth image to match the statistical properties of the reconstructed image) because the resulting scores were more consistent with the literature.

\paragraph*{Comparison to noisy ground truth} In some cases, the ground-truth image itself is corrupted with measurement artifacts. 
Hence, even if the reconstructed image is of high quality, the score might not reflect that and this might result in difficulties of comparing different reconstructions. 

\paragraph*{Volume- vs. image-based comparison} Finally, all the afore-mentioned metrics are sensitive to whether the comparison is done on an image-based level, or is volume-based. Specifically, each scan of a knee or brain consists of a number of slices. Each slice is an image and together those images form a volume. 
All afore-mentioned metrics depend on the dynamic range of the volumes, and computing scores in a volume- or image-wise fashion gives different values. 
In the fastMRI challenge, scores are computed in a volume-based manner, i.e., the dynamic range of the volume is considered for computing the scores~\citep{wang2019pyramid,ramzi2020benchmarking,sriram2019grappanet,sriram2020end}.
Since images within each volume are analyzed independently, we consider the dynamic range of each image separately. 

We refer the reader to the supplementary material for further discussion on the mentioned evaluation challenges.


\subsection{Evaluation results}\label{sec:knee4x}
We evaluate the performance of ConvDecoder along with other methods on the 4x accelerated multi-coil knee measurements of the fastMRI dataset. 
We consider seven methods (five un-trained and two trained), specifically
(i) ConvDecoder, 
(ii) Deep Decoder~\citep{heckel2019deep},
(iii) DIP~\citep{ulyanov2018deep},
(iv) a standard un-trained 
TV-norm minimization method~\citep{block2007undersampled}, 
(v) the recently published calibration-less un-trained method ENLIVE~\citep{holme2019enlive},
(vi) the U-net~\citep{ronneberger2015u}, a standard trained method,
and finally
(vii) the end-to-end variational network (VarNet)~\citep{sriram2020end}.
We did an extensive grid search to select the best parameters for each un-trained neural network. See the supplementary material for details and for the parameter setup for each method.

\begin{table}[h]
\centering
\begin{adjustbox}{width=0.7\textwidth}
\begin{tabular}{|c|c|c|c|c|}
\hline 
Method & VIF & MS-SSIM & SSIM & PSNR \\
\hline
    ConvDecoder & \textbf{0.6717 $\pm$ 0.0411} & \textbf{0.9443 $\pm$ 0.0152} & \textbf{0.7775 $\pm$ 0.0258} & \textbf{31.81 $\pm$ 0.96} \\
    DIP & 0.6311 $\pm$ 0.0391 & 0.8981 $\pm$ 0.0122 & 0.5938 $\pm$ 0.0264 & 28.40 $\pm$ 1.02 \\
    DD & 0.6359 $\pm$ 0.0421 & 0.8599 $\pm$ 0.0176 & 0.6991 $\pm$ 0.0287 & 29.16 $\pm$ 1.13\\
\hline
\end{tabular}
\end{adjustbox}
\caption{Average image-based scores
for the ConvDecoder, DIP, and Deep Decoder (DD) on the mid-slice images of 20 randomly-chosen volumes in the multi-coil knee measurements from the fastMRI validation set (4x accelerated). 
ConvDecoder significantly outperforms DIP and DD according to all metrics. Marginal errors denote 95\% confidence interval.
}
\label{tab:validation_results_untrained}
\end{table}

\begin{figure}[th]
\centering
  \begin{subfigure}[t]{0.14\textwidth}
  \caption*{ConvDecoder}
  \centering\includegraphics[scale=0.3]{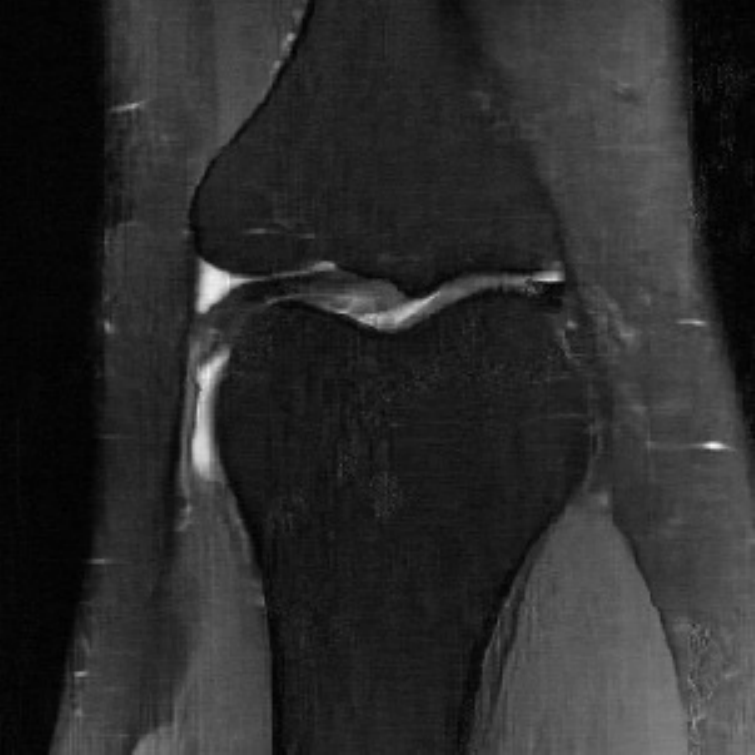}
  \end{subfigure}
  \begin{subfigure}[t]{0.14\textwidth}
  \caption*{DIP}
  \centering\includegraphics[scale=0.3]{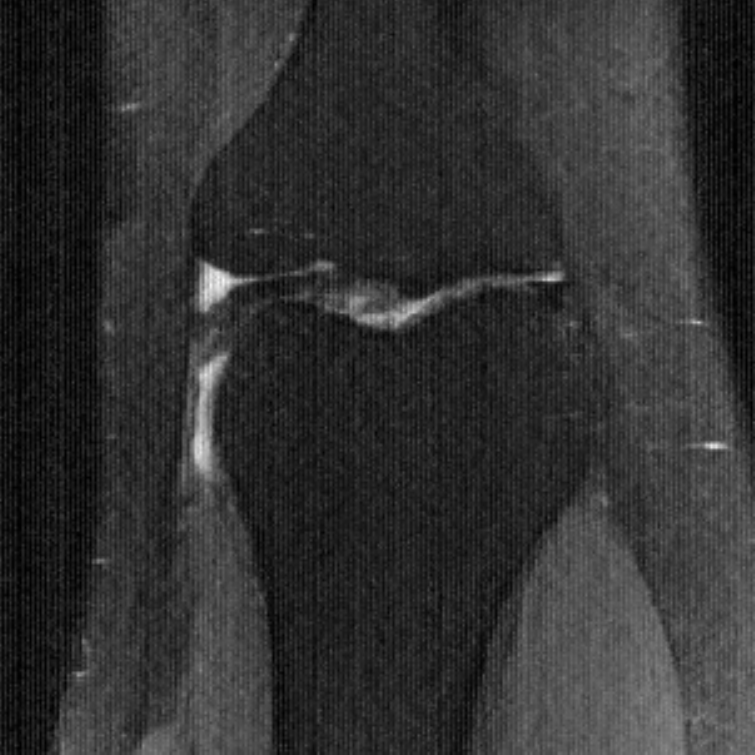}
  \end{subfigure}
  \begin{subfigure}[t]{0.14\textwidth}
  \caption*{DD}
  \centering\includegraphics[scale=0.3]{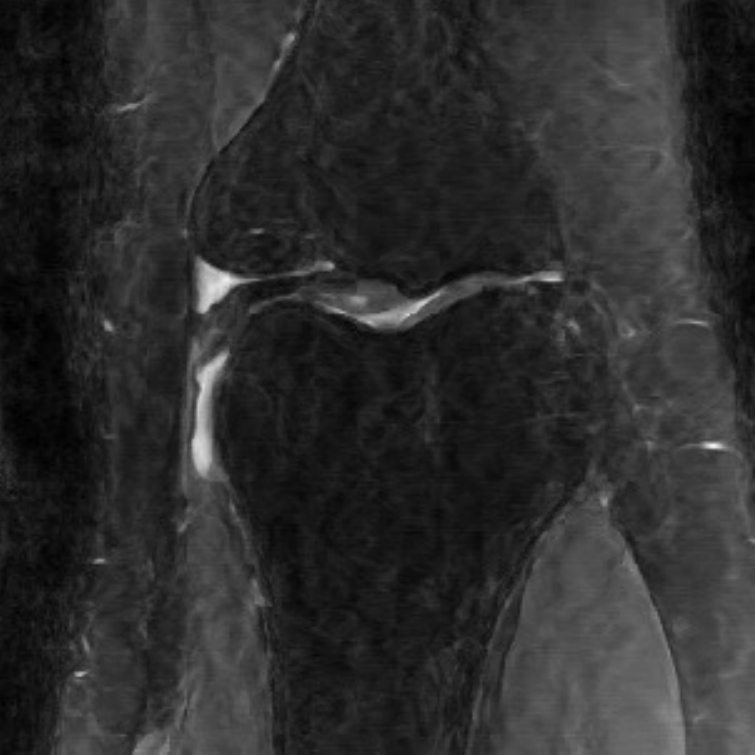}
  \end{subfigure}
  \begin{subfigure}[t]{0.14\textwidth}
  \caption*{ground truth}
  \centering\includegraphics[scale=0.3]{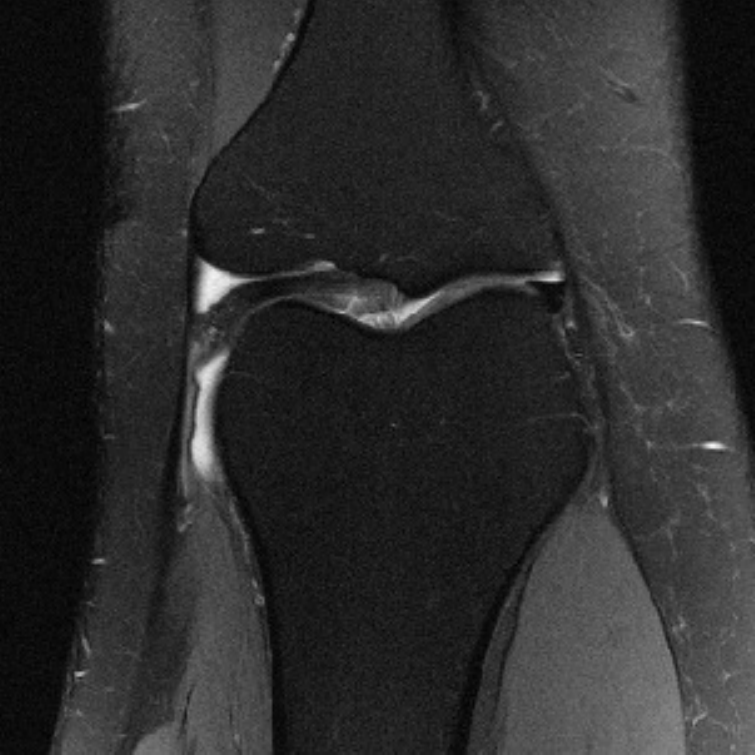}
  \end{subfigure}\par\medskip 
  \begin{subfigure}[t]{0.14\textwidth}
  \centering\includegraphics[scale=0.3]{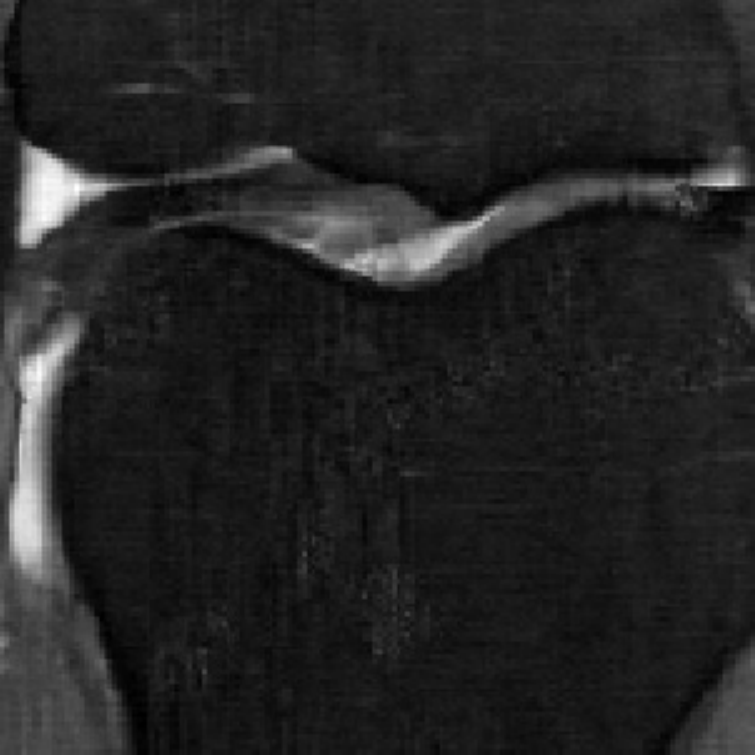}
  \end{subfigure}
  \begin{subfigure}[t]{0.14\textwidth}
  \centering\includegraphics[scale=0.3]{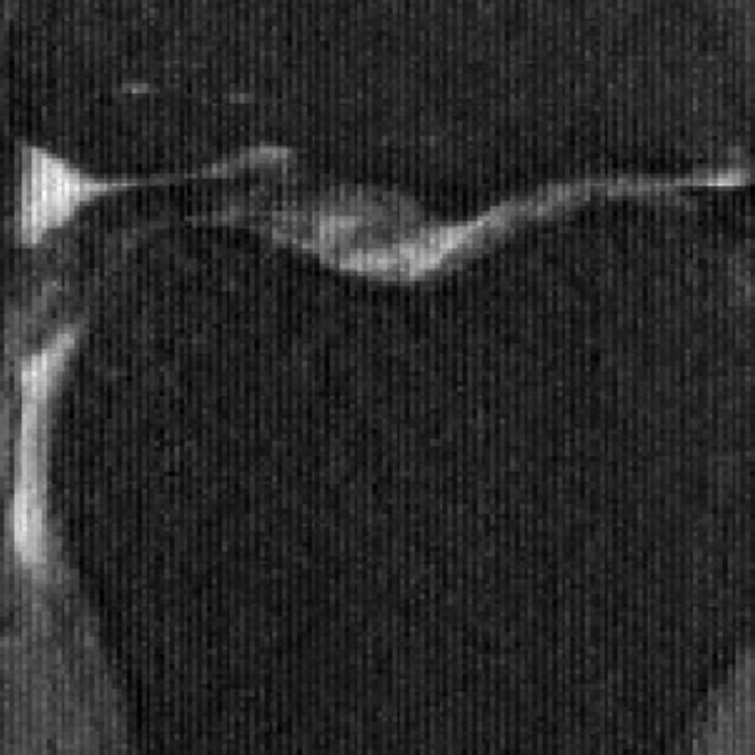}
  \end{subfigure}
  \begin{subfigure}[t]{0.14\textwidth}
  \centering\includegraphics[scale=0.3]{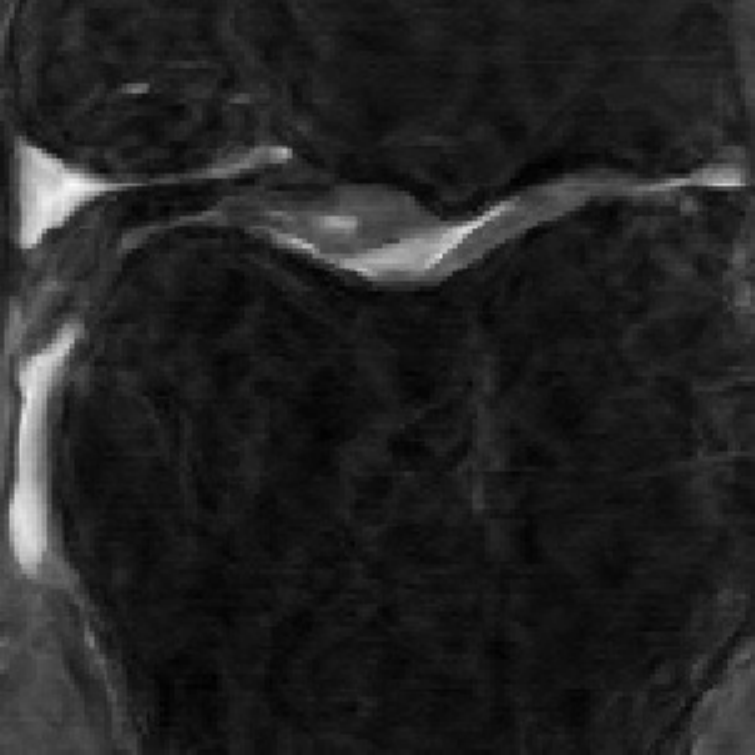}
  \end{subfigure}
  \begin{subfigure}[t]{0.14\textwidth}
  \centering\includegraphics[scale=0.3]{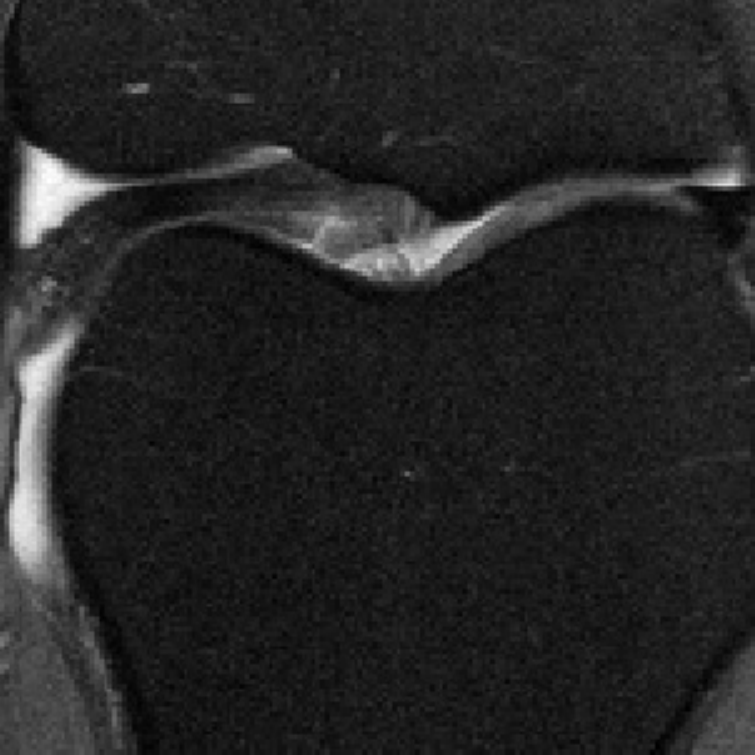}
  \end{subfigure}
\caption{Sample reconstructions for ConvDecoder, DIP, and Deep Decoder (DD) for a validation image from multi-coil knee measurements (4x accelerated). The top row represents zoomed-in version of the bottom row. ConvDecoder gives the best reconstruction for this image.}
\label{fig:untrained}
\end{figure}

We start by comparing the ConvDecoder with Deep Decoder and DIP architectures, and find that ConvDecoder performs best for knee MRI images.
Specifically, to compare the afore-mentioned un-trained networks, we computed the scores by averaging over 20 randomly-chosen mid-slice (i.e., the middle slice of the volume) images of different volumes in the validation set, and as mentioned earlier, the scores are computed in an image-based manner; see Table~\ref{tab:validation_results_untrained} for those scores and Figure~\ref{fig:untrained} for sample reconstructions. 

From Figure~\ref{fig:untrained}, it can be seen that DIP induces a noticeable amount of vertical artifacts. Moreover, Deep Decoder--in addition to having reconstruction artifacts--tends to generate rather smooth images and therefore misses texture details. 
These differences are present in other images as well, and hence are reflected in the scores in  Table~\ref{tab:validation_results_untrained}.

From Table~\ref{tab:validation_results_untrained}, we conclude that ConvDecoder performs best for knee MRI images. Therefore, in the rest of the knee experiments, we consider ConvDecoder and compare it to the baselines.

\paragraph*{ConvDecoder without training performs as well as a trained U-net and significantly better than un-trained baselines}
Next, we compare ConvDecoder with U-net, VarNet, TV, and ENLIVE. We trained a standard U-net with 8 layers and width factor equal to 32 on the whole training set (974 volumes). We also trained an end-to-end variational network with 12 cascades and width factor equal to 18 on the training set\footnote{Both U-net and VarNet are trained by following the exact instructions outlined in the \href{https://github.com/facebookresearch/fastMRI/tree/master/fastmri}{fastMRI} repository.}. To compute the results on the whole validation set, we ran the four mentioned methods on the mid-slice image of each volume in the validation set (a set of 200 images). We also randomly chose the mask for each run, but the marginal errors shown in Table~\ref{tab:baselines} denote that even with the randomness associated with masks, confidence intervals are sufficiently tight.

\begin{table}[th!]
\centering
\begin{adjustbox}{width=0.7\textwidth}
\begin{tabular}{|c|c|c|c|c|}
\hline 
Method & VIF & MS-SSIM & SSIM & PSNR \\
\hline
    ConvDecoder & \textbf{0.6823 $\pm$ 0.0217} & 0.9387 $\pm$ 0.0059 & 0.7753 $\pm$ 0.0145 & 31.67 $\pm$ 0.39\\
    ConvDecoder-noDC & 0.6323 $\pm$ 0.0236 & 0.9265 $\pm$ 0.0061 & 0.7107 $\pm$ 0.0153 & 30.46 $\pm$ 0.42\\
    U-net & 0.5955 $\pm$ 0.0147 & 0.9489 $\pm$ 0.0035 & 0.7883 $\pm$ 0.0097 & 32.04 $\pm$ 0.21 \\
    VarNet & 0.6456 $\pm$ 0.0122 & \textbf{0.9592 $\pm$ 0.0028} & \textbf{0.8342 $\pm$ 0.0084} & \textbf{34.20 $\pm$ 0.18} \\
    TV & 0.4412 $\pm$ 0.0272 & 0.9262 $\pm$ 0.0061 & 0.6977 $\pm$ 0.0141 & 30.20 $\pm$ 0.41\\
    ENLIVE & 0.3906 $\pm$ 0.0346 & 0.8516 $\pm$ 0.0098 & 0.5531 $\pm$ 0.0213 & 26.14 $\pm$ 0.29\\
\hline
\end{tabular}
\end{adjustbox}
\caption{
Performance for reconstructing 200 mid slice knee images of the fastMRI validation set (4x accelerated).
ConvDecoder performs best in the VIF metric deemed most relevant by Clinicians. In the other metrics, ConvDecoder and U-net are extremely similar. Marginal errors denote 95\% confidence interval.
}
\label{tab:baselines}
\end{table}

\begin{figure*}[th!]
\centering
  \begin{subfigure}[t]{0.16\textwidth}
  \caption*{ConvDecoder}
  \centering\includegraphics[scale=0.35]{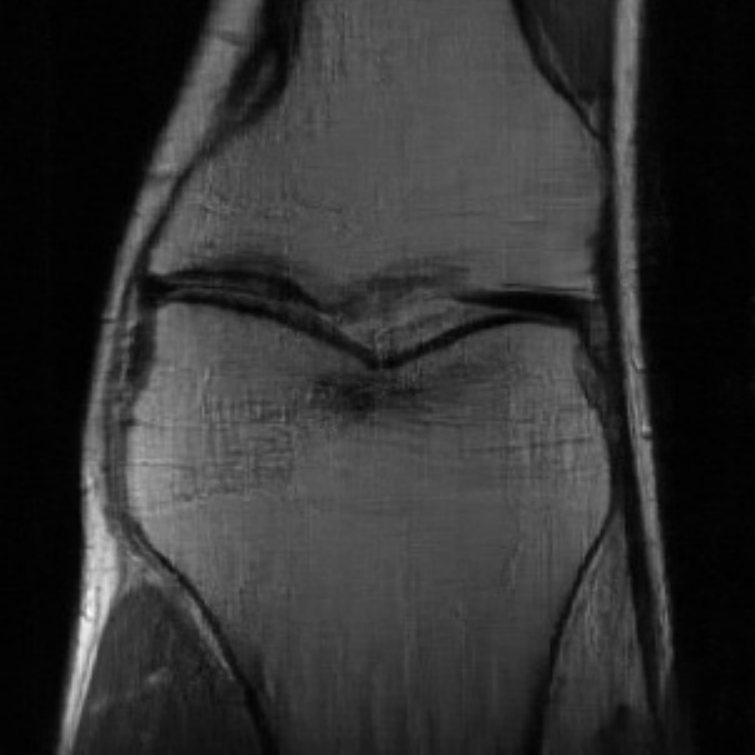}
  \end{subfigure}
  \begin{subfigure}[t]{0.16\textwidth}
  \caption*{TV}
  \centering\includegraphics[scale=0.35]{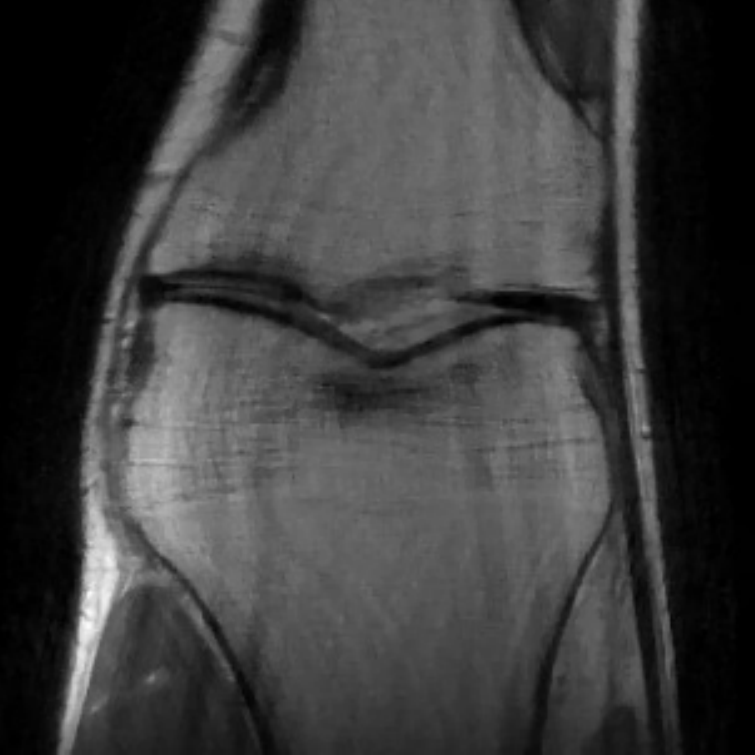}
  \end{subfigure}
  \begin{subfigure}[t]{0.16\textwidth}
  \caption*{ENLIVE}
  \centering\includegraphics[scale=0.35]{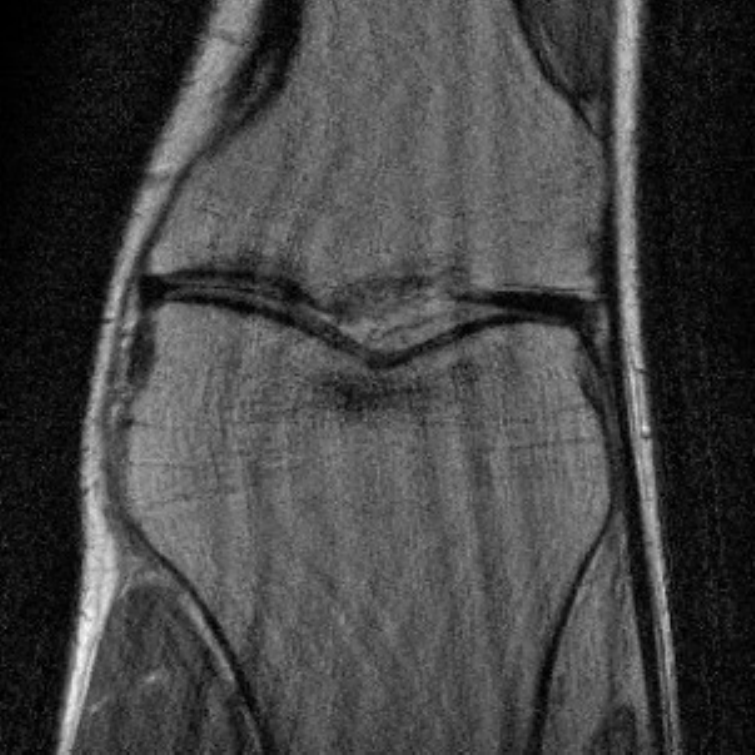}
  \end{subfigure}
  \begin{subfigure}[t]{0.16\textwidth}
  \caption*{U-net}
  \centering\includegraphics[scale=0.35]{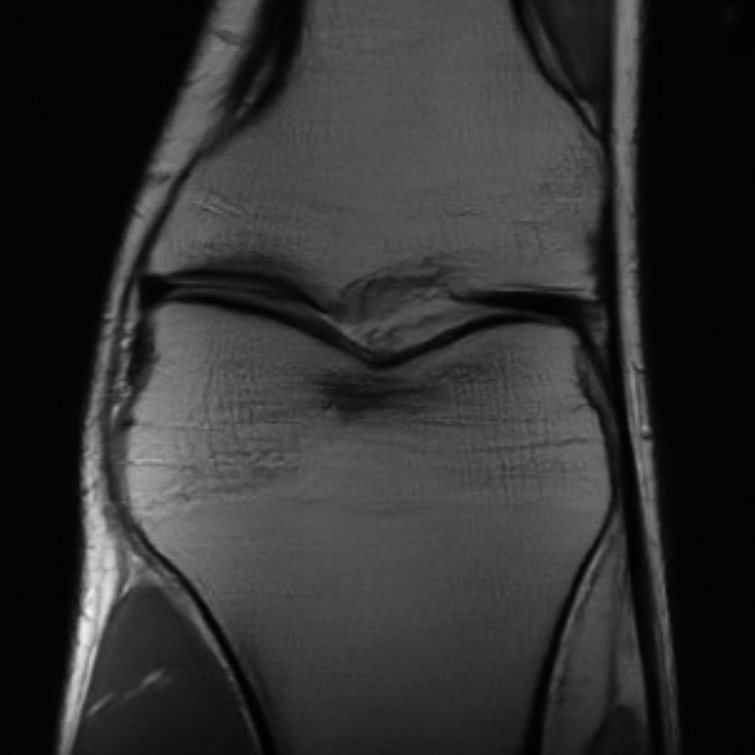}
  \end{subfigure}
  \begin{subfigure}[t]{0.16\textwidth}
  \caption*{VarNet}
  \centering\includegraphics[scale=0.35]{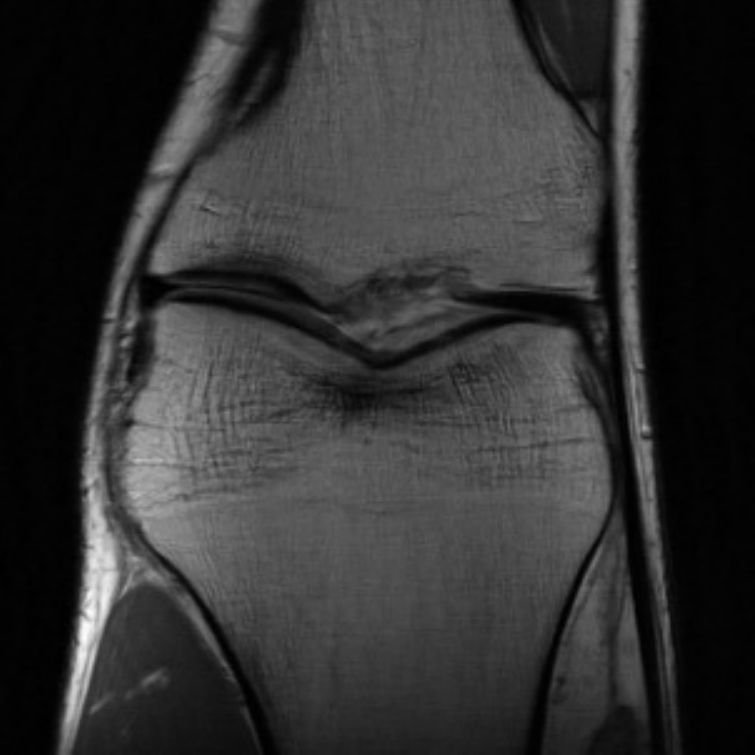}
  \end{subfigure}
  \begin{subfigure}[t]{0.16\textwidth}
  \caption*{ground truth}
  \centering\includegraphics[scale=0.35]{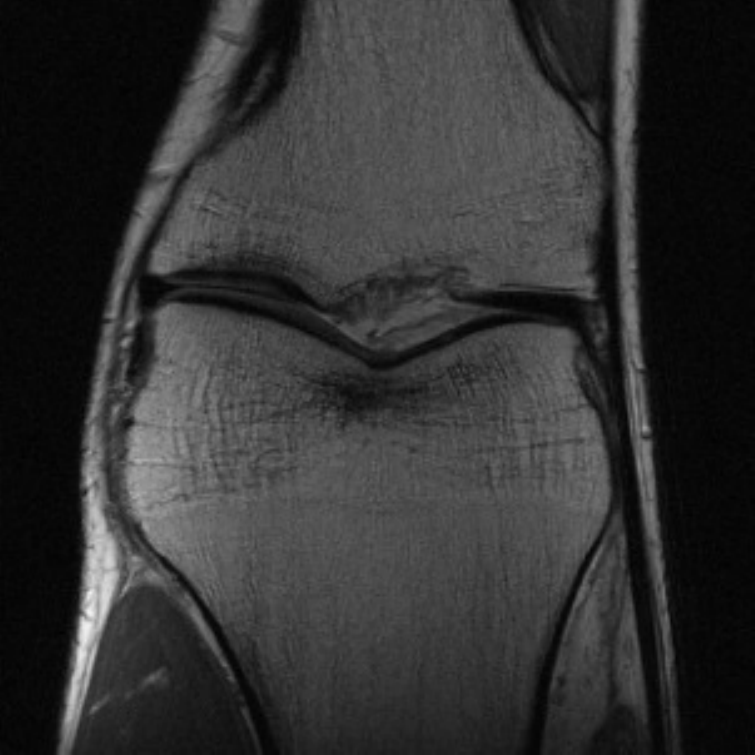}
  \end{subfigure}\par\medskip 
  \begin{subfigure}[t]{0.16\textwidth}
  \centering\includegraphics[scale=0.35]{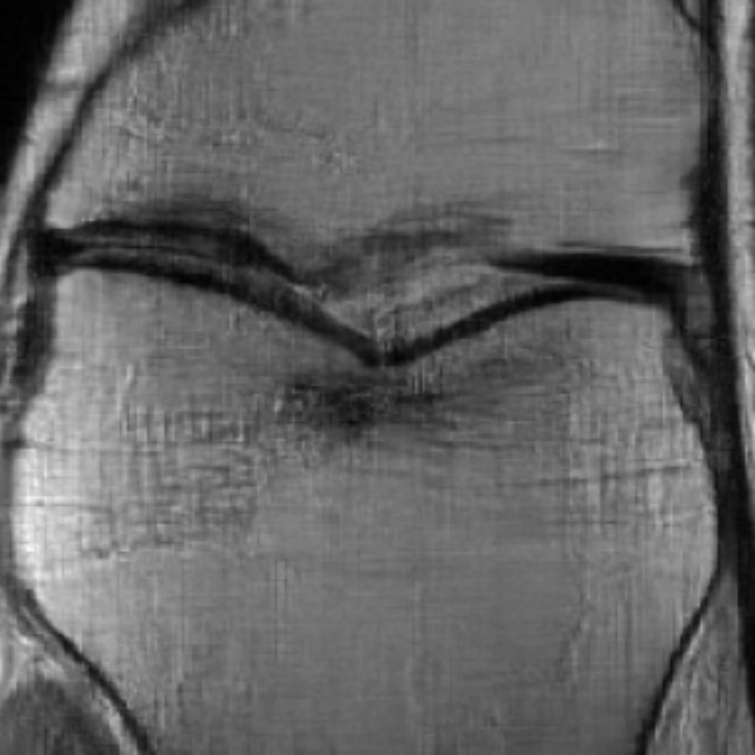}
  \end{subfigure}
  \begin{subfigure}[t]{0.16\textwidth}
  \centering\includegraphics[scale=0.35]{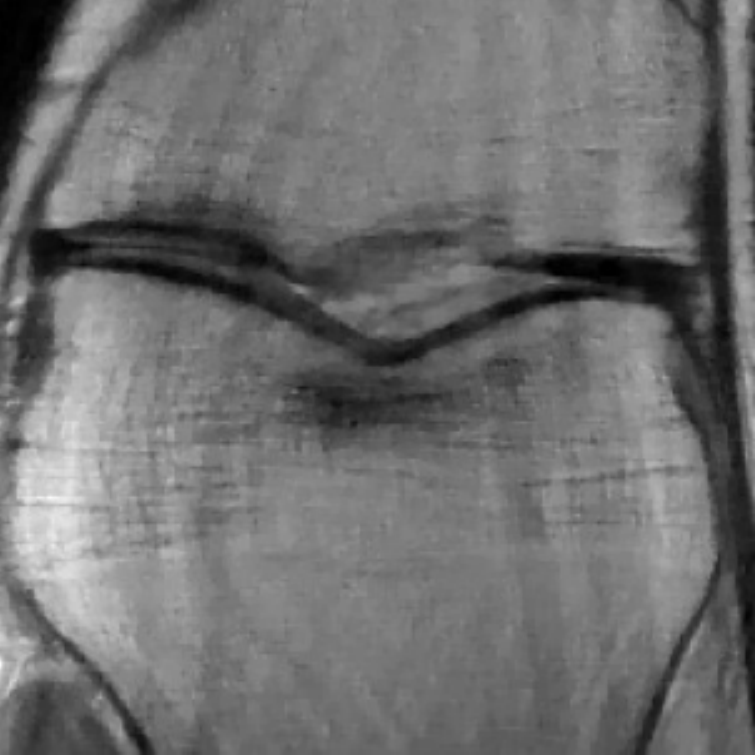}
  \end{subfigure}
  \begin{subfigure}[t]{0.16\textwidth}
  \centering\includegraphics[scale=0.35]{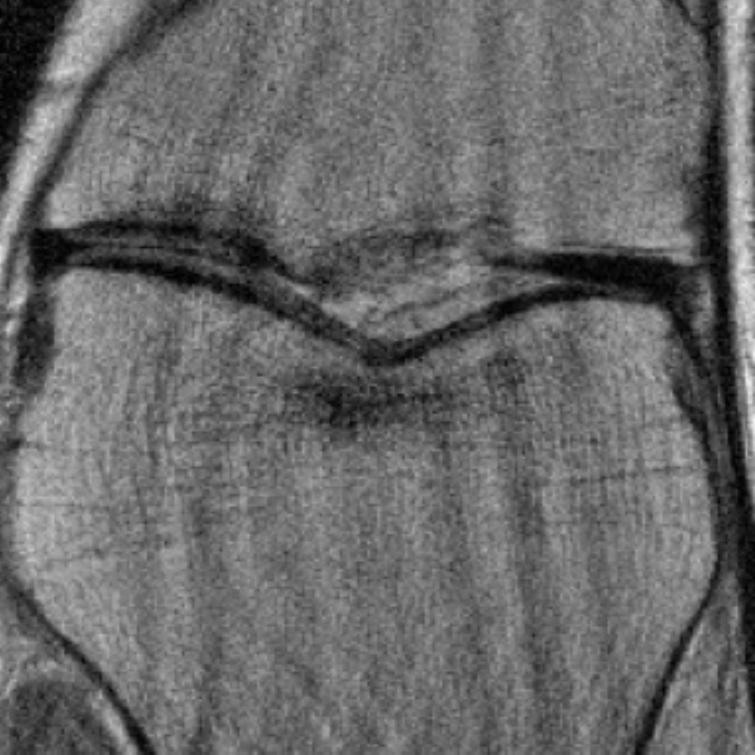}
  \end{subfigure}
  \begin{subfigure}[t]{0.16\textwidth}
  \centering\includegraphics[scale=0.35]{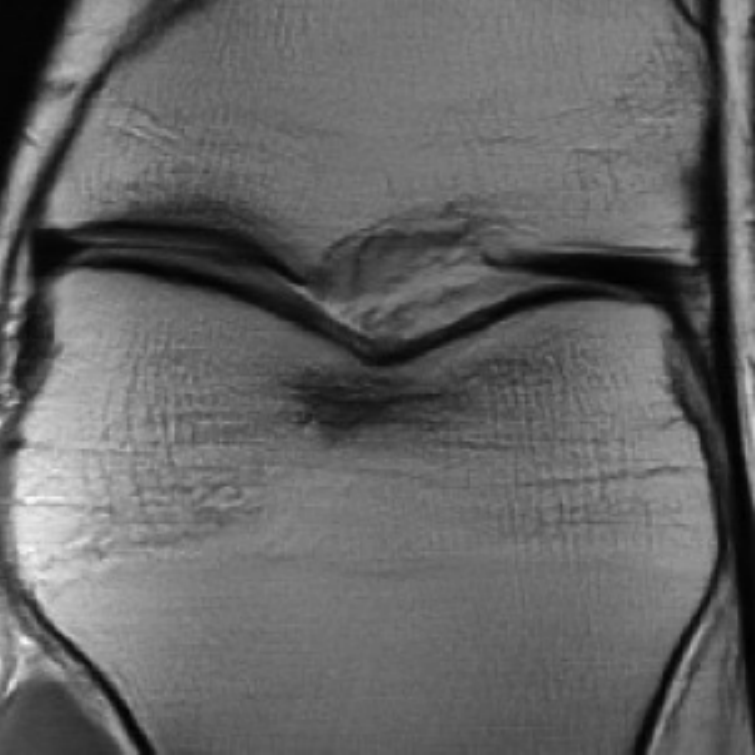}
  \end{subfigure}
  \begin{subfigure}[t]{0.16\textwidth}
  \centering\includegraphics[scale=0.35]{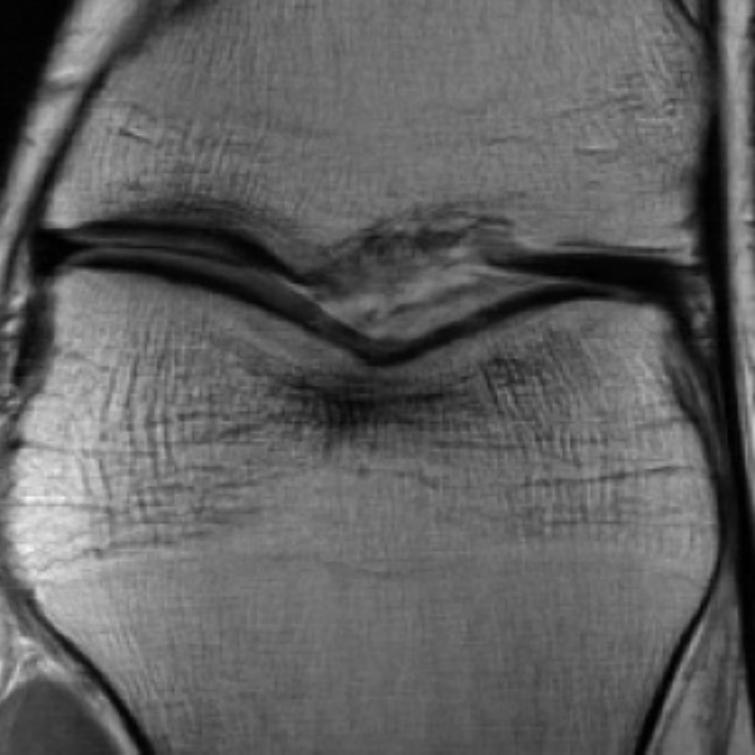}
  \end{subfigure}
  \begin{subfigure}[t]{0.16\textwidth}
  \centering\includegraphics[scale=0.35]{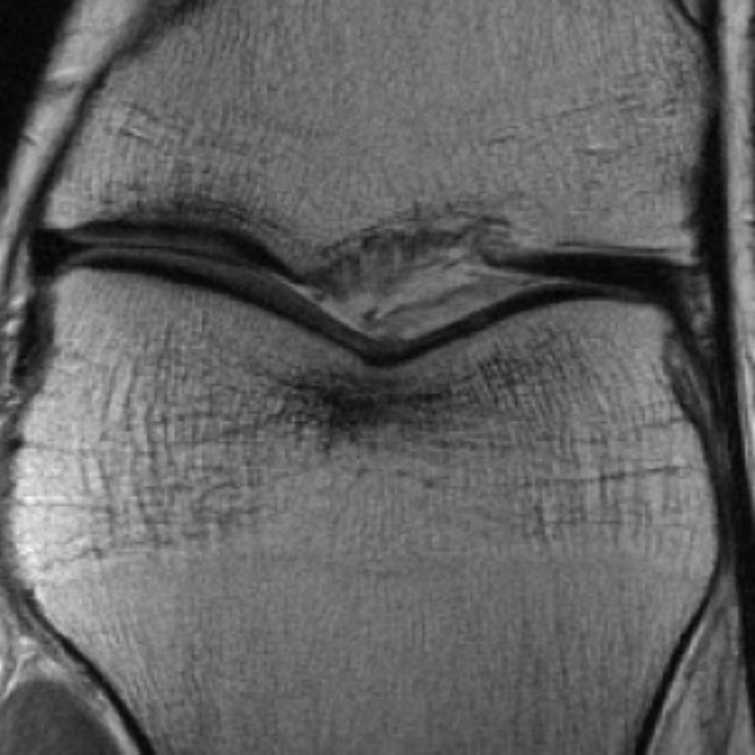}
  \end{subfigure}
\caption{Sample reconstructions for ConvDecoder, TV, ENLIVE, U-net, and the end-to-end variational network (VarNet) for a validation image from multi-coil knee measurements (4x accelerated). The bottom row represents zoomed-in version of the top row. ConvDecoder and the end-to-end variational network (VarNet) find the best reconstructions for this image (slightly better than U-net and significantly better than TV and ENLIVE).
}
\label{fig:baselines}
\end{figure*}

Table~\ref{tab:baselines} shows the results for each method. The scores show that the ConvDecoder has higher VIF performance than U-net and VarNet (recall that VIF is deemed most relevant by clinicians), and achieves on-par performance with U-net on all other metrics. Moreover, it significantly outperforms TV and ENLIVE. Figure~\ref{fig:baselines} shows a sample reconstruction for an image from the validation set for the mentioned methods. As we also noted before, the state-of-the-art VarNet outperforms both ConvDecoder and the U-net on the other metrics.

\section{Computational cost and making un-trained networks 10x faster}\label{sec:runtime}

We next show that un-trained neural networks are relatively slow, but in order to mitigate that, we propose a domain-specific initialization that accelerates the method by a factor of 10.
Table~\ref{tab:runtime} shows runtimes for testing each method on a single validation image, averaged over 10 runs.
All runtime values denote CPU clock times and the experiments were run on a single TITAN V GPU.
As those numbers demonstrate, and as is well known, un-trained methods, and in particular un-trained networks, are significantly slower than trained neural networks. That is because un-trained methods solve an optimization problem via an iterative algorithm, whereas neural networks only require a forward pass through the network.

\begin{table}[h]
\centering
\begin{adjustbox}{width=0.45\textwidth}
\begin{tabular}{|c|c|c|c|}
\hline 
Method & test & train & tune\\
\hline
    ConvDecoder & 63.5 $\pm$ 0.3  minutes & -       & 4 days     \\
    DIP         & 119.7 $\pm$ 0.2 minutes & -       & 8 days    \\
    DD          & 56 $\pm$ 0.2    minutes & -       & 4 days     \\
    U-net       & 0.2 $\pm$ 0.1   seconds & 8 days  & -          \\
    VarNet      & 0.7 $\pm$ 0.3   seconds & 2 weeks & -          \\
    TV          & 20 $\pm$ 3      seconds & -       & 5 minutes  \\
    ENLIVE      & 230 $\pm$ 10    seconds & -       & 60 minutes \\
\hline
\end{tabular}
\end{adjustbox}
\caption{
Runtimes when applying each method to a validation knee image (a single slice). The numbers are averaged over 10 runs and marginal errors are for the 95\% confidence interval.
}
\label{tab:runtime}
\end{table}

We next propose an approach to accelerate un-trained networks by a factor of 10.
It is well known that for iterative optimization methods (e.g., gradient descent), the initialization affects the time it takes to get a solution of desired accuracy. 
Our acceleration approach is based on starting from a good initialization. 

Specifically, we exploit that MRI knee images fall into two categories that are statistically similar: 
fat-suppressed and non-fat-suppressed images.
Images in each class share statistical properties such as contrast, pixel-wise histogram, etc. 
It turns out that a decoder optimized to represent (or reconstruct) a non-fat-suppressed image provides a good initialization for other non-fat-suppressed images.

\begin{figure*}
\captionsetup[subfigure]{justification=centering}
\centering
  \begin{subfigure}[t]{0.22\textwidth}
  \caption*{ground truth \vspace{5pt}}
  \centering\includegraphics[scale=0.38]{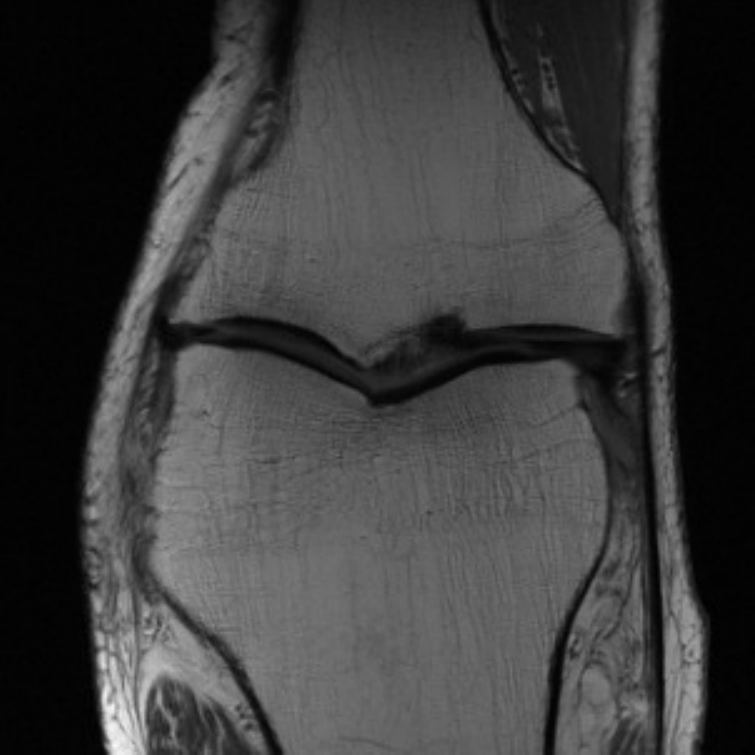}
  \end{subfigure}
  \begin{subfigure}[t]{0.22\textwidth}
  \caption*{ConvDecoder \\[-2pt] (slow) \vspace{-3pt}}
  \centering\includegraphics[scale=0.38]{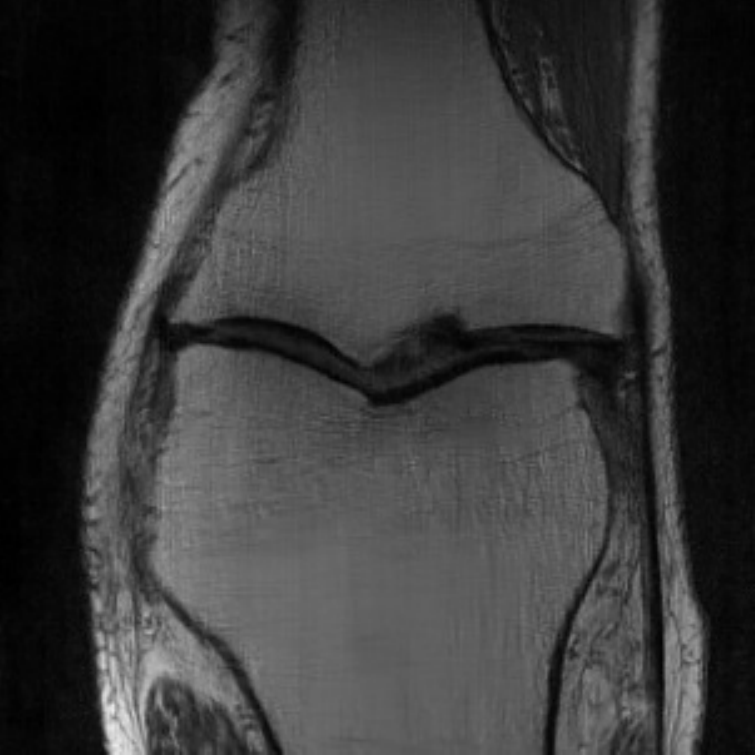}
  \end{subfigure}
  \begin{subfigure}[t]{0.22\textwidth}
  \caption*{ConvDecoder \\[-2pt] (10x faster - random init) \vspace{-3pt}}
  \centering\includegraphics[scale=0.38]{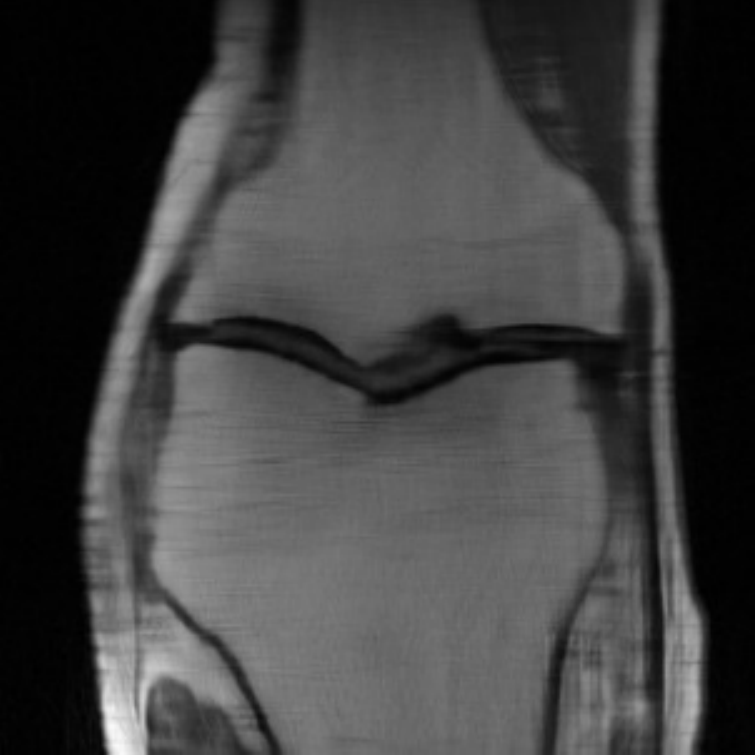}
  \end{subfigure}
  \begin{subfigure}[t]{0.22\textwidth}
  \caption*{ConvDecoder \\[-2pt] (10x faster - guided init) \vspace{-3pt}}
  \centering\includegraphics[scale=0.38]{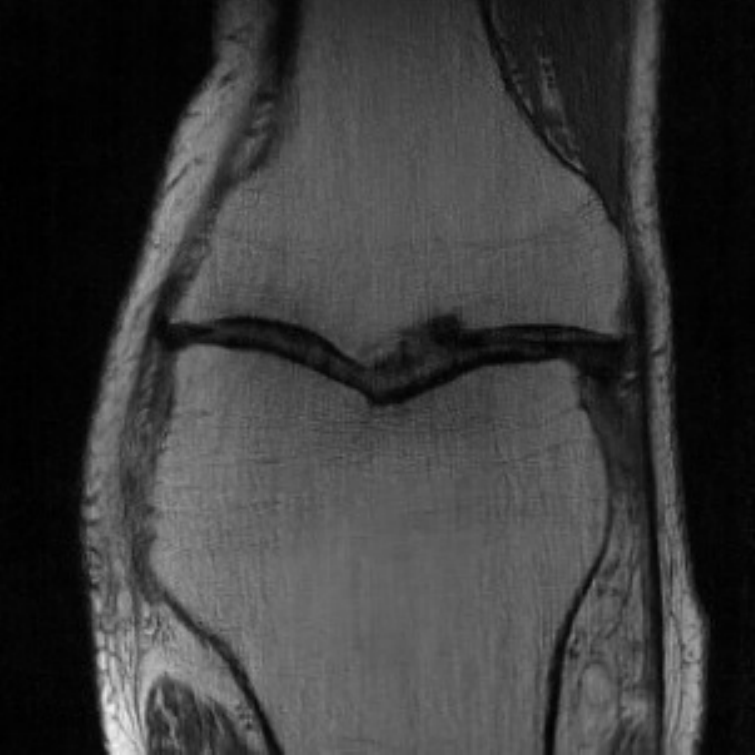}
  \end{subfigure}\par\medskip 
  \begin{subfigure}[t]{0.22\textwidth}
  \centering\includegraphics[scale=0.38]{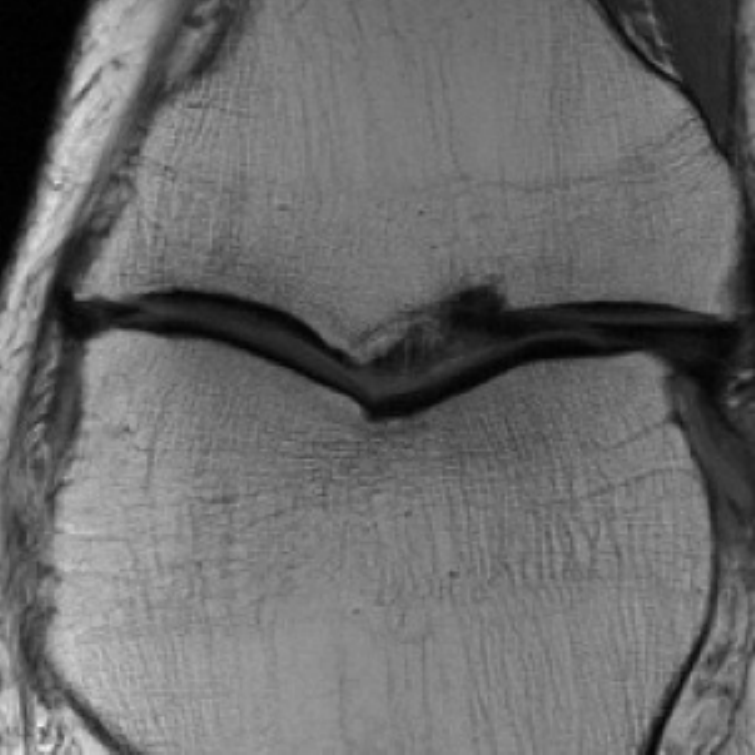}
  \end{subfigure}
  \begin{subfigure}[t]{0.22\textwidth}
  \centering\includegraphics[scale=0.38]{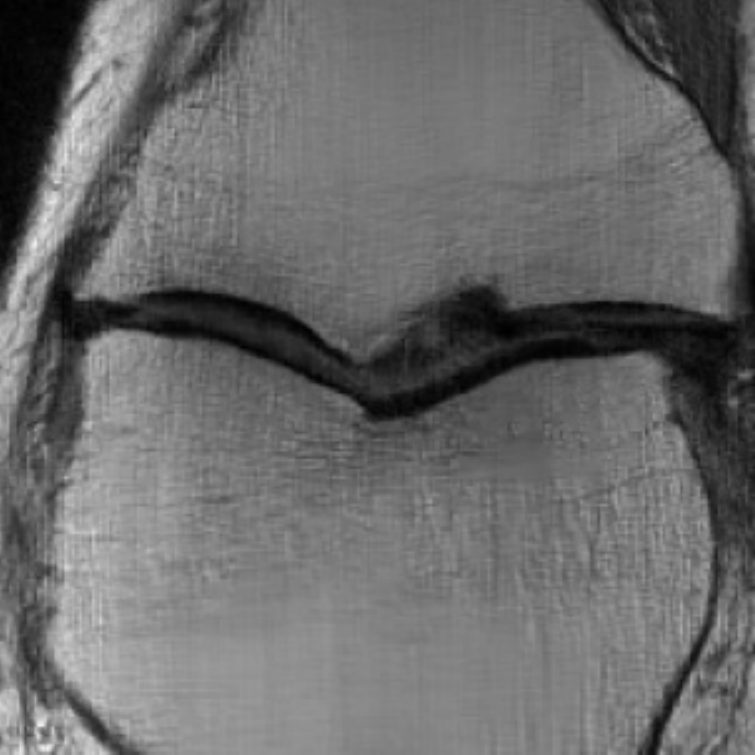}
  \end{subfigure}
  \begin{subfigure}[t]{0.22\textwidth}
  \centering\includegraphics[scale=0.38]{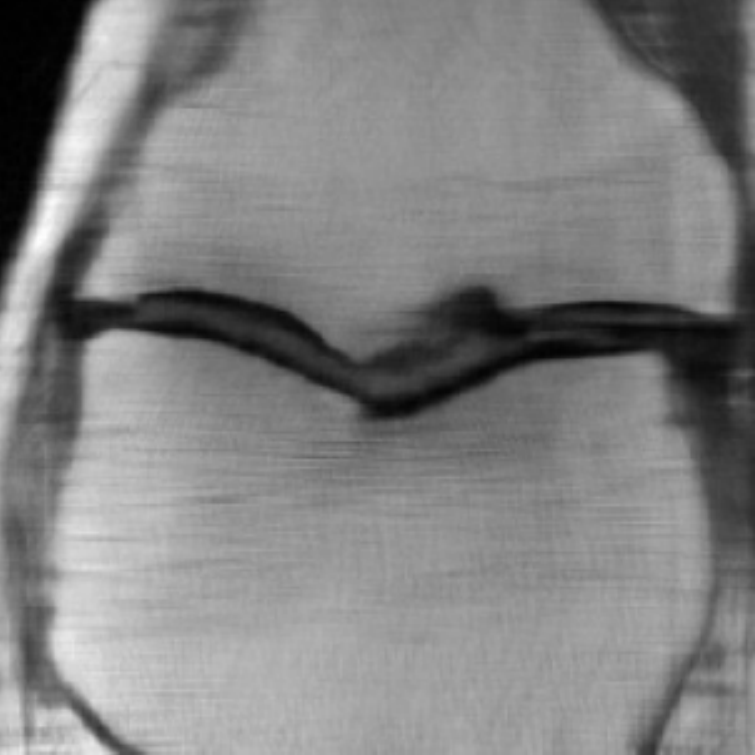}
  \end{subfigure}
  \begin{subfigure}[t]{0.22\textwidth}
  \centering\includegraphics[scale=0.38]{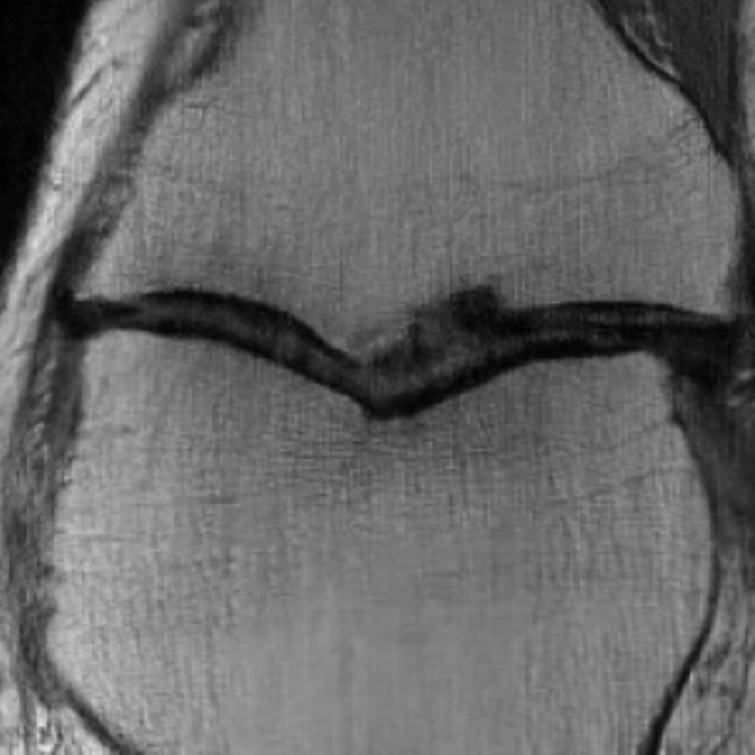}
  \end{subfigure}
\caption{
Guided initialization results in the same reconstruction accuracy as full convergence from random initialization. ``10x faster - random init'' denotes fitting ConvDecoder from random initialization, but for 10 times less number of full convergence iterations. 
}
\label{fig:fast}
\end{figure*}

Accordingly, we first select one non-fat-suppressed knee slice randomly and fit ConvDecoder (based on loss function \eqAutoref{eq:loss}) until full convergence to reconstruct that image from its 4x under-sampled measurements. As a result, we obtain a set of parameters $\mC_1$ for the decoder. 
We then use $\mC_1$ as an initialization to reconstruct other non-fat-suppressed images using ConvDecoder. Table~\ref{tab:fast} and Figure~\ref{fig:fast} show that this approach enables achieving the same reconstruction accuracy in 10 times less iterations. Thus, 60 minutes in Table~\ref{tab:runtime} reduces to 6 minutes at the cost of no performance loss. This result shows that interestingly, only one sample gives sufficient information to represent a specific category of knee images.

We hasten to add that the idea of initializing generative models through training has been used before with the goal of enhancing reconstruction quality, see \citep{kohli2020semantic,hussein2020image} for examples where a generative model is initialized by training on some training set. 


\begin{table}[h]
\centering
\begin{adjustbox}{width=0.7\textwidth}
\begin{tabular}{|c|c|c|c|c|}
\hline 
Method & VIF & MS-SSIM & SSIM & PSNR \\
\hline
    ConvDecoder & \textbf{0.8449 $\pm$ 0.0108} & \textbf{0.9612 $\pm$ 0.0032} & 0.8244 $\pm$ 0.0097 & 32.94 $\pm$ 0.28\\
    ConvDecoder-Fast & 0.8317 $\pm$ 0.0115 & 0.9598 $\pm$ 0.0036 & \textbf{0.8297 $\pm$ 0.0092} & \textbf{33.05 $\pm$ 0.26}\\
\hline
\end{tabular}
\end{adjustbox}
\caption{
Guided domain-specific initialization results in achieving the same reconstruction accuracy, yet 10x faster. Scores are averaged over all 100 non-fat-suppressed mid-slice images from the fastMRI validation set for 4x acceleration. Marginal errors denote 95\% confidence interval.
}
\label{tab:fast}
\end{table}

\section{Better performance at the cost of more computations}\label{sec:better-performance}


According to Section \ref{sec:knee4x}, ConvDecoder significantly outperforms TV and achieves performance close to U-net. 
In this section, we study an approach to obtain an even better performance with ConvDecoder. 
Consider an under-sampled measurement $\vy$ and $k$-many ConvDecoders with the same hyper-parameters, but initialized independently. 
After fitting each decoder to $\vy$, we average the resulting $k$ reconstructed images. This, shown to be effective for denoising in~\citep{ulyanov2018deep}, also results in a higher reconstruction accuracy in our setup, in terms of SSIM and PSNR scores, but makes very little visible difference.

Table~\ref{tab:do-better} shows PSNR scores when using this averaging technique. The ConvDecoder's output is obtained through 20 runs based on the mentioned averaging technique. Also, the numbers are averaged over 18 random images (9 proton-density and 9 fat suppressed) from the 4x accelerated knee measurements from the fastMRI validation set. According to Table~\ref{tab:do-better}, ConvDecoder marginally outperforms U-net and achieves performance close to the end-to-end variational network.

How many runs are enough to achieve this higher performance? For a given image, Figure~\ref{fig:do-better} shows how PSNR changes based on the number of ConvDecoders whose outputs are averaged together to form the final reconstruction (the numbers are averaged over the same 18 images as in Table~\ref{tab:do-better}).
Even averaging over the output of two decoders has a noticeable impact (more than 0.5 dB) on the reconstruction quality. Note that the improvement rate decreases as we increase the number of decoders. A sample reconstruction of our ensemble trick is included in the supplementary material. The sample reconstruction shows that the visible performance increase over the ConvDecoder without the averaging technique is small.

\begin{table}[ht]
\begin{minipage}[b]{0.5\linewidth}
\centering
\begin{adjustbox}{width=0.65\textwidth}
\tiny 
\begin{tabular}{|c|c|c|}
\hline 
Method & SSIM & PSNR \\
\hline
    ConvDecoder-A & 0.8127 & 32.72 \\ 
    ConvDecoder   & 0.7946 & 32.13 \\
    U-net         & 0.8094 & 32.71\\
    VarNet        & \textbf{0.8365} & \textbf{34.50}\\
    TV            & 0.7012 & 30.35\\
    ENLIVE        & 0.5763 & 26.72 \\
\hline
\end{tabular}
\end{adjustbox}
\caption{Using our averaging technique, ConvDecoder's performance (ConvDecoder-A) exceeds U-net's and becomes close to the performance of the state-of-the-art reconstruction method, the end-to-end variational network (VarNet).
}
\label{tab:do-better}
\end{minipage}\hfill
\begin{minipage}[b]{0.47\linewidth}
\centering
\begin{tikzpicture}
\begin{groupplot}[
y tick label style={/pgf/number format/.cd,fixed,precision=4},
scaled y ticks = false, 
         group
         style={group size= 1 by 1, xlabels at=edge bottom, ylabels at=edge left,
         xticklabels at=edge bottom,
         horizontal sep=1.5cm, vertical sep=0.8cm,
         }, 
           try min ticks=4,
         width=0.75\linewidth,height=0.45\linewidth,
         ]
\nextgroupplot[title={},ylabel={PSNR},xlabel={\#ConvDecoders}]
	\addplot [mark=*,steelblue] table[x index=0,y index=1]{./files/do_better.csv};
\end{groupplot}
\end{tikzpicture}
\captionsetup{type=figure}

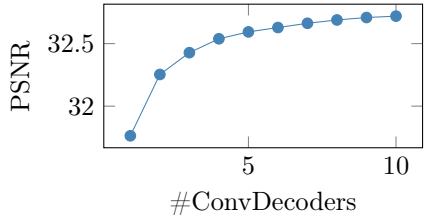
\captionof{figure}{Averaging the output of $k \in \{1,2,...10\}$ ConvDecoders monotonically enhances the PSNR score for the resulting image.}
\label{fig:do-better}
\end{minipage}
\end{table}


\section{Robustness to out-of-distribution samples}\label{sec:distribution_shift}


In Section \ref{sec:performance}, we found that ConvDecoder achieves on-par performance with U-net when evaluated on in-distribution data (i.e., fastMRI validation set). 
This is a setup where trained methods shine because the test and train distribution match perfectly. However, this might not reflect performance in practice, where it is difficult to impossible to train on a distribution that matches the test distribution perfectly. 
Un-trained networks only mildly rely on the training distribution through hyper-parameter tuning and we therefore expect them to potentially be less sensitive to a distribution shift. 

In this section, we first show that, perhaps surprisingly, even un-trained methods suffer from distribution shifts through parameter tuning. However, we demonstrate that for un-trained methods, this can be mitigated (i) either through tuning on a few images on the test domain, if available, or (ii) through an auto parameter tuning approach introduced here.  
With either technique, un-trained networks become robust to distribution, and even perform on par with the best trained method, VarNet, when evaluated under distribution shifts. 
In practice we suggest to choose option (i) if it is possible to obtain a very small tuning dataset from the test domain, and the auto-tuning technique (ii) otherwise.

\subsection{Performance loss under distribution shifts}\label{sec:dshift-no-data}

We start by studying the effect of a distribution shift on reconstruction performance.
Even un-trained methods are sensitive to such distribution shifts, because its hyper-parameters are tuned on the training distribution.

We consider a distribution shift from knee to brain images, i.e., we train/tune on knees and test on brains. 
Specifically, we train the U-net and VarNet models on both knee and brain images, one model for each set of images. 
Moreover, we tune the un-trained methods on 10 randomly-chosen knee images, which results in an 8-layer ConvDecoder with 256 channels per layer. 
We also tune the un-trained methods on  10 randomly-chosen brain images, which results in a 5-layer ConvDecoder with 64 channels per layer.

Next, we evaluate those methods on 100 randomly-chosen mid-slice brain images from the fastMRI validation set. Table~\ref{tab:dist-shift-no-data} demonstrates that both trained and un-trained methods suffer a similar performance loss under this distribution shift.

\begin{table}[h]
\centering
\begin{adjustbox}{width=0.7\textwidth}
\begin{tabular}{|c|cc|cc|}
\hline 
& \multicolumn{2}{c|}{\textbf{train/tune on knee}} & \multicolumn{2}{c|}{\textbf{train/tune on brain}}\\
& \multicolumn{2}{c|}{\textbf{test on brain}} & \multicolumn{2}{c|}{\textbf{test on brain}} \\
 Method & PSNR & SSIM & PSNR & SSIM \\
\hline
    ConvDecoder & 31.75 $\pm$ 0.22 & 0.8589 $\pm$ 0.0056 & 34.46 $\pm$ 0.13 & 0.9097 $\pm$ 0.0049 \\
    U-net          & 31.21 $\pm$ 0.14 & 0.8842 $\pm$ 0.0043 & 34.29 $\pm$ 0.10 & 0.9173 $\pm$ 0.0036 \\
    VarNet         & \textbf{33.96 $\pm$ 0.11} & \textbf{0.9067 $\pm$ 0.0039} & \textbf{37.46 $\pm$ 0.07} & \textbf{0.9421 $\pm$ 0.0031}\\
    TV             & 26.02 $\pm$ 0.48 & 0.7659 $\pm$ 0.0078 & 27.10 $\pm$ 0.57 & 0.7846 $\pm$ 0.0082 \\
    ENLIVE         & 24.86 $\pm$ 0.42 & 0.7534 $\pm$ 0.0081 & 26.51 $\pm$ 0.55 & 0.7780 $\pm$ 0.0087 \\
\hline
\end{tabular}
\end{adjustbox}
\caption{Validation results for a set of 100 mid-slice brain images from the fastMRI validation set. All methods suffer a significant performance loss under a distribution shift.
\textbf{Left.} PSNR and SSIM scores when evaluating on the brain validation set (training U-net and VarNet as well as tuning ConvDecoder, TV, and ENLIVE is done on the knee training set).
\textbf{Right.} PSNR and SSIM scores when evaluating on the brain validation set (training U-net and VarNet as well as tuning ConvDecoder, TV, and ENLIVE is done on the brain training set). Marginal errors denote 95\% confidence intervals.}
\label{tab:dist-shift-no-data}
\end{table}

\subsection{Mitigating performance loss under distribution shifts with meta learning}\label{sec:dshift-with-data}

A straightforward approach to eliminate the performance loss of un-trained methods under distribution shifts is to record a few ground-truth samples from the test domain for hyper-parameter tuning. To illustrate this approach for the distribution shift from knee to brain images, we tune un-trained methods on 10 randomly-chosen brain images. 

The results in Table~\ref{tab:dist-shift-data} show that un-trained methods are robust against a distribution shift provided that a few samples from the test domain are available for meta learning (i.e., tuning their hyper-parameters). In this setup, our un-trained network achieves on-par performance with the state-of-the art VarNet and it outperforms the baseline U-net.

\begin{table}[h]
\centering
\begin{adjustbox}{width=0.7\textwidth}
\begin{tabular}{|c|cc|cc|}
\hline 
& \multicolumn{2}{c|}{\textbf{train on knee}} & \multicolumn{2}{c|}{\textbf{train on brain}}\\
& \multicolumn{2}{c|}{\textbf{test on brain}} & \multicolumn{2}{c|}{\textbf{test on brain}} \\
 Method & PSNR & SSIM & PSNR & SSIM \\
\hline
    ConvDecoder-SM & \textbf{34.46 $\pm$ 0.13} & \textbf{0.9097 $\pm$ 0.0049} & 34.46 $\pm$ 0.13 & 0.9097 $\pm$ 0.0049 \\
    U-net          & 31.21 $\pm$ 0.14 & 0.8842 $\pm$ 0.0043 & 34.29 $\pm$ 0.10 & 0.9173 $\pm$ 0.0036 \\
    VarNet         & 33.96 $\pm$ 0.11 & 0.9067 $\pm$ 0.0039 & \textbf{37.46 $\pm$ 0.07} & \textbf{0.9421 $\pm$ 0.0031}\\
    TV             & 27.10 $\pm$ 0.57 & 0.7846 $\pm$ 0.0082 & 27.10 $\pm$ 0.57 & 0.7846 $\pm$ 0.0082 \\
    ENLIVE         & 26.51 $\pm$ 0.55 & 0.7780 $\pm$ 0.0087 & 26.51 $\pm$ 0.55 & 0.7780 $\pm$ 0.0087 \\
\hline
\end{tabular}
\end{adjustbox}
\caption{Validation results for a set of 100 mid-slice brain images from the fastMRI validation set. Trained neural networks suffer a significant performance loss under a distribution shift.
\textbf{Left.} PSNR and SSIM scores for evaluating on the brain validation set (training is done on the knee training set only for VarNet and U-net).
\textbf{Right.} PSNR and SSIM scores when evaluating on the brain validation set (training is done on the brain training set only for VarNet and U-net). Marginal errors denote 95\% confidence interval.}
\label{tab:dist-shift-data}
\end{table}

\subsection{
Mitigating performance loss under distribution shifts with auto-tuning}\label{sec:autotune}

Next, we consider again the setup in Section~\ref{sec:dshift-no-data} where no data points from the test distribution are available for parameter tuning. 
Here, we show that it is possible to auto-tune the parameters of the un-trained neural networks to adjust their hyper-parameters for a given reconstruction problem. 

The idea of our new auto-tuning approach is to measure how well ConvDecoder with a set of hyper-parameters interpolates missing information in the $k$-space. 
Specifically, given an under-sampled measurement $\vy$, we first randomly remove (i.e., set to zero) a fraction $q$ of the given $k$-space measurements, denoted by the hold-out set $\mathcal{S}$. The measurement vector without those frequencies is denoted by $\vy_{-\mathcal{S}}$. 
Then, for a specific hyper-parameter configuration $h$, we reconstruct an image with ConvDecoder based on the measurement $\vy_{-\mathcal{S}}$. 
The Fourier representation of the reconstructed image is denoted by $\hat \vy$. We then compute the MSE between $\vy$ and $\hat \vy$ over $\mathcal{S}$, the set of hold-out values in the $k$-space. 
The resulting number is a good proxy on how well the given ConvDecoder configuration performs on all the missing frequency elements, even those not in the hold-out set $\mc S$. 

At the cost of computation, this procedure might be performed $k$ times for each parameter setup, similar to performing $k$-fold cross validation.
The described auto-tuning framework selects a set of hyper-parameters whenever a new sample appears, therefore it does not rely on any training or tuning data except the given sample itself. 
As mentioned before, this approach is computationally expensive as it increases the computational cost of reconstructing an image by a factor of $|\mathcal H| \cdot k$, where $k$ is the number of folds and $|\mathcal H|$ is the number of hyper-parameter configurations. We set $k=2$ in our experiment and also use any combination from $\#$layers = $\{5,8\}$, $\#$channels = $\{64,256\}$, and sens = $\{0,1\}$ for the hyper-parameters (sens = 1 denotes incorporating coil sensitivity estimates in the optimization as shown in loss function~\eqref{eq:optim4}).


We next evaluate the performance of ConvDecoder with the auto-tuning approach under the distribution shift from knee to brain. 
We compare: (i) tuning ConvDecoder on a few brain images, which is the best-performing option but relies on training data (cf. Section~\ref{sec:dshift-with-data}), (ii) simply applying knee-tuned ConvDecoder to brain images (cf.  Section~\ref{sec:dshift-no-data}), and (iii) using our auto-tuning approach.
Table~\ref{tab:autotune} shows that our auto-tuning approach significantly mitigates the performance loss due to distribution shifts, yet still incurs a slight performance loss through this distribution shift when relative to tuning on the test distribution.


\begin{table}[h]
\centering
\begin{adjustbox}{width=0.6\textwidth}
\begin{tabular}{|c|cc|}
\hline 
& \multicolumn{2}{c|}{\textbf{Score}} \\
 Setup & PSNR & SSIM \\
\hline
    Tune on brain - test on brain & \textbf{34.46 $\pm$ 0.13} & \textbf{0.9097 $\pm$ 0.0049} \\
    Tune on knee - test on brain & 31.75 $\pm$ 0.22 & 0.8589 $\pm$ 0.0056 \\
    Auto-tune - test on brain & 33.54 $\pm$ 0.16 & 0.8944 $\pm$ 0.0052  \\
\hline
\end{tabular}
\end{adjustbox}
\caption{Validation results for a group of 100 mid-slice brain images from the fastMRI validation set. Un-trained methods lose performance through a restricted distribution shift. Our auto-tuning framework significantly improves over naive usage of hyper-parameters obtained from the original domain. Marginal errors denote 95\% confidence interval.}
\label{tab:autotune}
\end{table}

\section{Discussion and Conclusion}\label{sec:conclusion}
MRI scans are typically accelerated by under-sampling the measurements, and classically the images are reconstructed with an un-trained method by solving an optimization problem such as $\ell_1$-minimization or total-variation minimization. Those classical algorithms are significantly outperformed by deep learning  approaches that learn to reconstruct an image from measurements. Thus, those learning based approaches enable higher accelerations. However, concerns have been raised over the stability of those methods and the ability to generalize to different scanners and distributions of images. For example, if trained on knees, a neural network might perform poorly on brains as demonstrated here.

In this paper we studied un-trained neural networks (from both reconstruction accuracy and reconstruction speed aspects) for accelerated MRI reconstruction and proposed an architecture that is most suitable for MRI. We find that this architecture significantly outperforms other un-trained methods--including traditional CS methods--and has image reconstruction performance close to that of trained neural networks, but without using any training data. We further demonstrated that under a distribution shift, un-trained networks suffer a similar performance than trained methods, but this loss can be mitigated through (i) meta learning with a small tuning dataset, or (ii) auto-tuning without any access to the test distribution. 

While our results show that the best trained neural networks still slightly outperform our un-trained methods if sufficient training data is available, the (i) robustness to distribution shift and (ii) not needing any training data (except a few images for meta learning) make un-trained networks an important tool in practice, especially in regimes with a lack of training data, and a need for robustness.

\section*{Reproducibility}

The code to reproduce all results in this paper is available at 
\href{https://github.com/MLI-lab/ConvDecoder}
{https://github.com/MLI-lab/ConvDecoder} and an online demonstration that uses ConvDecoder for reconstructing a sample is available at \href{https://drive.google.com/file/d/1xu_NS6ClikkOM1TTPL7EDqOjQZCvCvlL/view?usp=sharing}
{colab-ConvDecoder}.

\section*{Acknowledgment}\label{sec:ack}

M. Zalbagi Darestani and R. Heckel are (partially) supported by NSF award IIS-1816986, and R. Heckel acknowledges support of the IAS at TUM and the DFG.

The authors would like to thank Zalan Fabian for sharing his finding that averaging the outputs of multiple runs of deep decoders improves denoising performance, which led us to study the performance of averaging multiple ConvDecoder outputs (i.e., Section \ref{sec:better-performance}). The authors would also like to thank Lena Heidemann for helpful discussions and for running a grid search on the brain images for the Deep Decoder (i.e., Table~\ref{tab:params} in the supplementary material).



\printbibliography

\newpage
\noindent{\huge APPENDIX\par}
\appendix

\section{Details on evaluating the reconstruction performance}\label{sec:evaluation_dilemma_sup}

Evaluating the performance of different reconstruction methods is challenging for reasons related to the 
(i) choice of image comparison metrics, 
(ii) impact of the normalization of images,
(iii) the fact that we often compare to a noisy ground truth image, and 
(iv) because we can compare image wise or volume wise.
In this section, we further iterate points ii and iii.

\paragraph*{Normalization}
As for the image normalization method, which is typically required to fairly compare two images, we investigated three image normalization methods: min-max normalization, which transforms image $I$ to $\frac{I-\mathrm{min}(I)}{\mathrm{max}(I)-\mathrm{min}(I)}$; mean-std normalization on both ground-truth and reconstructed images, which transforms image $I$ to $\frac{I-\mathrm{mean}(I)}{\mathrm{std}(I)}$; and mean-std normalization which is only applied to the ground-truth image to match its histogram to that of the reconstructed image.

\begin{figure}[h!]
\begin{center}
\begin{tikzpicture}

\begin{groupplot}[xtick style={draw=none},y tick label style={/pgf/number format/.cd,fixed,precision=1},x tick label style={/pgf/number format/.cd,fixed,precision=5},
scaled y ticks = false,
legend style={at={(1,1)} , nodes={scale=0.6}, 
/tikz/every even column/.append style={column sep=-0.1cm}
 },grid=both, grid style={white},axis background/.style={fill=gray!12},
         group
         style={group size= 4 by 1, xlabels at=edge bottom,
         yticklabels at=edge left,
         horizontal sep=0.6cm, vertical sep=1.4cm,
         }, 
         width=0.3\textwidth,height=0.25\textwidth,
         ymin = 0,
         minor y tick num=1,
         ]
\nextgroupplot[title=without normalization,scaled x ticks = false,]
	\addplot[ybar,ybar legend,bar width=8pt,draw=none, fill=steelblue,opacity=1] 
  table[x index=0,y index=1] {./files/hists.csv};
  
  \addplot[ybar,ybar legend,bar width=8pt, draw = none, fill=amber,opacity=0.6] 
  table[x index=2,y index=3] {./files/hists.csv};
  
  \legend{ground truth, reconstruction}

\nextgroupplot[title style={yshift=-0.4ex,},title=mean-std (ground truth),scaled x ticks = false,]
	\addplot[ybar,ybar legend,bar width=8pt,draw=none, fill=steelblue,opacity=1] 
  table[x index=4,y index=5] {./files/hists.csv};
  
  \addplot[ybar,ybar legend,bar width=8pt, draw = none, fill=amber,opacity=0.6] 
  table[x index=6,y index=7] {./files/hists.csv};
  
  \legend{ground truth, reconstruction}
  
\nextgroupplot[title style={yshift=-0.4ex,},title=mean-std (both),scaled x ticks = false,]
	\addplot[ybar,ybar legend,bar width=8pt,draw=none, fill=steelblue,opacity=1] 
  table[x index=8,y index=9] {./files/hists.csv};
  
  \addplot[ybar,ybar legend,bar width=8pt, draw = none, fill=amber,opacity=0.6] 
  table[x index=10,y index=11] {./files/hists.csv};
  
  \legend{ground truth, reconstruction}
  
\nextgroupplot[title=min-max,scaled x ticks = false,]
	\addplot[ybar,ybar legend,bar width=8pt,draw=none, fill=steelblue,opacity=1] 
  table[x index=12,y index=13] {./files/hists.csv};
  
  \addplot[ybar,ybar legend,bar width=8pt, draw = none, fill=amber,opacity=0.6] 
  table[x index=14,y index=15] {./files/hists.csv};
  
  \legend{ground truth, reconstruction}

\end{groupplot}
\end{tikzpicture}
\caption{The effect of different image normalization techniques on the distribution of ground truth and reconstructed images.
}
\label{fig:distributions}
\end{center}
\end{figure}
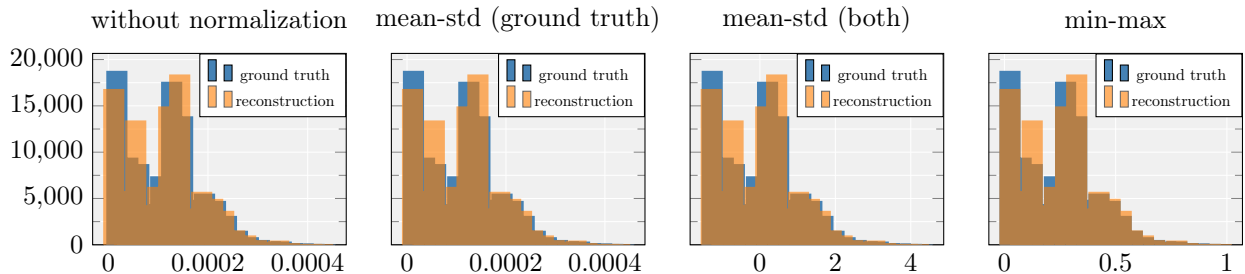

\begin{table}[h]
\centering
\begin{adjustbox}{width=0.8\textwidth}
\footnotesize
\begin{tabular}{ccccccccc}
\toprule 
Method & \multicolumn{2}{c}{no norm} & \multicolumn{2}{c}{min-max} & \multicolumn{2}{c}{mean-std (both))} & \multicolumn{2}{c}{mean-std (gt)}\\
& SSIM & PSNR & SSIM & PSNR & SSIM & PSNR & SSIM & PSNR \\
\midrule
    ConvDecoder & 0.7563 & 29.93 & \textbf{0.7464} & \textbf{30.09} & 0.6957 & \textbf{29.57} & 0.7753 & 31.67 \\
    U-net & \textbf{0.7867} & \textbf{32.01} & 0.7354 & 28.04 & \textbf{0.7091} & 29.37 & \textbf{0.7883} & \textbf{32.04}\\
    TV & 0.6592 & 27.21 & 0.6875 & 26.67 & 0.6565 & 28.43 & 0.6977 & 30.20\\
\bottomrule
\end{tabular}
\end{adjustbox}
\caption{Average image-based scores for the ConvDecoder, U-net, and TV on the mid-slice images of the volumes in the multi-coil knee measurements from the fastMRI validation set (4x accelerated). Scores are computed based on different image normalization methods. Surprisingly, different image normalization types result in a very different ranking of the reconstruction methods.}
\label{tab:norms}
\end{table}

Figure~\ref{fig:distributions} illustrates how each of the mentioned normalization methods affects the distribution of ground-truth and reconstructed images for a sample file from the multi-coil knee dataset. 
In addition, Table~\ref{tab:norms} shows the average SSIM and PSNR scores for ConvDecoder, Unet, and TV after running them on 200 mid-slice images from the multi-coil knee dataset (4x accelerated). 
The remarkable outcome of these results is that different image normalization methods can result in a totally different winning reconstruction method.

\begin{figure}
\centering
  \begin{subfigure}[t]{0.2\textwidth}
  \caption*{ground truth}
  \centering\includegraphics[scale=0.35]{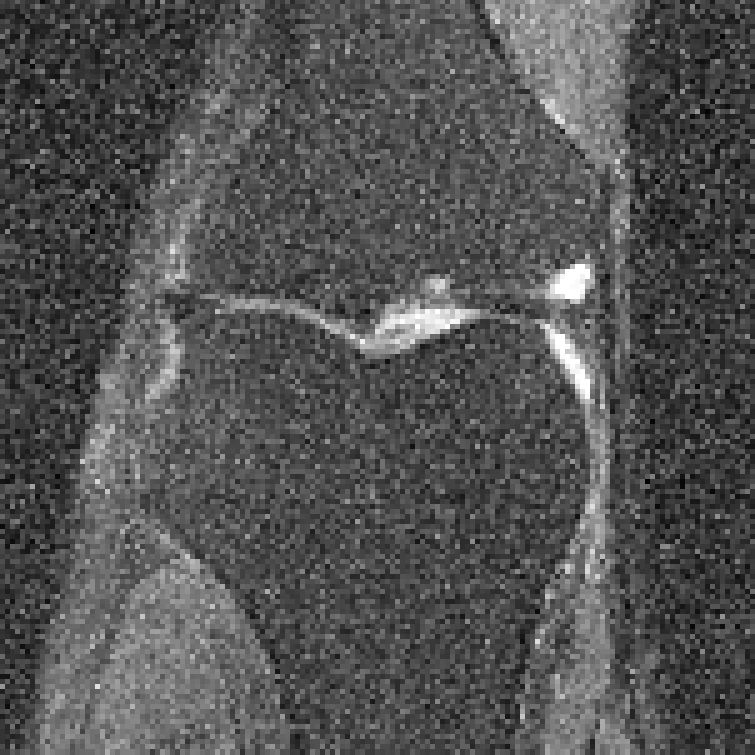}
  \end{subfigure}
  \begin{subfigure}[t]{0.2\textwidth}
  \caption*{reconstruction}
  \centering\includegraphics[scale=0.35]{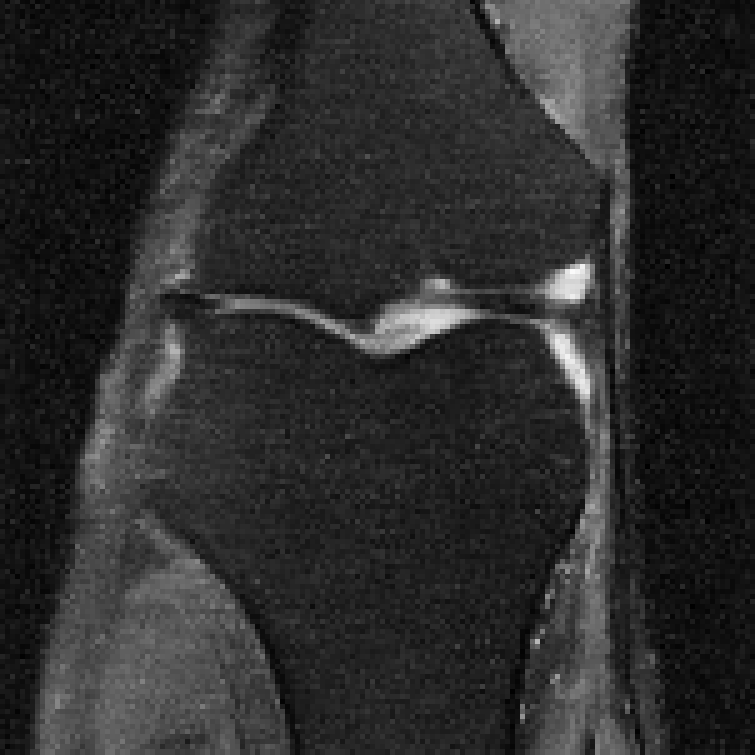}
  \end{subfigure}
\caption{A sample image from the multi-coil knee dataset where the ground-truth image is very noisy. 
In the evaluation, we compare a reconstructed image to such a noisy image, and even if the reconstruction is a subjectively sharp and clear image, the error metric is large.}
\label{fig:artifact}
\end{figure}

\paragraph*{Comparison to noisy ground truth}
As mentioned, in some cased the ground-truth image itself is corrupted with measurements images, as illustrated in Figure~\ref{fig:artifact} where the ground truth image is very noisy.
In such cases, the scores may not reflect the true quality of the reconstructed image. Note that the reconstruction is almost free from noise, demonstrating the image prior also denoises the image.

\paragraph{Volume-based vs. image-based evaluation} To illustrate the point that volume- and image-based evaluations give very different numbers, Table~\ref{tab:volume-image} provides average volume-based as well as image-based SSIM and PSNR scores for ConvDecoder, U-net, the end-to-end variational network, and TV. Note that the results are averaged over 20 randomly-chosen volumes (640 slice images in total) from the validation set. The numbers show that the image-based score computation results in a lower range of numbers compared with the volume-based one, yet we employ the former because we think it better reflects the performance. In any case, the ranking of algorithms is typically roughly preserved when transitioning from image- to volume-based evaluation. 

\begin{table}[h]
\centering
\begin{adjustbox}{width=0.5\textwidth}
\footnotesize
\begin{tabular}{ccccc}
\toprule
Method & \multicolumn{2}{c}{Volume-based} & \multicolumn{2}{c}{Image-based} \\
& SSIM & PSNR & SSIM & PSNR \\
\midrule
     ConvDecoder & 0.8713 & 35.61 & 0.7868 & 31.81 \\
     U-net & 0.8753 & 35.68 & 0.7992 & 32.14 \\
     VarNet & \textbf{0.9062} & \textbf{38.21} & \textbf{0.8432} & \textbf{34.10} \\
     TV & 0.7214 & 33.21 & 0.6832 & 30.12 \\
\bottomrule
\end{tabular}
\end{adjustbox}
\caption{Volume-based vs. image-based score compution leads to different numbers:
Average scores for ConvDecoder, U-net, the end-to-end variational network (VarNet), and TV. For comparing volume- vs. image-based evaluation, SSIM and PSNR scores are averaged over 20 randomly-chosen volumes (640 slices).}
\label{tab:volume-image}
\end{table}

\section{Parameter setup and optimization}\label{sec:setup}

We tuned the parameters of ConvDecoder, DIP, and Deep Decoder by performing a grid search over a group of 10 randomly-chosen training images. We also tried different architectures (e.g., adding skip connections, or stacking multiple decoders to form a multi-resolution network) but found the plain ConvDecoder to perform best.

Table~\ref{tab:params} shows the grid parameters for each network.
For the U-net and the end-to-end variational network (VarNet), we chose the set of parameters used in the fastMRI challenge\footnote{\href{https://github.com/facebookresearch/fastMRI/tree/master/models}{\color{black}https://github.com/facebookresearch/fastMRI/tree/master/models}} \citep{zbontar2018fastmri} and trained them accordingly. 
The parameter setup is provided in Table~\ref{tab:params}. 



\begin{table}[h]
\centering
\begin{adjustbox}{width=1\textwidth}
\footnotesize
\begin{threeparttable}
\begin{tabular}{cccccccc}
\toprule
Method & \#layers & \#channels & convolutional  & \#pools & \#sens-pools &\#sens-channels & \#cascades\\
       &          & (or width) & kernel size    &         &              &                &           \\
\midrule
    ConvDecoder & \{4, 5**, 6, 7, 8*, 9\}  & \{32,64**,128, 160, 256*, 480\} & 3 & 0  & - & - & - \\
    U-net       & 8  & 32\tnote{***}  & 3 & 4 & - & - & - \\
    Varnet & -  & 18 & 3 & 4 & 4 & 8 & 12 \\
    DIP         & \{10, 12**, 14, 16*, 18\} & \{64**,160, 256*, 360\} & 3 & 2  & - & - & -\\
    DD          & \{6, 7, 8, 9, 10$^{*,**}$,11\} & \{64**,128,256, 368*, 512\} & 1 & 0  & - & - & -\\
\bottomrule
\end{tabular}
\begin{tablenotes}
\item[*] {\footnotesize{Knee chosen parameters.}}
\item[**] {\footnotesize{Brain chosen parameters.}}
\item[***] {\footnotesize{This is the number of channels for the first layer of U-net. For the 8-layer U-net that we used, the number of channels are [32, 64, 128, 256, 512, 256, 128, 64, 32] including a non-pooling layer in the middle.}}
\end{tablenotes}
\end{threeparttable}
\end{adjustbox}
\caption{Model parameters for ConvDecoder, U-net, Varnet, DIP, and Deep Decoder (DD).
}
\label{tab:params}
\end{table}


In order to fit the un-trained methods to the under-sampled measurements, we used the Adam optimizer~\citep{kingma2014adam} with constant stepsize 0.01 (without early stopping) for optimizing the loss function (which is MSE loss function in our experiments). 

Regarding the output dimension of un-trained methods, as an example, for a $15\times640\times368$ (15 is the number of coils) under-sampled measurement, ConvDecoder (also DIP or DD) generates an image of size $30\times640\times368$, because it recovers the real and complex pixel values of an image separately with two separate channels. 
Finally, we fixed the input of the un-trained methods which is sampled from $\mathcal{N}(0,I)$ and has dimension $256\times10\times5$ (256 being the number of channels).

\subsection{Sensitivity to initialization and choice of hyper-parameters}\label{sec:hyper-params}

We next discuss (i) how the width of the network affects the reconstruction quality, and (ii) demonstrate that there is little variance in the scores given a specific setup when fitting the ConvDecoder starting from a random initialization to a given under-sampled measurement over multiple runs on the same problem.

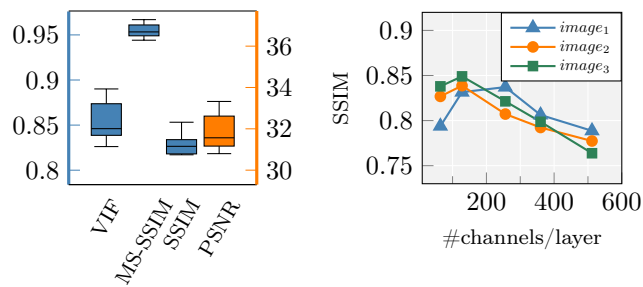
\begin{figure}[h!]
\begin{center}
\begin{tikzpicture}

\begin{groupplot}[
y tick label style={/pgf/number format/.cd,fixed,precision=2},
scaled y ticks = false,
legend style={at={(1,1)} , nodes={scale=0.6}, 
/tikz/every even column/.append style={column sep=-0.1cm}
 },
         group
         style={group size= 2 by 1, xlabels at=edge bottom,
         xticklabels at=edge bottom,
         horizontal sep=2.2cm, vertical sep=0.4cm,
         }, 
         width=0.26\textwidth,height=0.275\textwidth,
         ]
\nextgroupplot[boxplot/draw direction = y,every y tick/.style={steelblue},xtick = {1, 2, 3},xticklabels = {\footnotesize VIF,\footnotesize MS-SSIM,\footnotesize SSIM},xtick style = {draw=none},axis x line* = bottom, axis y line* = left, enlarge y limits, xticklabel style = {align=center, font=\small, rotate=60}, separate axis lines, first y axis line style={very thick,steelblue}, scale only axis, x = 0.5cm, y=12cm, xmin=0, xmax=5,ymin=0.8,ymax=0.96,]
        \addplot+[boxplot, fill=steelblue, draw=black] table[y index=0] {./files/var.csv};
		\addplot+[boxplot, fill=steelblue, draw=black] table[y index=1] {./files/var.csv};
		\addplot+[boxplot, fill=steelblue, draw=black] table[y index=2] {./files/var.csv};

\nextgroupplot[ylabel={\footnotesize SSIM},xlabel={\footnotesize \#channels/layer}, xtick pos=bottom, minor x tick num=1, ytick pos=left, xmax =600, ymin=0.73, ymax=0.92,grid=both, grid style={white},axis background/.style={fill=gray!10},x=0.045mm,y=12cm,]
	\addplot +[mark=triangle*,mark options={solid,fill=steelblue,scale=1.3},steelblue,thick] table[x index=0,y index= 1]{./files/width.csv};
	\addlegendentry{$image_1$}
	\addplot +[mark=*,mark options={solid,fill=amber,scale=0.9},amber,thick] table[x index=0,y index=2]{./files/width.csv};
	\addlegendentry{$image_2$}
	\addplot +[mark=square*,mark options={solid,fill=darkspringgreen!90,scale=0.8},darkspringgreen!90,thick] table[x index=0,y index=3]{./files/width.csv};
	\addlegendentry{$image_3$}
\end{groupplot}

\begin{groupplot}[
y tick label style={/pgf/number format/.cd,fixed,precision=1},
scaled y ticks = false,
legend style={at={(1,0.35)} , 
/tikz/every even column/.append style={column sep=-0.1cm}
 },
         group
         style={group size= 2 by 1, xlabels at=edge bottom,
         xticklabels at=edge bottom,
         horizontal sep=1.6cm, vertical sep=0.4cm,
         }, 
         width=0.5\textwidth,height=0.44\textwidth,
         ]
\nextgroupplot[boxplot/draw direction = y,every y tick/.style={amber},xtick = {1},xticklabels = {\footnotesize PSNR},xtick style = {draw=none}, axis y line* = right, enlarge y limits, xticklabel style = {align=center, font=\small, rotate=60}, separate axis lines, second y axis line style={very thick,amber}, scale only axis, x = 0.5cm, y=0.275cm, xmin=-3, xmax=2,ymin=30,ymax=37,]
        \addplot+[boxplot, fill=amber, draw=black] table[y index=3] {./files/var.csv};
	
\end{groupplot}

\end{tikzpicture}
\caption{Effect of hyper-parameters on the ConvDecoder's performance. \textbf{Left}: fluctuations of different scores during 10 runs on a single data point. \textbf{Right}: effect of network width on the SSIM score for three data points from the validation set.}
\label{fig:hyper-params}
\end{center}
\end{figure}

A key hyper-parameter of the ConvDecoder is the width of the network (the number of channels per layer). 
In order to check how different wideness factors affect the performance, we ran a seven-layer ConvDecoder on three under-sampled measurements (again from the multi-coil accelerated knee measurements of the fastMRI dataset) and computed the SSIM score. We performed this experiment four times to average the results. 
Figure~\ref{fig:hyper-params} (right) shows the SSIM score based on network width for the three mentioned data points. It can be seen that if the network width is either too small or too large, it does not perform well. A width parameter around 200 performs well across images.

Recall that to recover an image, we run gradient descent starting from a random initialization. It is natural to ask whether the reconstruction quality varies significantly as a function of the random initialization. 
We find that the reconstruction quality is relatively insensitive to the particular initialization. Specifically, for the general setup we used in Section \ref{sec:setup}, we ran the ConvDecoder 10 times on an under-sampled measurement and averaged the scores. 
Figure~\ref{fig:hyper-params} (left) depicts the variances of different scores over several runs of the algorithm, and illustrates that the scores vary relatively mildly (VIF as well as PSNR and MS-SSIM tend to have the highest and lowest variations, respectively).

\section{How does ConvDecoder represent an image?}\label{sec:how-it-works}

Despite the notable empirical success of un-trained neural networks for solving inverse problems, there is still little knowledge about why these methods work so well in practice. 
In this section, we illustrate \emph{how} un-trained networks functions as image priors. 
Specifically we demonstrate (i) how different layers play a role in forming successively higher resolution versions of an image and (ii) how different layers are fitted in the optimization, and why this matters.

\subsection{Successive approximation of an image}
\label{sec:resolution-enhancement}

To understand the role of different layers in reconstructing the image, it is instructive to see how an un-trained network generates an image by visualizing the outputs of each layer. 
Visualizing the layers' 256 channels, however, is not informative.
Instead, we visualize the best representation that can be achieved by linearly combining the channels in each layer to the re-scaled ground-truth image.
For example, if the image size (omitting the number of channels) in layer $i$'s output is $(w_i,h_i)$, then we down-sample the ground-truth image to match this size.

\begin{figure*}
\centering
  \begin{subfigure}[t]{0.12\textwidth}
  \centering\includegraphics[scale=0.28]{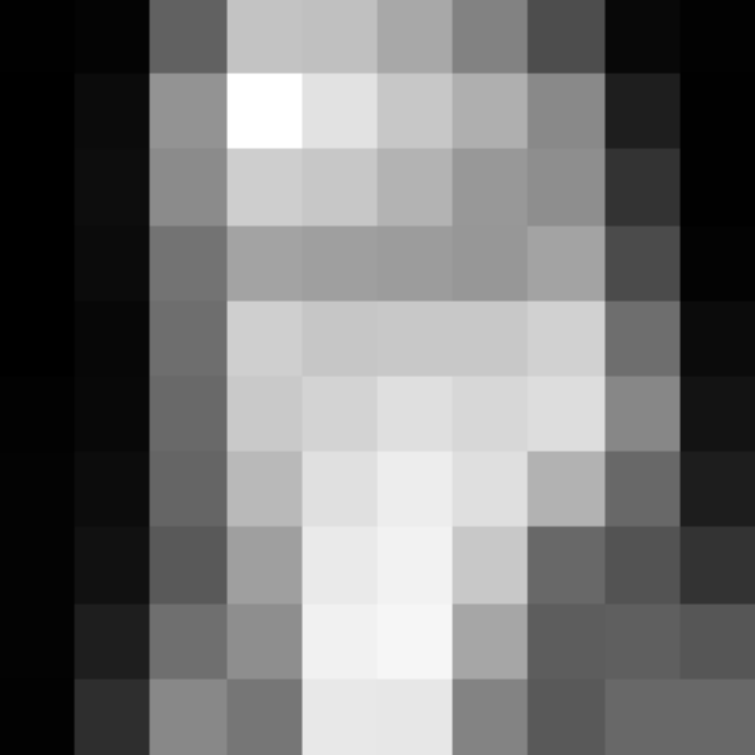}
  \end{subfigure}
  \begin{subfigure}[t]{0.12\textwidth}
  \centering\includegraphics[scale=0.28]{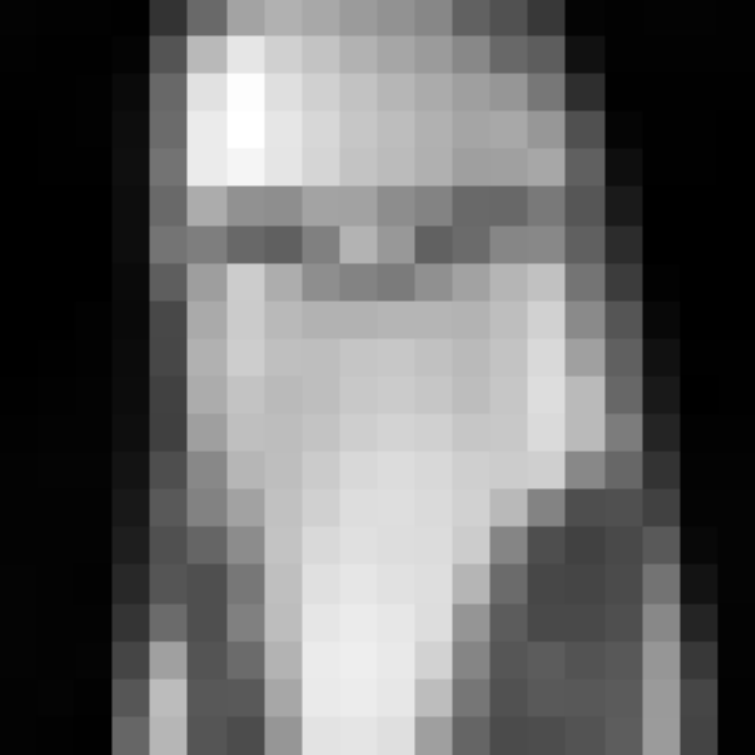}
  \end{subfigure}
  \begin{subfigure}[t]{0.12\textwidth}
  \centering\includegraphics[scale=0.28]{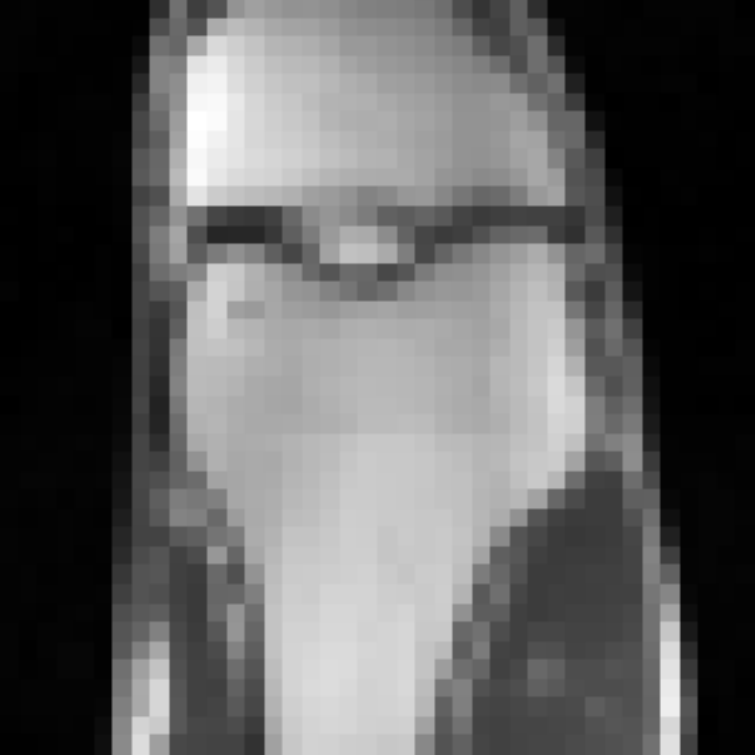}
  \end{subfigure}
  \begin{subfigure}[t]{0.12\textwidth}
  \centering\includegraphics[scale=0.28]{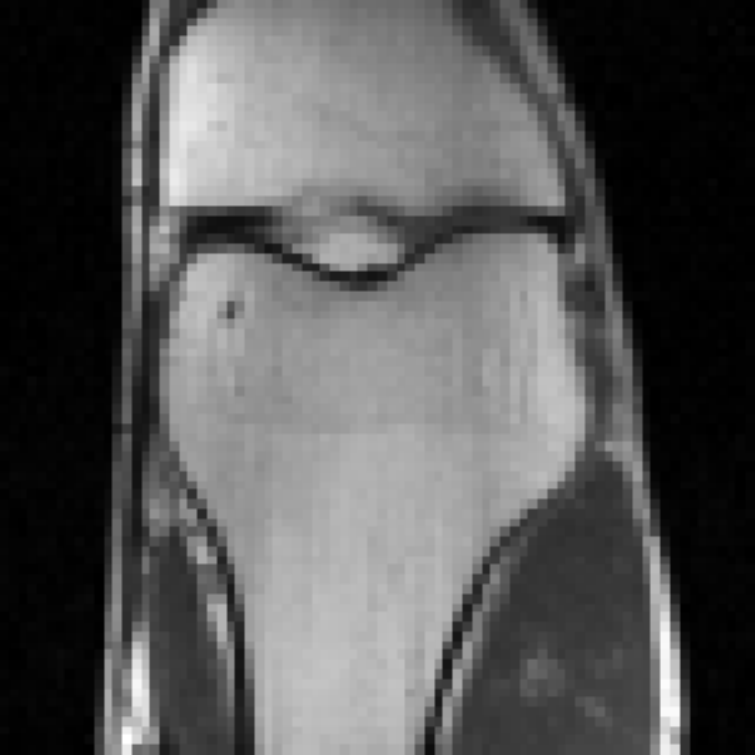}
  \end{subfigure}
  \begin{subfigure}[t]{0.12\textwidth}
  \centering\includegraphics[scale=0.28]{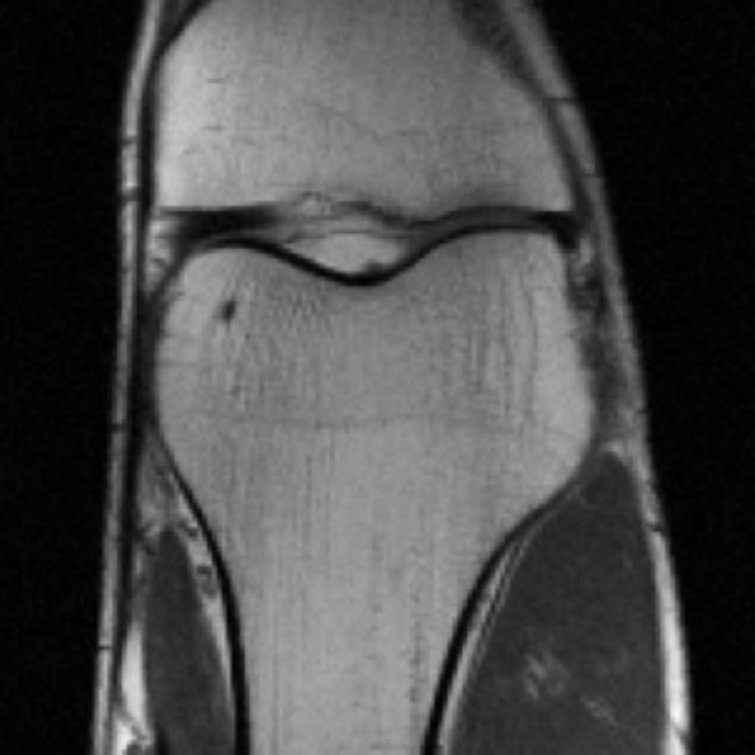}
  \end{subfigure}
  \begin{subfigure}[t]{0.12\textwidth}
  \centering\includegraphics[scale=0.28]{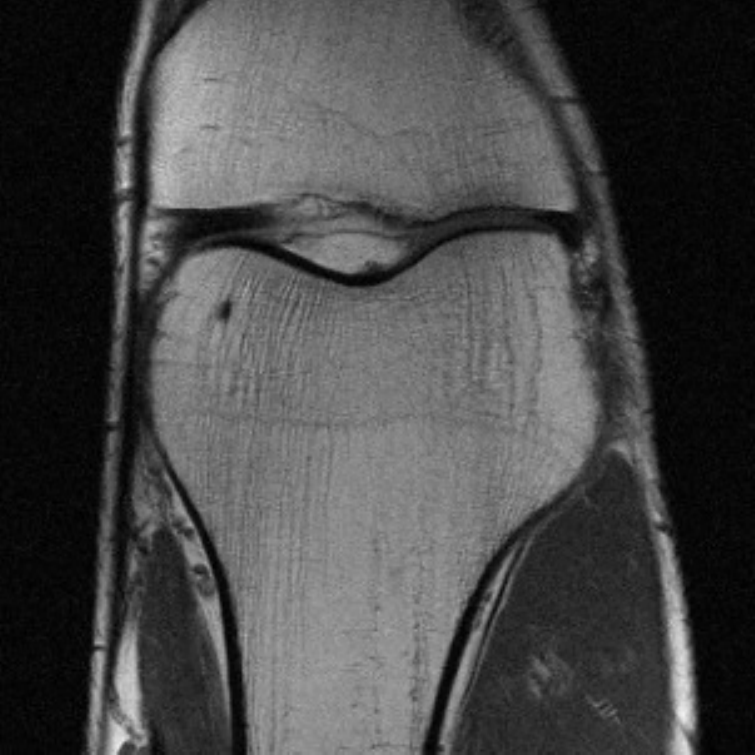}
  \end{subfigure}
  \begin{subfigure}[t]{0.12\textwidth}
  \centering\includegraphics[scale=0.28]{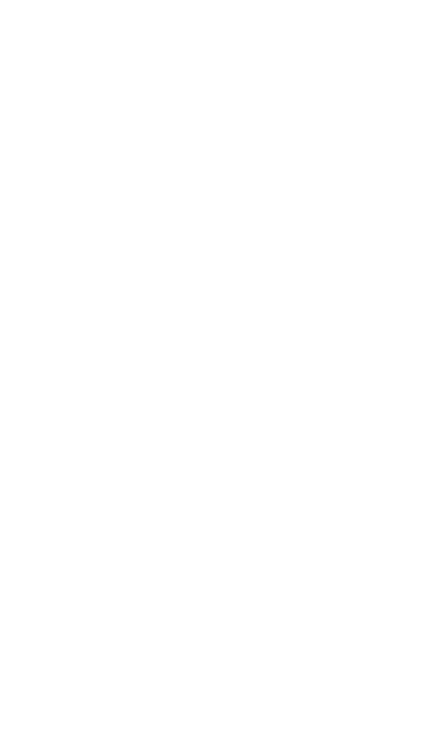}
  \end{subfigure}\par\medskip 
  \begin{subfigure}[t]{0.12\textwidth}
  \centering\includegraphics[scale=0.28]{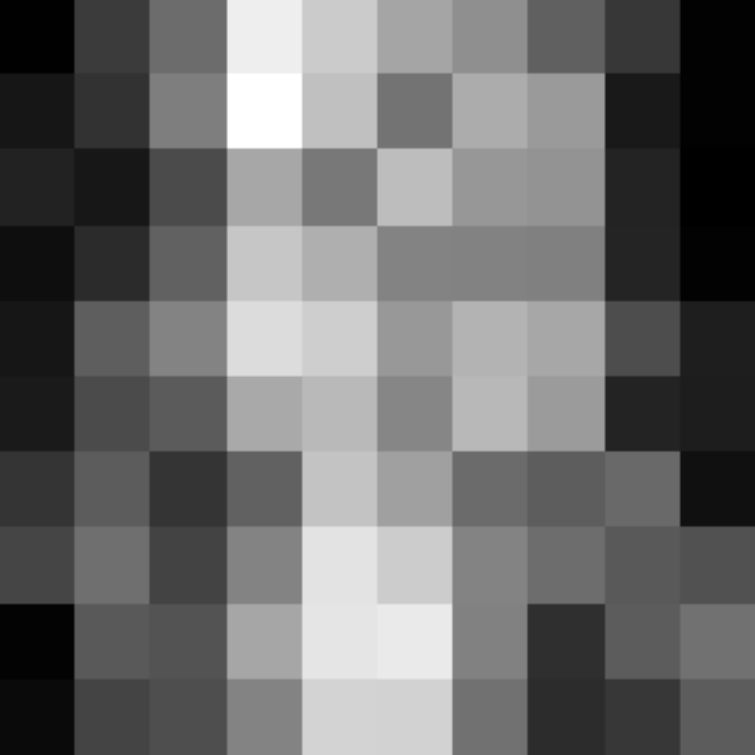}
  \caption*{(20,11)}
  \end{subfigure}
  \begin{subfigure}[t]{0.12\textwidth}
  \centering\includegraphics[scale=0.28]{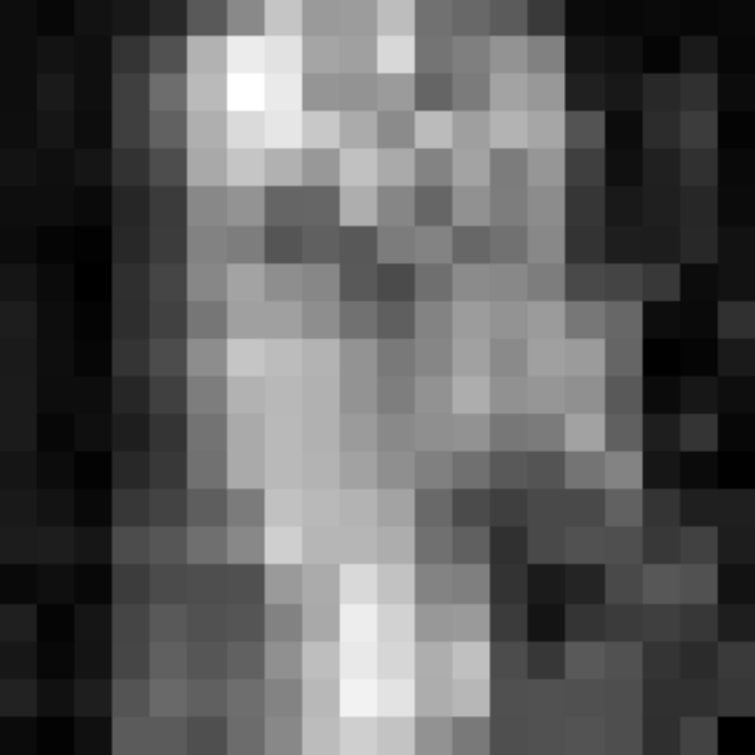}
  \caption*{(40,21)}
  \end{subfigure}
  \begin{subfigure}[t]{0.12\textwidth}
  \centering\includegraphics[scale=0.28]{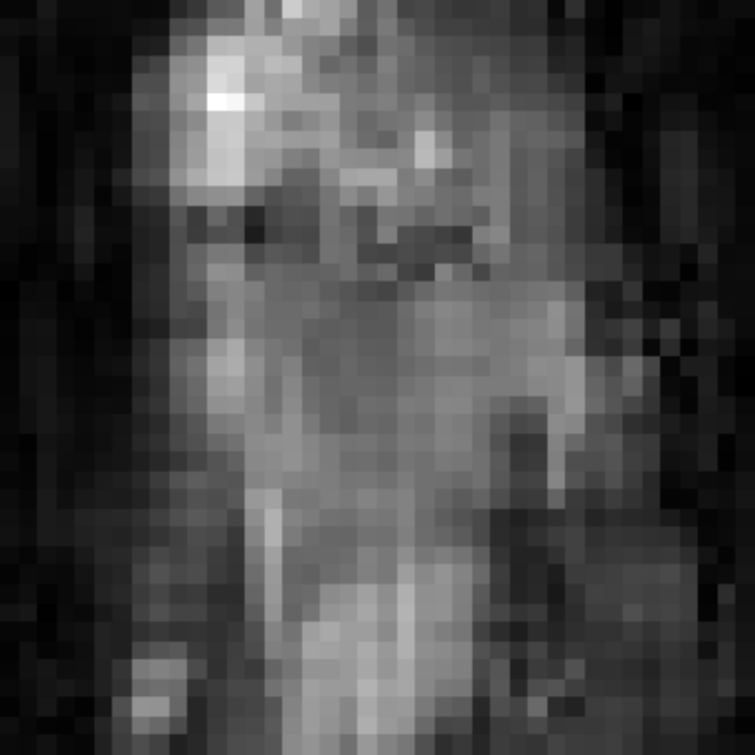}
  \caption*{(80,43)}
  \end{subfigure}
  \begin{subfigure}[t]{0.12\textwidth}
  \centering\includegraphics[scale=0.28]{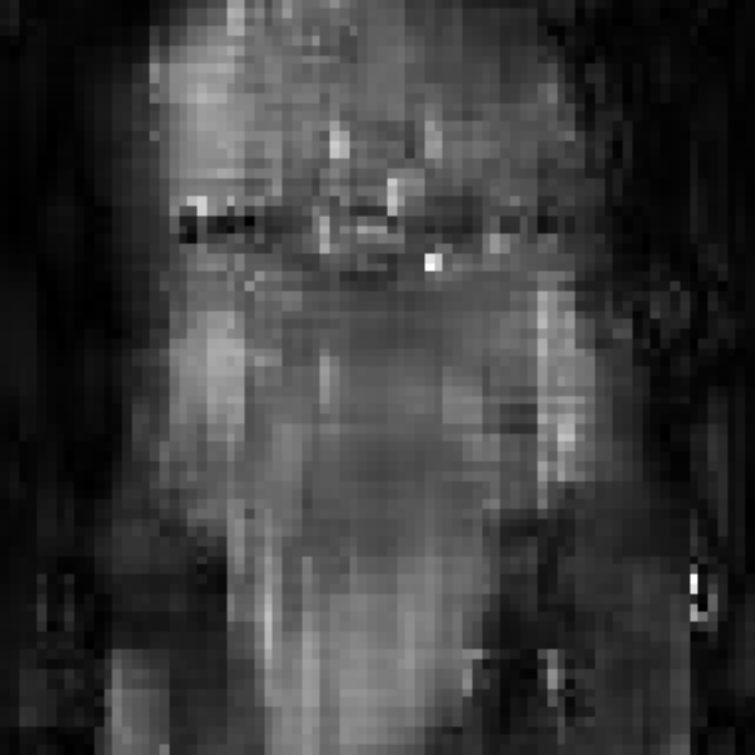}
  \caption*{(160,88)}
  \end{subfigure}
  \begin{subfigure}[t]{0.12\textwidth}
  \centering\includegraphics[scale=0.28]{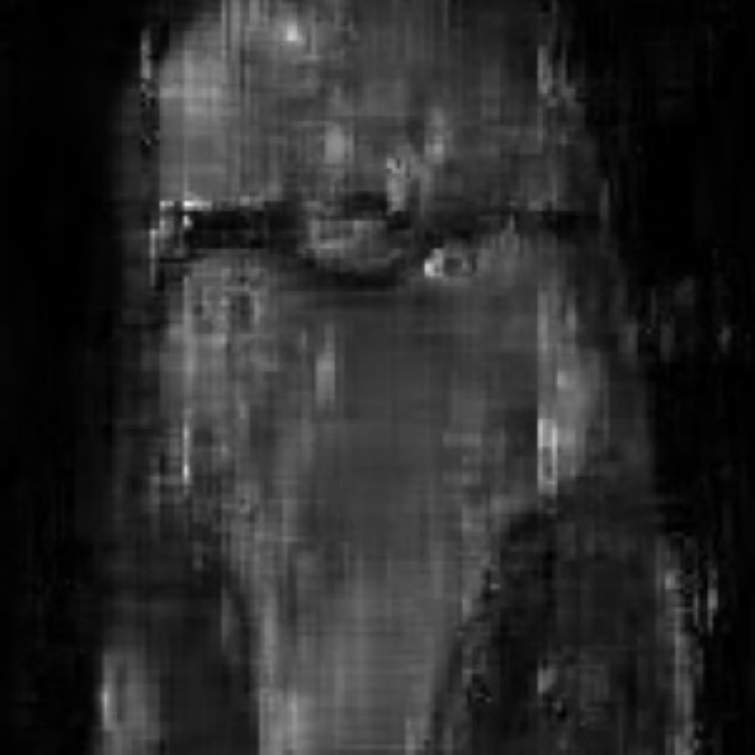}
  \caption*{(320,180)}
  \end{subfigure}
  \begin{subfigure}[t]{0.12\textwidth}
  \centering\includegraphics[scale=0.28]{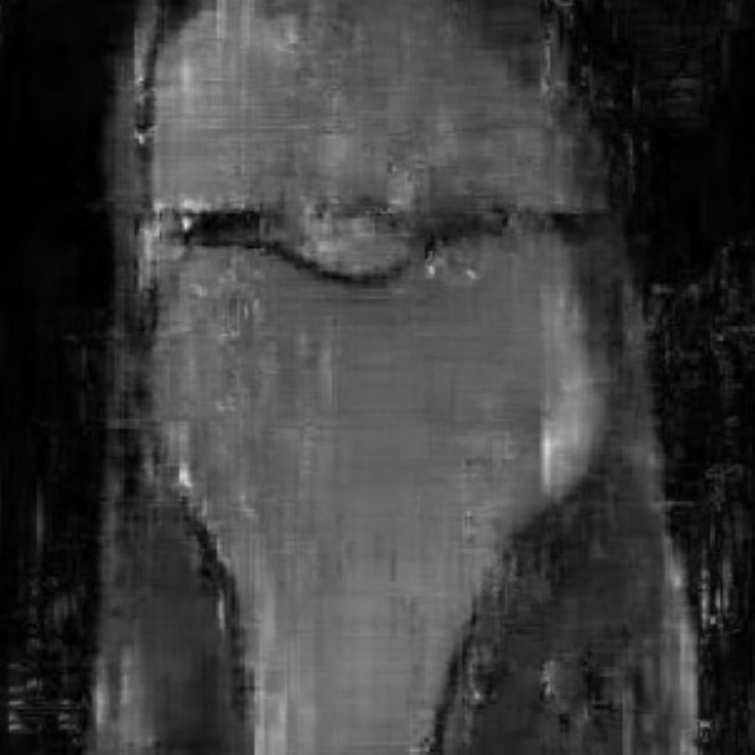}
  \caption*{(640,168)}
  \end{subfigure}
  \begin{subfigure}[t]{0.12\textwidth}
  \centering\includegraphics[scale=0.28]{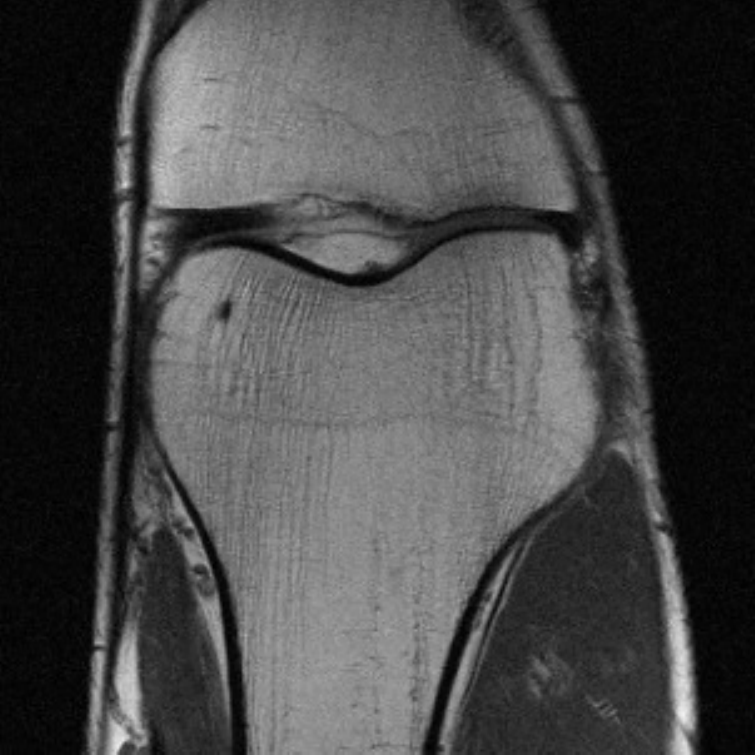}
  \caption*{ground truth}
  \end{subfigure}
\caption{ConvDecoder finds an image representation by constructing fine details per each layer. The network is fitted for representing an image from the 4x under-sampled $k$-space of the ground-truth image in the mid row. \textbf{Top row:} different resolutions of the ground-truth image in the bottom row to each of which we fit the layer outputs. \textbf{Bottom row:} output visualization for each of the six layers.
}
\label{fig:representation}
\end{figure*}

Figure~\ref{fig:representation} shows the results of our visualization method. 
The top row shows different resolutions of the ground-truth image $\vx_1$. 
The bottom row shows the visualization results for a network fitted to reconstruct $\vx_1$ from the 4x under-sampled measurements $\vy_1$.
We observe that:
(i) ConvDecoder finds a fine representation of an image by adding more detail in each layer. 
(ii) As shown in Figure~\ref{fig:representation}, we observe the role of up-sampling blocks in inducing the notion of resolution to the network, in that different layers represent reconstructions of different resolutions.

\subsection{It is critical to fit different layers at different speeds}\label{sec:fitting-speed}

The optimization method employed to minimize the loss function of an un-trained network has a significant impact on the quality of the reconstructed image because it determines which layer is fitted at which speed. The Adam optimizer yields a significantly better reconstruction than Gradient Descent (GD) both qualitatively and quantitatively (an approximately 3\% higher SSIM score). 

The reason behind this discrepancy in performance lies in the fact that Adam chooses different stepsizes (and hence different fitting speeds) for each parameter in the network, whereas GD treats all network parameters the same. 
It is important to (i) choose larger stepsizes for earlier layers and (ii) employ an increasing stepsize schedule, but it is not critical to choose the stepsize adaptively or differently within a layer.

\begin{figure}[ht!]
\begin{center}
\begin{tikzpicture}

\begin{groupplot}[
y tick label style={/pgf/number format/.cd,fixed,precision=4},
scaled y ticks = false, 
         group
         style={group size= 2 by 1, xlabels at=edge bottom, ylabels at=edge left,
         xticklabels at=edge bottom,
         horizontal sep=1.8cm, vertical sep=0.8cm,
         }, 
         width=0.3\textwidth,height=0.27\textwidth,
         title style={yshift=-7pt,},
         ]

\nextgroupplot[y tick label style={/pgf/number format/.cd,fixed,precision=4},xmin=0.5,xmax=3.5,
scaled y ticks = false,xticklabels={Adam, GD, GD-A},xtick={1,2,3}, ylabel=SSIM, xlabel={optimization method}]

\addplot +[mark size=1.5,draw=none,steelblue,thick, error bars/.cd, y dir = both, y explicit, error mark options={rotate=90, steelblue, mark size=2, line width=1.2}] table[x index=0, y index=1, y error index=2]{./files/growing_mulruns.csv};

\addplot[dashed,amber!60,ultra thin] coordinates {(0.5,0.809) (4.5,0.809)};


\addplot[dashed,amber!60,ultra thin] coordinates {(0.5,0.831) (4.5,0.831)};

\addplot[dashed,amber!60,ultra thin] coordinates {(0.5 ,0.837) (4.5,0.837)};

\nextgroupplot[title={},ylabel={stepsize},xlabel={iteration},ylabel shift = -5 pt,scaled x ticks=false,ymode=log,ymax = 10000000,legend cell align=left, legend image post style={scale=0.5},
legend style={draw={none},at={(0.4,1)} , fill opacity=0, text opacity=1, nodes={scale=0.6}, 
/tikz/every even column/.append style={column sep=-0.1cm}
 },]
	\addplot [mark=none,steelblue,thick] table[x index=0,y index=1]{./files/adam_adag_avg.csv};
	\addlegendentry{layer1}
	\addplot [mark=none,amber,thick] table[x index=0,y index=3]{./files/adam_adag_avg.csv};
	\addlegendentry{layer3}
	\addplot [mark=none,black,solid,thick] table[x index=0,y index=5]{./files/adam_adag_avg.csv};
	\addlegendentry{layer5}
	\addplot [mark=none,ash,thick] table[x index=0,y index=9]{./files/adam_adag_avg.csv};
	\addlegendentry{layer8}
	
	\addplot +[name path=upper,draw=none, mark=none] table[x index=0,y expr=\thisrowno{1}+\thisrowno{10}*2] {./files/adam_adag_avg.csv};
    \addplot +[name path=lower,draw=none,mark=none] table[x index=0,y expr=\thisrowno{1}-\thisrowno{10}*2] {./files/adam_adag_avg.csv};
    \addplot +[fill=blue!10] fill between[of=upper and lower];
    
	\addplot +[name path=upper,draw=none, mark=none] table[x index=0,y expr=\thisrowno{3}+\thisrowno{12}*2] {./files/adam_adag_avg.csv};
    \addplot +[name path=lower,draw=none,mark=none] table[x index=0,y expr=\thisrowno{3}-\thisrowno{12}*2] {./files/adam_adag_avg.csv};
    \addplot +[fill=amber!20] fill between[of=upper and lower];
    
	\addplot +[name path=upper,draw=none, mark=none] table[x index=0,y expr=\thisrowno{5}+\thisrowno{14}*2] {./files/adam_adag_avg.csv};
    \addplot +[name path=lower,draw=none,mark=none] table[x index=0,y expr=\thisrowno{5}-\thisrowno{14}*2] {./files/adam_adag_avg.csv};
    \addplot +[fill=black!20] fill between[of=upper and lower];
    
	\addplot +[name path=upper,draw=none, mark=none] table[x index=0,y expr=\thisrowno{9}+\thisrowno{18}*2] {./files/adam_adag_avg.csv};
    \addplot +[name path=lower,draw=none,mark=none] table[x index=0,y expr=\thisrowno{9}-\thisrowno{18}*2] {./files/adam_adag_avg.csv};
    \addplot +[fill=ash!20] fill between[of=upper and lower];
\end{groupplot}

\end{tikzpicture}
\caption{\textbf{Left.} Assigning larger stepsizes to shallower layers and utilizing an increasing schedule enhances the performance of GD relative to Adam. SSIM scores are shown for different optimizers over 10 runs on a sample image from the 4x accelerated fastMRI validation set. Adam, GD, and GD-A (GD with the same stepsize schedule as Adam) are considered.
\textbf{Right.} Average layer-wise stepsizes for different layers of ConvDecoder based on the iteration number when using the Adam optimizer. All values are averaged over 6 randomly-chosen images from the 4x accelerated fastMRI validation set.}
\label{fig:fitting-speed}
\end{center}
\end{figure}
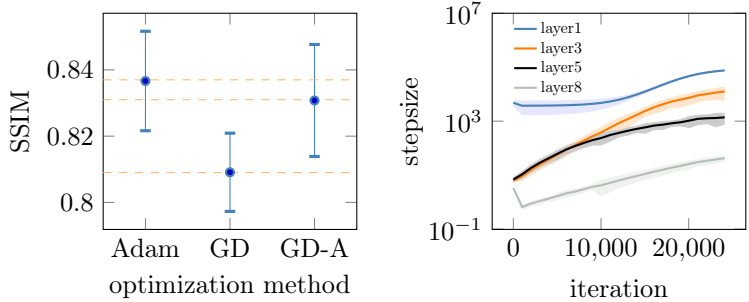

To illustrate this point, we recorded layer-wise stepsizes assigned by Adam over the number of iterations which is shown in Figure \ref{fig:fitting-speed} (right). We then ran GD with the same stepsize schedule for each layer as we found Adam to use (we refer to this setup as GD-A), and observed that it performs essentially the same as Adam (See Figure~\ref{fig:fitting-speed} (left)). This demonstrates that it is critical to learn the layers with different stepsizes, but the adaptive gradients chosen by Adam are not critical nor relevant for performance.

\section{8x accelerated multi-coil knee measurements}\label{sec:higher-sampling-rate}


In this section, we report the results for 8x accelerated knee measurements. For 8x acceleration, we found different hyper-parameters of the ConvDecoder (relative to 4x acceleration) to work best. To find good hyper-parameters for the ConvDecoder, we again performed a grid search---the details of which is provided in Section~\ref{sec:setup}---which resulted in a ConvDecoder with 6 layers, 64 channels, and $(4,4)$ as the input size.

\begin{table}[h]
\centering
\begin{adjustbox}{width=0.7\textwidth}
\begin{tabular}{ccccc}
\toprule 
Method & VIF & MS-SSIM & SSIM & PSNR \\
\midrule
    ConvDecoder & 0.5234 $\pm$ 0.0214 & 0.8827 $\pm$ 0.0070 & 0.6815 $\pm$ 0.0142 & 28.49 $\pm$ 0.31\\
    U-net & 0.5233 $\pm$ 0.0146 & 0.9148 $\pm$ 0.0029 & 0.7115 $\pm$ 0.0088 & 29.25 $\pm$ 0.15 \\
    VarNet & \textbf{0.5821 $\pm$ 0.0115} & \textbf{0.9432 $\pm$ 0.0025} & \textbf{0.7812} $\pm$ 0.0076& \textbf{31.65 $\pm$ 0.13} \\
    TV & 0.3119 $\pm$ 0.0232 & 0.8340 $\pm$ 0.0075 & 0.5986 $\pm$ 0.0139 & 26.55 $\pm$ 0.35\\
    ENLIVE & 0.3636 $\pm$ 0.0283 & 0.7889 $\pm$ 0.0086 & 0.4986 $\pm$ 0.0183 & 23.11 $\pm$ 0.26\\
\bottomrule
\end{tabular}
\end{adjustbox}
\caption{Average image-based scores for reconstructing the mid-slice images of 200 volumes in the multi-coil knee measurements from the fastMRI validation set (8x accelerated). ConvDecoder achieves on-par performance with U-net and outperforms TV as well as ENLIVE, yet the end-to-end variational network (VarNet) performs significantly better for 8x acceleration. Marginal errors denote 95\% confidence interval.}
\label{tab:8x}
\end{table}

\begin{figure}[ht!]
\centering
  \begin{subfigure}[t]{0.16\textwidth}
  \caption*{ConvDecoder}
  \centering\includegraphics[scale=0.32]{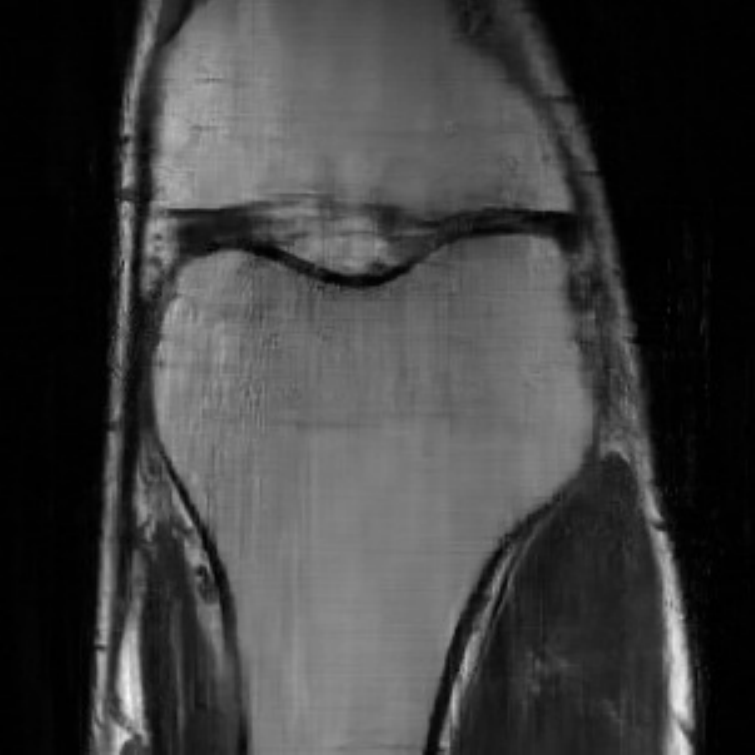}
  \end{subfigure}
  \begin{subfigure}[t]{0.16\textwidth}
  \caption*{TV}
  \centering\includegraphics[scale=0.32]{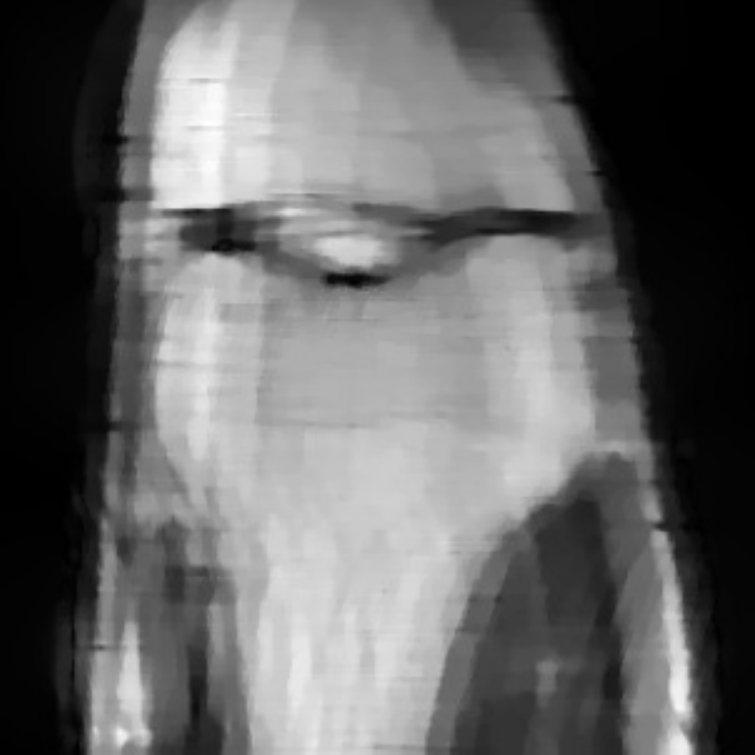}
  \end{subfigure}
  \begin{subfigure}[t]{0.16\textwidth}
  \caption*{ENLIVE}
  \centering\includegraphics[scale=0.32]{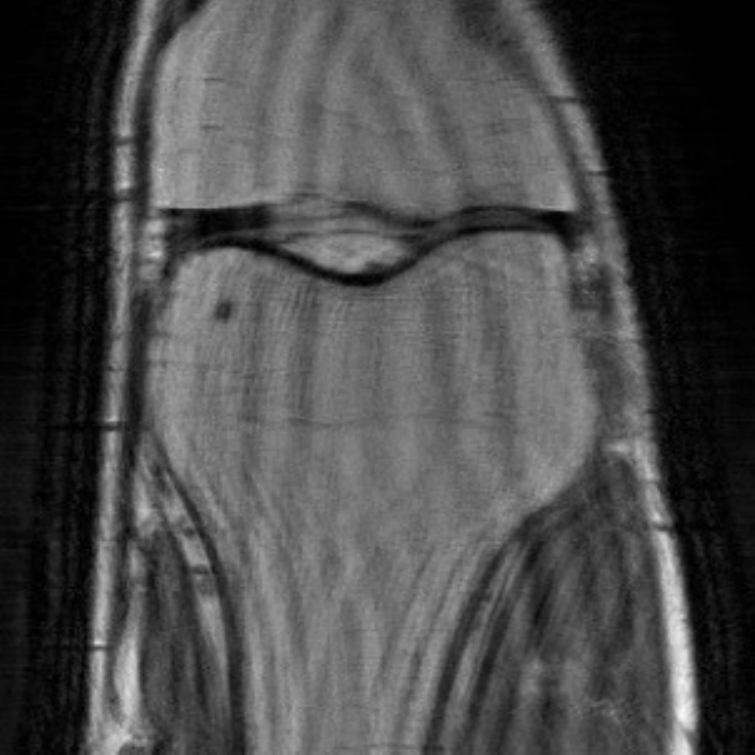}
  \end{subfigure}
  \begin{subfigure}[t]{0.16\textwidth}
  \caption*{U-net}
  \centering\includegraphics[scale=0.32]{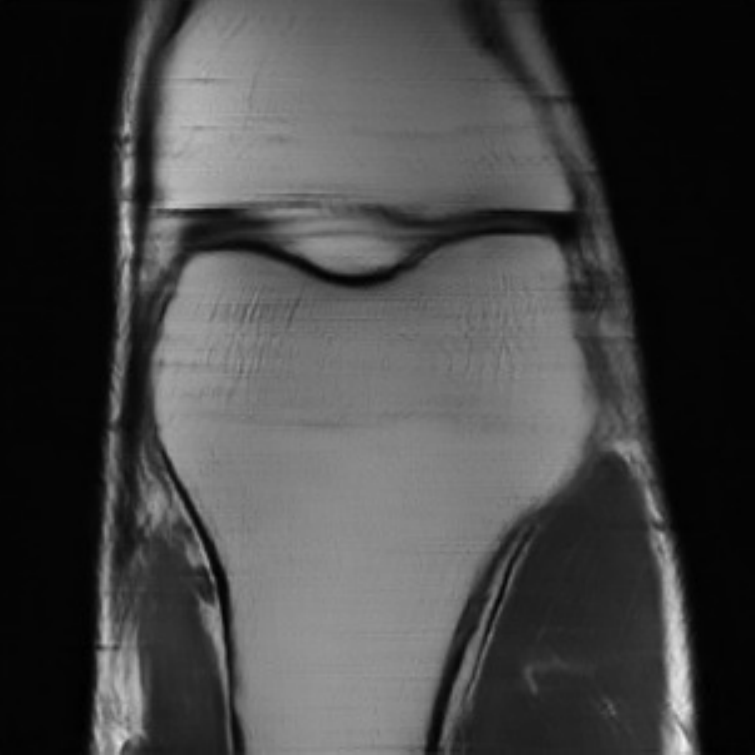}
  \end{subfigure}
  \begin{subfigure}[t]{0.16\textwidth}
  \caption*{VarNet}
  \centering\includegraphics[scale=0.32]{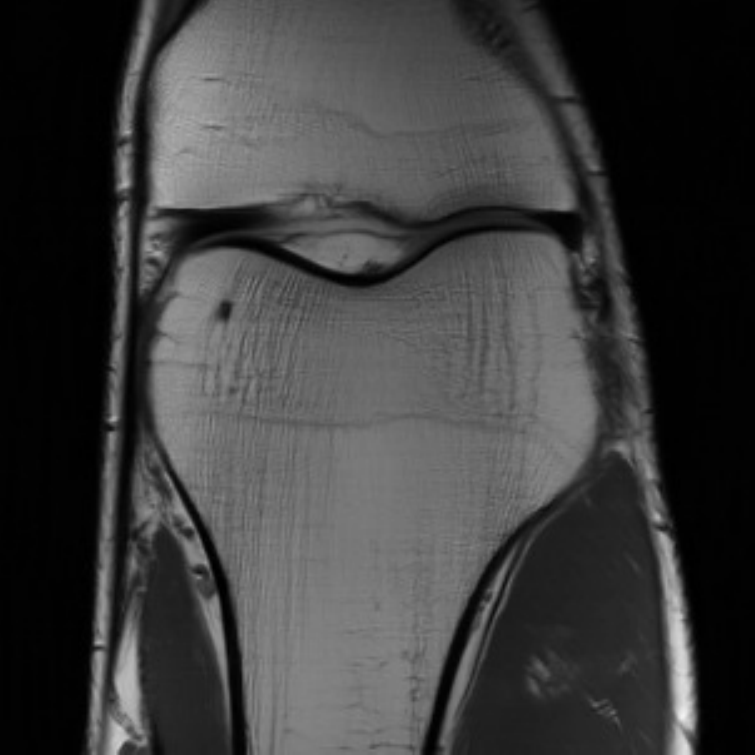}
  \end{subfigure}
  \begin{subfigure}[t]{0.16\textwidth}
  \caption*{ground truth}
  \centering\includegraphics[scale=0.32]{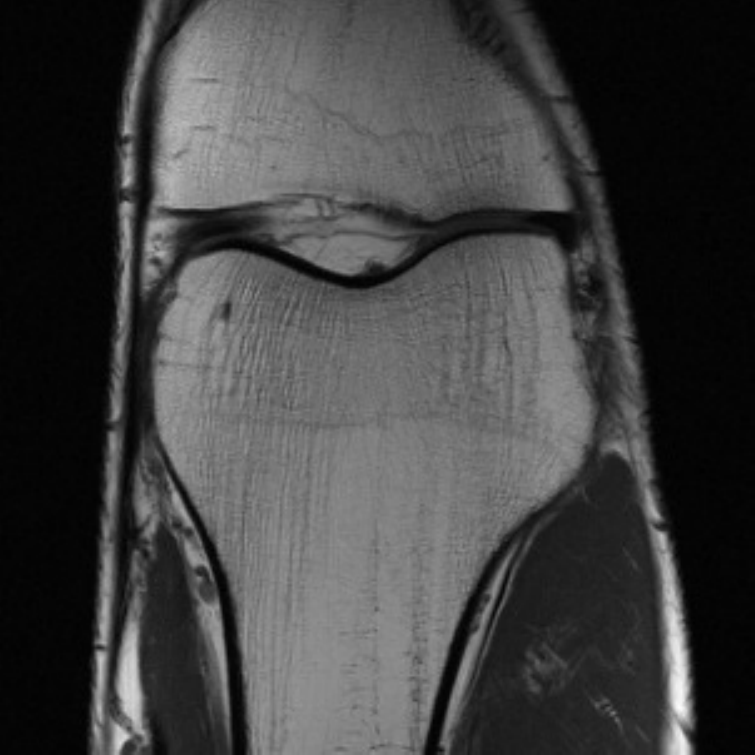}
  \end{subfigure}\par\medskip 
  \begin{subfigure}[t]{0.16\textwidth}
  \centering\includegraphics[scale=0.32]{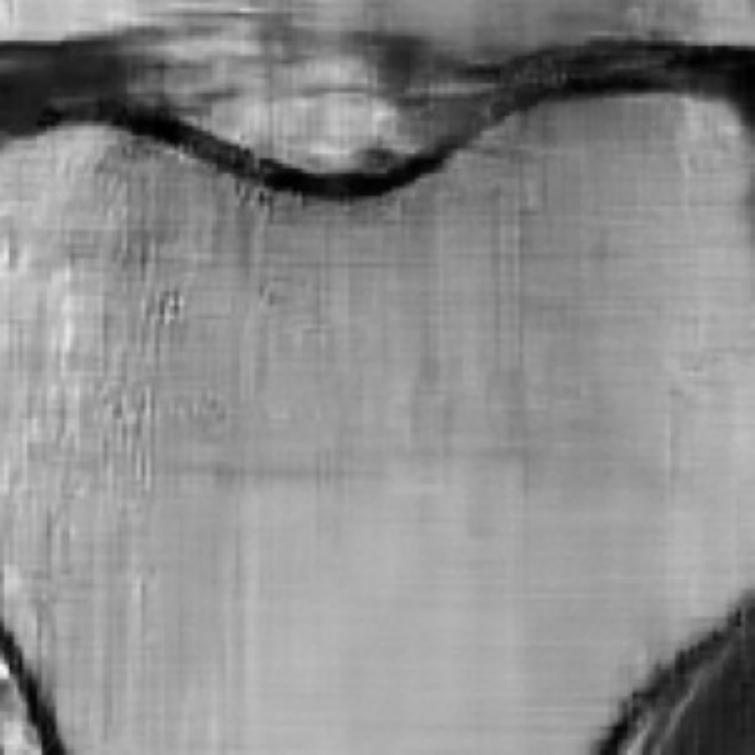}
  \end{subfigure}
  \begin{subfigure}[t]{0.16\textwidth}
  \centering\includegraphics[scale=0.32]{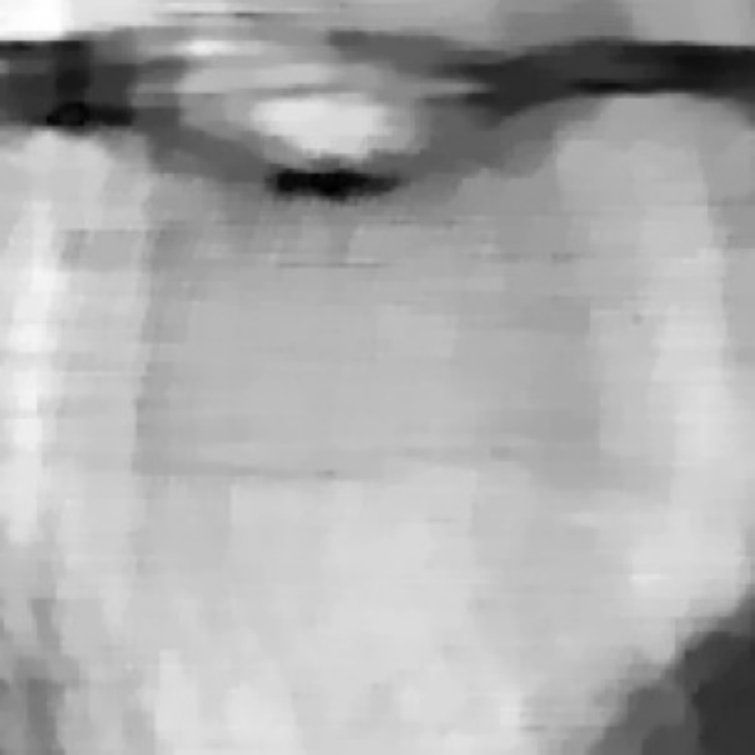}
  \end{subfigure}
  \begin{subfigure}[t]{0.16\textwidth}
  \centering\includegraphics[scale=0.32]{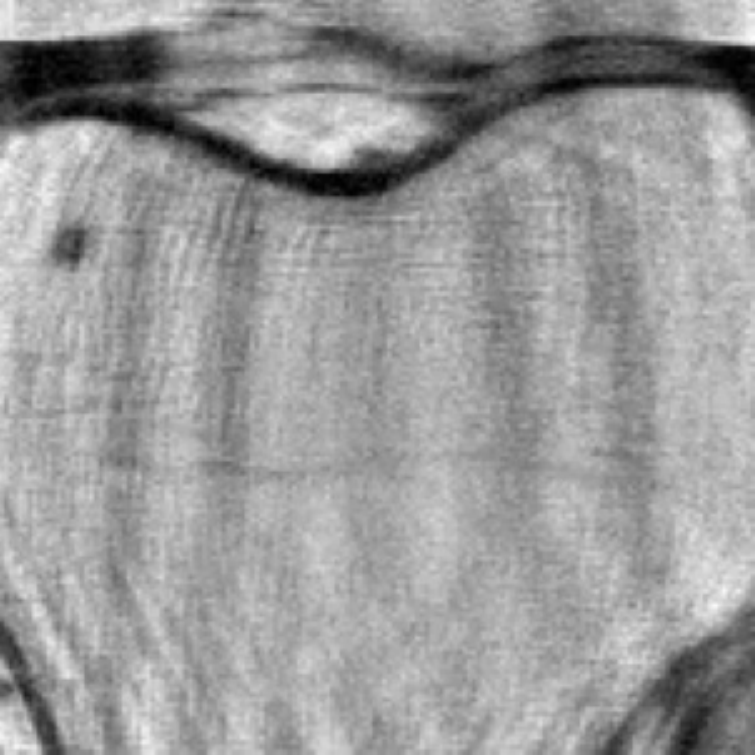}
  \end{subfigure}
  \begin{subfigure}[t]{0.16\textwidth}
  \centering\includegraphics[scale=0.32]{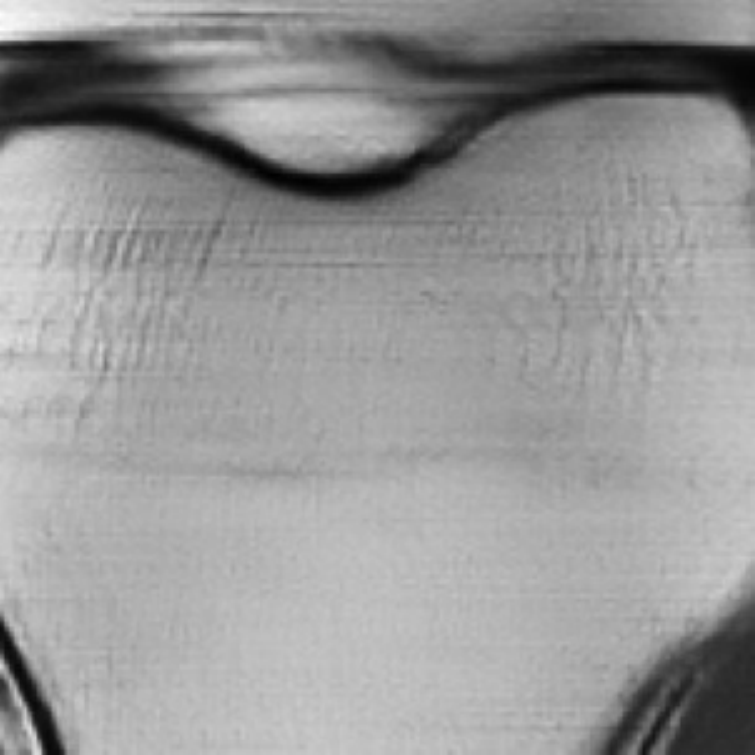}
  \end{subfigure}
  \begin{subfigure}[t]{0.16\textwidth}
  \centering\includegraphics[scale=0.32]{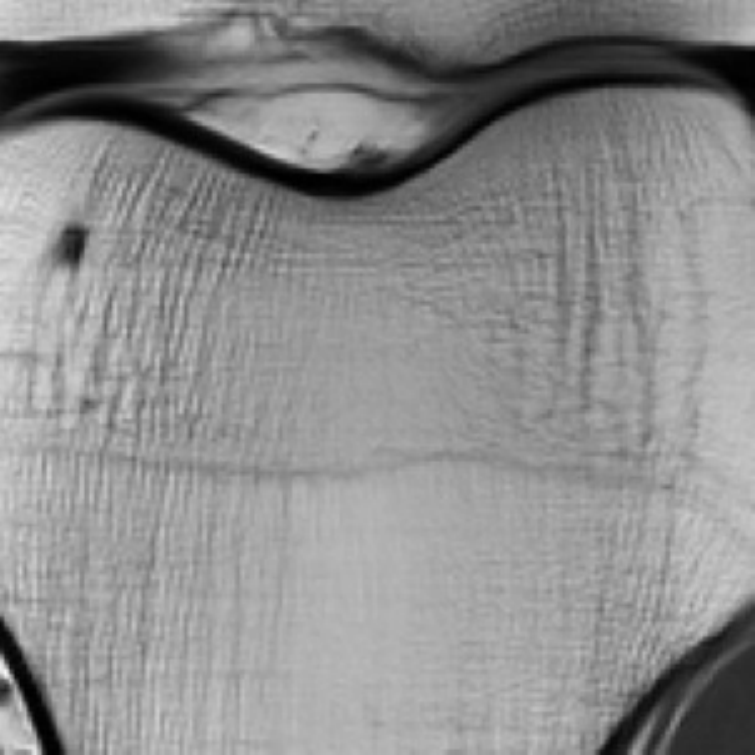}
  \end{subfigure}
  \begin{subfigure}[t]{0.16\textwidth}
  \centering\includegraphics[scale=0.32]{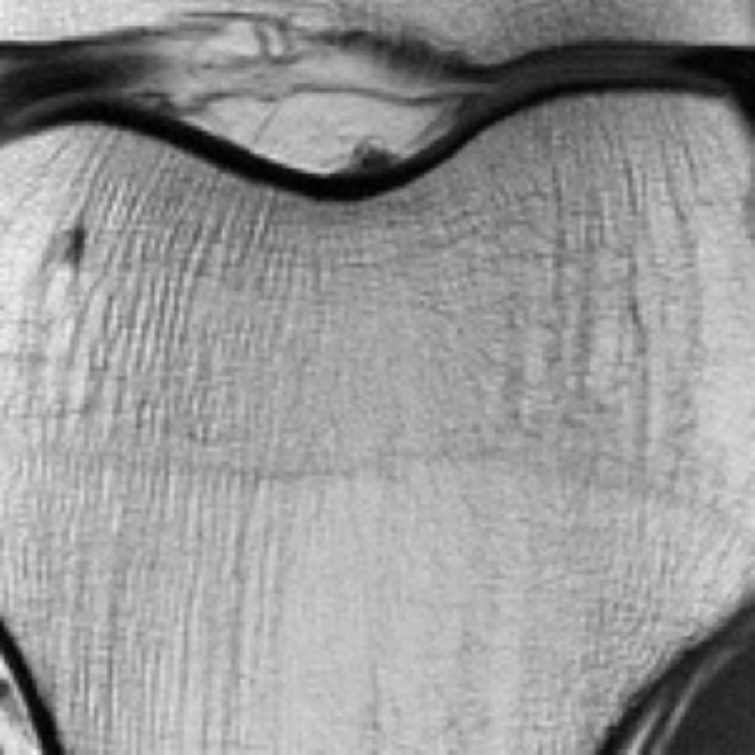}
  \end{subfigure}
\caption{Sample reconstructions for ConvDecoder, TV, ENLIVE, U-net, VarNet for a validation image from multi-coil knee measurements (8x accelerated). The bottom row represents zoomed-in version of the top row. ConvDecoder finds a similar reconstruction as U-net and outperforms TV. The state-of-the-art VarNet yields the best reconstruction for this image.
}
\label{fig:8x_baselines}
\end{figure}

Table~\ref{tab:8x} shows that ConvDecoder achieves similar performance
to U-net (according to all metrics except SSIM and
MS-SSIM which U-net slightly outperforms ConvDecoder)
and significantly outperforms TV as well as ENLIVE. Figure~\ref{fig:8x_baselines} shows a sample reconstruction for all considered methods.

\section{4x accelerated multi-coil brain measurements}\label{sec:brain}

So far, we have shown that un-trained neural networks perform surprisingly well for knee MRI reconstruction. Specifically, they perform on-par with a baseline trained neural network and significantly better than a baseline un-trained method (TV) as well as a modern un-trained method (ENLIVE). In this section, we consider un-trained networks for the reconstruction of brain images to fortify our claims.


For the sake of consistency with the rest of the paper, we first compare different un-trained neural networks and then compare the best-performing one to competing methods. We again performed an extensive grid search and the resulting hyper-parameters are shown in Section~\ref{sec:setup}.

Table~\ref{tab:validation-results-untrained-brain} shows average scores for the ConvDecoder, DIP, and DeepDecoder. Interestingly, all three un-trained networks perform similar on the brain images and unlike knee, there is not a significant difference among them. However, we emphasize the role of our new data consistency step which resulted in approximately 8\% SSIM score improvement for these un-trained networks.

\begin{table}[h]
\centering
\begin{adjustbox}{width=0.7\textwidth}
\begin{tabular}{ccccc}
\toprule
Method & VIF & MS-SSIM & SSIM & PSNR \\
\midrule
    ConvDecoder & 0.8002 $\pm$ 0.0168 & \textbf{0.9743  $\pm$ 0.0055} & 0.9018  $\pm$ 0.0121 & 34.58  $\pm$ 0.41 \\
    DIP         & \textbf{0.8223 $\pm$ 0.0144} & 0.9736 $\pm$ 0.0051  & 0.8918  $\pm$ 0.0131 & \textbf{34.96  $\pm$ 0.35} \\
    DD          & 0.7952  $\pm$ 0.0184 & 0.9713 $\pm$ 0.0063 & \textbf{0.9023  $\pm$ 0.0138} & 34.52  $\pm$ 0.51 \\
\bottomrule
\end{tabular}
\end{adjustbox}
\caption{Average image-based scores for the ConvDecoder, DIP, and Deep Decoder (DD) on the mid-slice images of 20 randomly-chosen volumes in the multi-coil brain measurements from the fastMRI validation set (4x accelerated). All three networks perform similar on the brain images. Marginal errors denote 95\% confidence interval.}
\label{tab:validation-results-untrained-brain}
\end{table}

\begin{figure*}[h!]
\centering
  \begin{subfigure}[t]{0.14\textwidth}
  \caption*{\tiny ConvDecoder}
  \centering\includegraphics[scale=0.3]{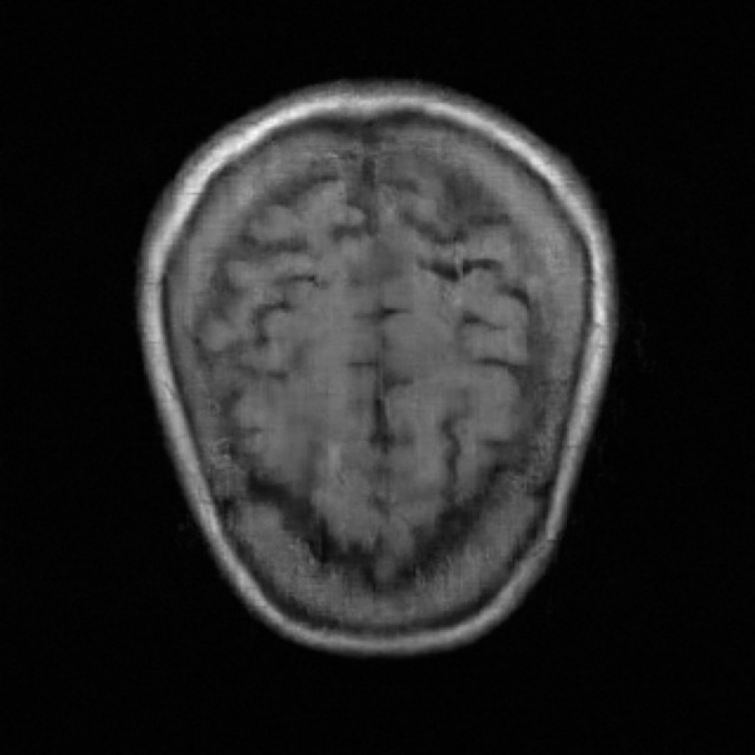}
  \end{subfigure}
  \begin{subfigure}[t]{0.14\textwidth}
  \caption*{\tiny DD}
  \centering\includegraphics[scale=0.3]{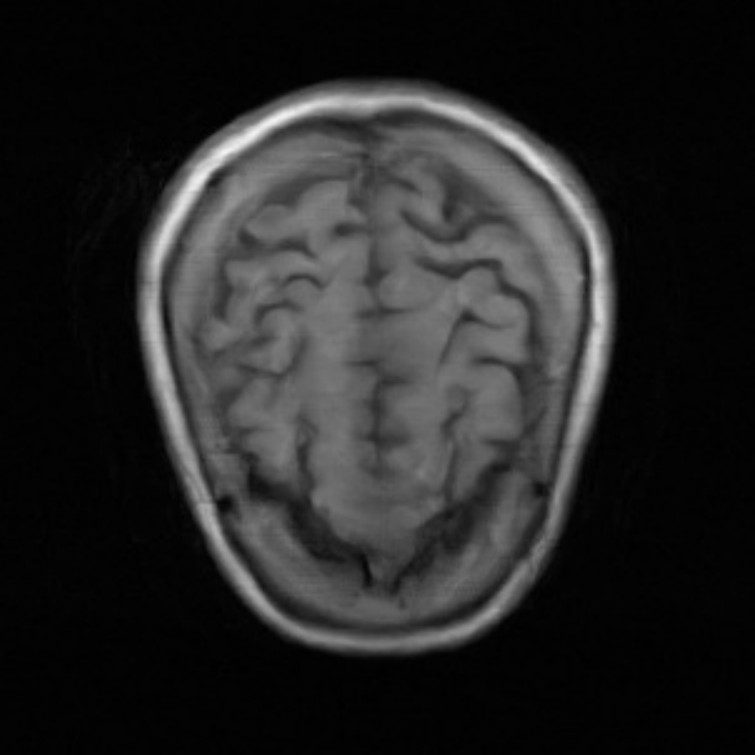}
  \end{subfigure}
  \begin{subfigure}[t]{0.14\textwidth}
  \caption*{\tiny DIP}
  \centering\includegraphics[scale=0.3]{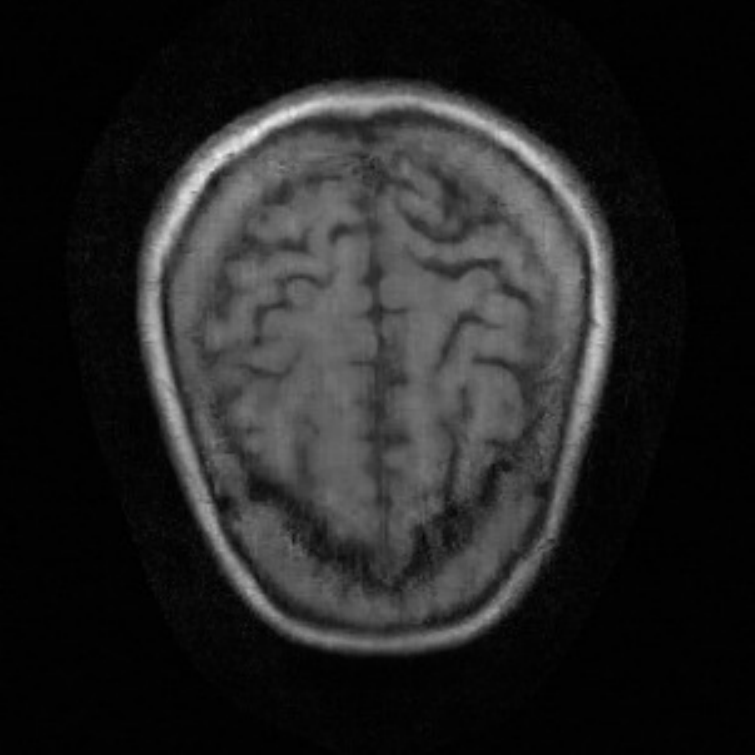}
  \end{subfigure}
  \begin{subfigure}[t]{0.14\textwidth}
  \caption*{\tiny ground truth}
  \centering\includegraphics[scale=0.3]{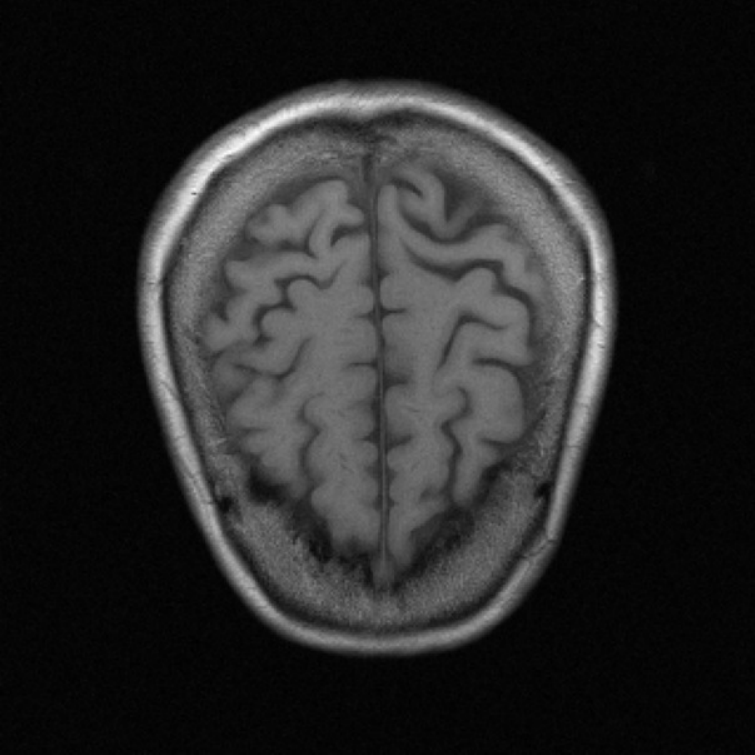}
  \end{subfigure}\par\medskip 
  \begin{subfigure}[t]{0.14\textwidth}
  \centering\includegraphics[scale=0.243]{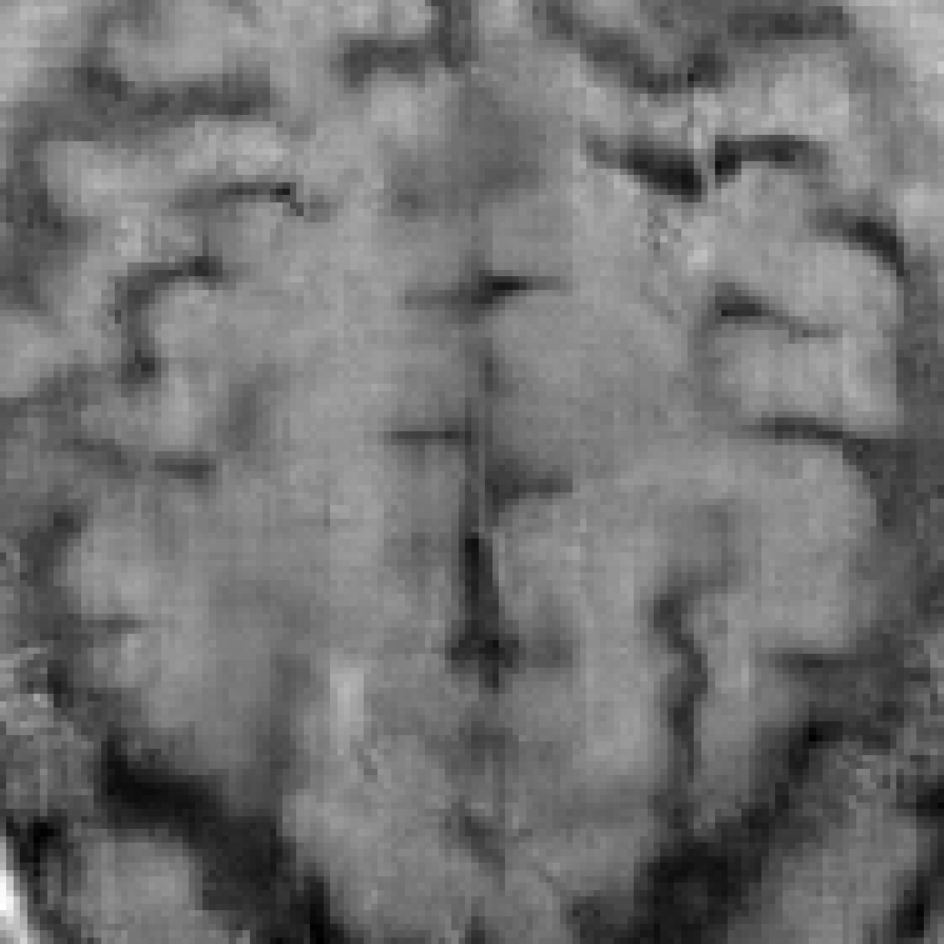}
  \end{subfigure}
  \begin{subfigure}[t]{0.14\textwidth}
  \centering\includegraphics[scale=0.303]{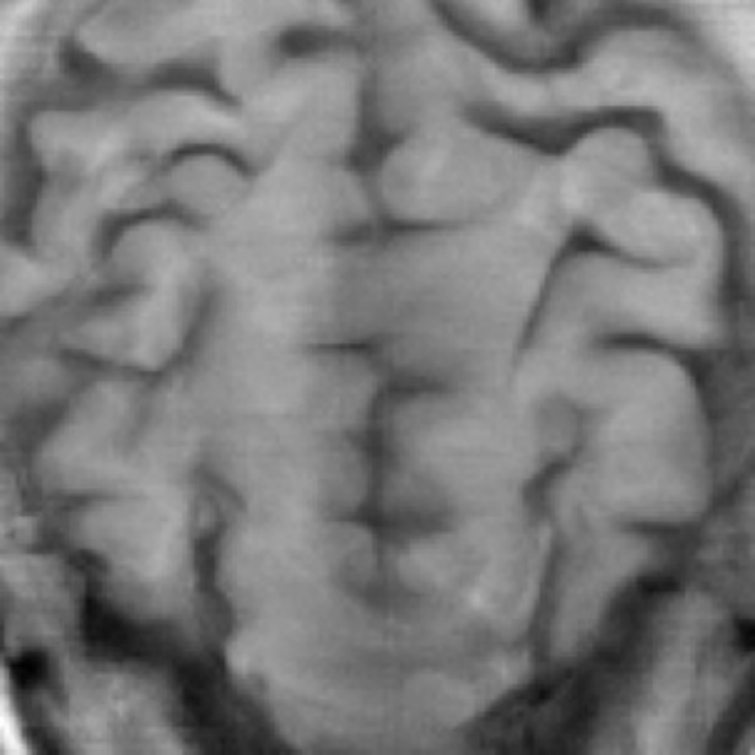}
  \end{subfigure}
  \begin{subfigure}[t]{0.14\textwidth}
  \centering\includegraphics[scale=0.303]{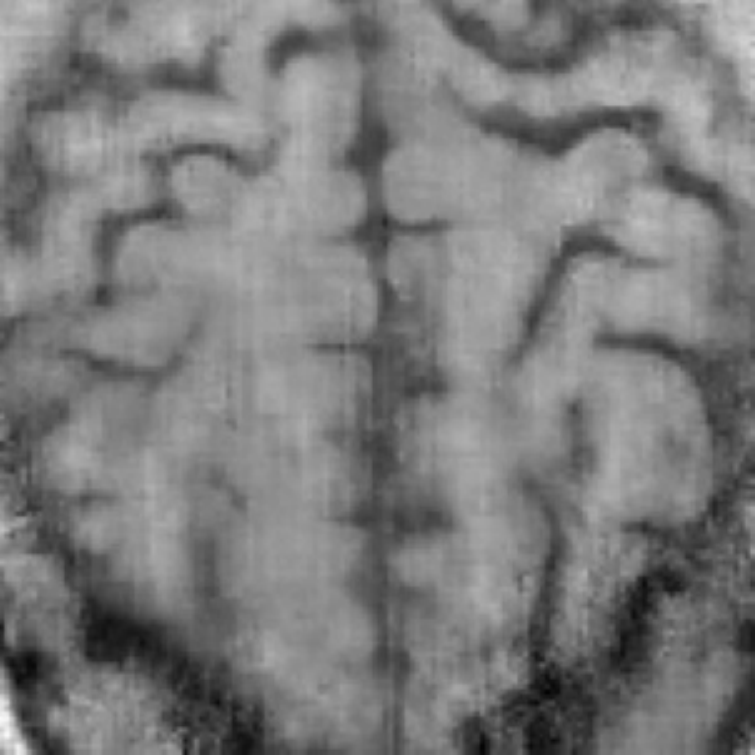}
  \end{subfigure}
  \begin{subfigure}[t]{0.14\textwidth}
  \centering\includegraphics[scale=0.303]{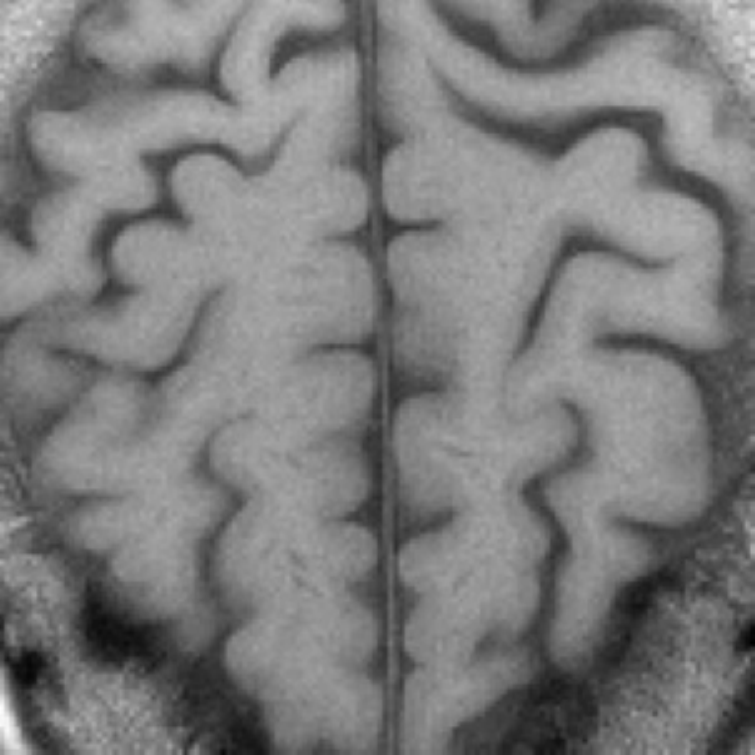}
  \end{subfigure}\par\medskip  
\caption{ConvDecoder, DD, and DIP yield similar reconstructions for brain images. The artifacts can be mitigated by incorporating coil sensitivity maps as shown in Fugure~\ref{fig:brain}. Images in the bottom row are zoomed-in versions of the top row.
}
\label{fig:brain-untrained}
\end{figure*}

Since there is not a noticeable difference among ConvDecoder, DIP, and Deep Decoder on the brain images, we proceed with the ConvDecoder for comparison to baselines, to be consistent with the previous sections. However, ConvDecoder can be replaced with the Deep Decoder or DIP architectures in the following comparison with similar results; although the DIP has a larger computational overhead.

During tuning, we noticed that Step 1-B---described in Section~\ref{sec:multi-coil}, which takes the estimates of the coil sensitivity maps into account instead of using Step 1-A which does not, works  significantly better for brain images.
To quantify the gains of employing the estimated coil sensitivity maps, we include ConvDecoder scores with/without sensitivity maps along with U-net, the end-to-end variational network, TV, and ENLIVE in Table~\ref{tab:brain}.  
The numbers are averaged over 100 validation brain images.

\begin{table}[h]
\centering
\begin{adjustbox}{width=0.7\textwidth}
\begin{tabular}{ccccc}
\toprule
Method & VIF & MS-SSIM & SSIM & PSNR \\
\midrule
    ConvDecoder & 0.8178 $\pm$ 0.0091 & 0.9722 $\pm$ 0.0024 & 0.8892 $\pm$ 0.0054 & 33.06 $\pm$ 0.11 \\
    ConvDecoder-SM & 0.8707 $\pm$ 0.0087& 0.9804 $\pm$ 0.0021& 0.9097 $\pm$ 0.0049 & 34.46 $\pm$ 0.13 \\
    U-net & 0.8714 $\pm$ 0.0062 & 0.9792 $\pm$ 0.0014 & 0.9173 $\pm$ 0.0036 & 34.29 $\pm$ 0.10 \\
    VarNet & \textbf{0.9173 $\pm$ 0.0051 } & \textbf{0.9912 $\pm$ 0.0009} & \textbf{0.9421 $\pm$ 0.0031} & \textbf{37.46 $\pm$ 0.07} \\
    TV & 0.6734 $\pm$ 0.0379 & 0.8748 $\pm$ 0.0113 & 0.7846 $\pm$ 0.0082 & 27.10 $\pm$ 0.57 \\
    ENLIVE & 0.6431 $\pm$ 0.0371 & 0.8660 $\pm$ 0.0108 & 0.7780 $\pm$ 0.0087 & 26.51 $\pm$ 0.55\\
\bottomrule
\end{tabular}
\end{adjustbox}
\caption{
Average image-based scores for the ConvDecoder, ConvDecoder-SM (ConvDecoder + sensitivity maps) vs U-net, VarNet, TV, and ENLIVE on 100 mid-slice images of the multi-coil brain measurements from the fastMRI validation set (4x accelerated). ConvDecoder with sensitivity maps achieves on-par performance with U-net and outperforms TV as well as ENLIVE. However, both U-net and ConvDecoder are slightly outperformed by the state of the art (VarNet). Marginal errors denote 95\% confidence interval.}
\label{tab:brain}
\end{table}

\begin{figure*}[h!]
\centering
  \begin{subfigure}[t]{0.135\textwidth}
  \caption*{\tiny ConvDecoder}
  \centering\includegraphics[scale=0.3]{./brain_files/im_cd-eps-converted-to.pdf}
  \end{subfigure}
  \begin{subfigure}[t]{0.135\textwidth}
  \caption*{\tiny TV}
  \centering\includegraphics[scale=0.3]{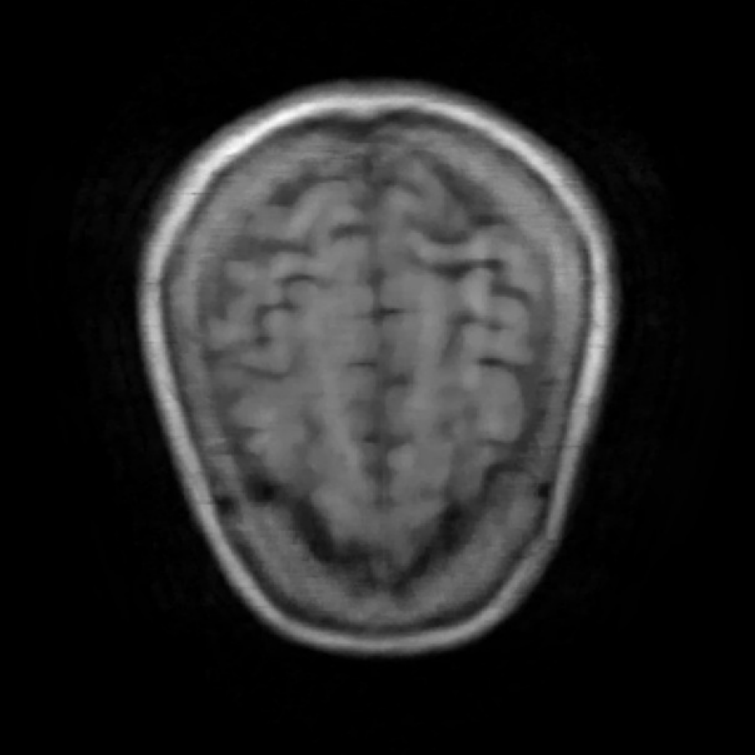}
  \end{subfigure}
  \begin{subfigure}[t]{0.135\textwidth}
  \caption*{\tiny ENLIVE}
  \centering\includegraphics[scale=0.3]{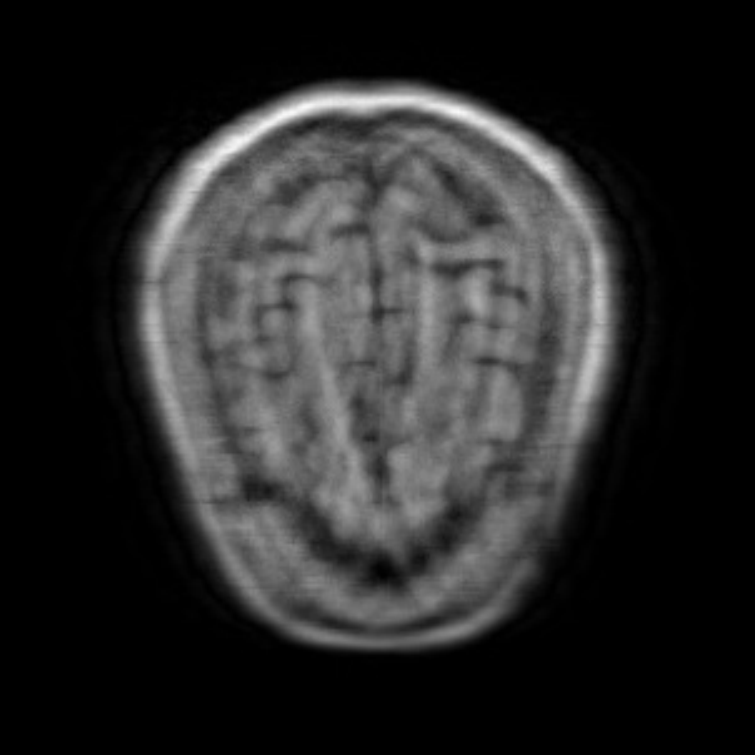}
  \end{subfigure}
  \begin{subfigure}[t]{0.135\textwidth}
  \caption*{\tiny ConvDecoder-SM}
  \centering\includegraphics[scale=0.3]{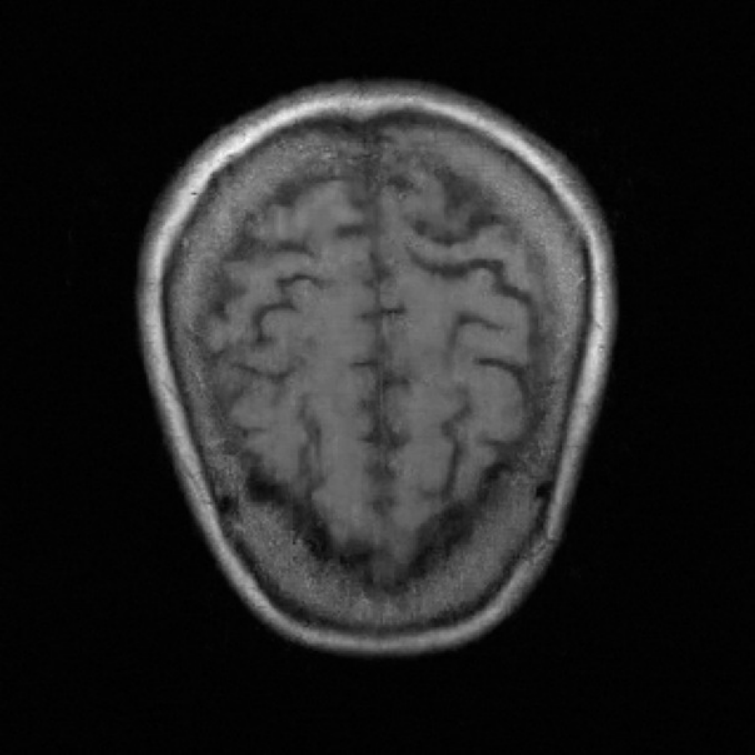}
  \end{subfigure}
  \begin{subfigure}[t]{0.135\textwidth}
  \caption*{\tiny U-net}
  \centering\includegraphics[scale=0.3]{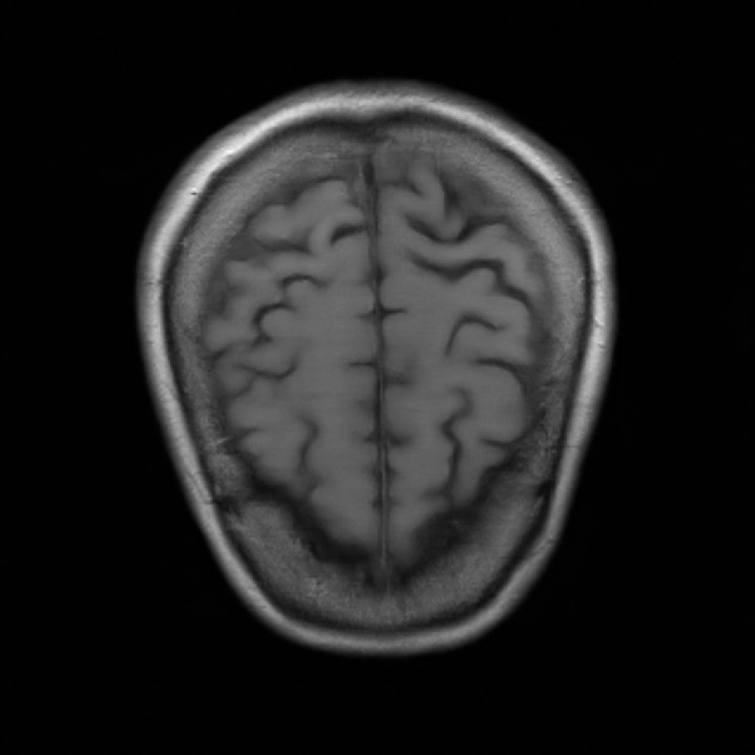}
  \end{subfigure}
  \begin{subfigure}[t]{0.135\textwidth}
  \caption*{\tiny VarNet}
  \centering\includegraphics[scale=0.3]{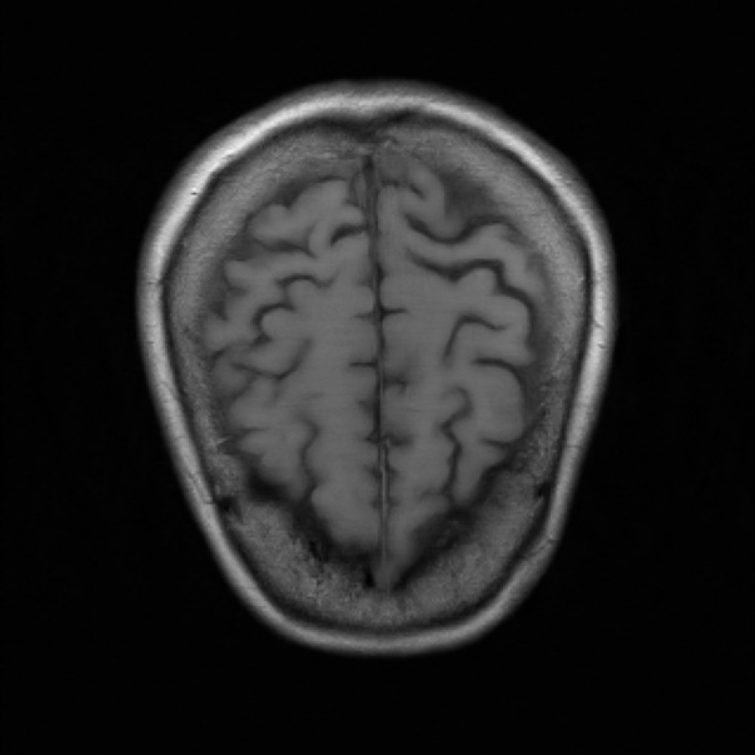}
  \end{subfigure}
  \begin{subfigure}[t]{0.135\textwidth}
  \caption*{\tiny ground truth}
  \centering\includegraphics[scale=0.3]{./brain_files/im_org-eps-converted-to.pdf}
  \end{subfigure}\par\medskip 
  \begin{subfigure}[t]{0.135\textwidth}
  \centering\includegraphics[scale=0.243]{./brain_files/anot_zim_cd-eps-converted-to.pdf}
  \end{subfigure}
  \begin{subfigure}[t]{0.135\textwidth}
  \centering\includegraphics[scale=0.243]{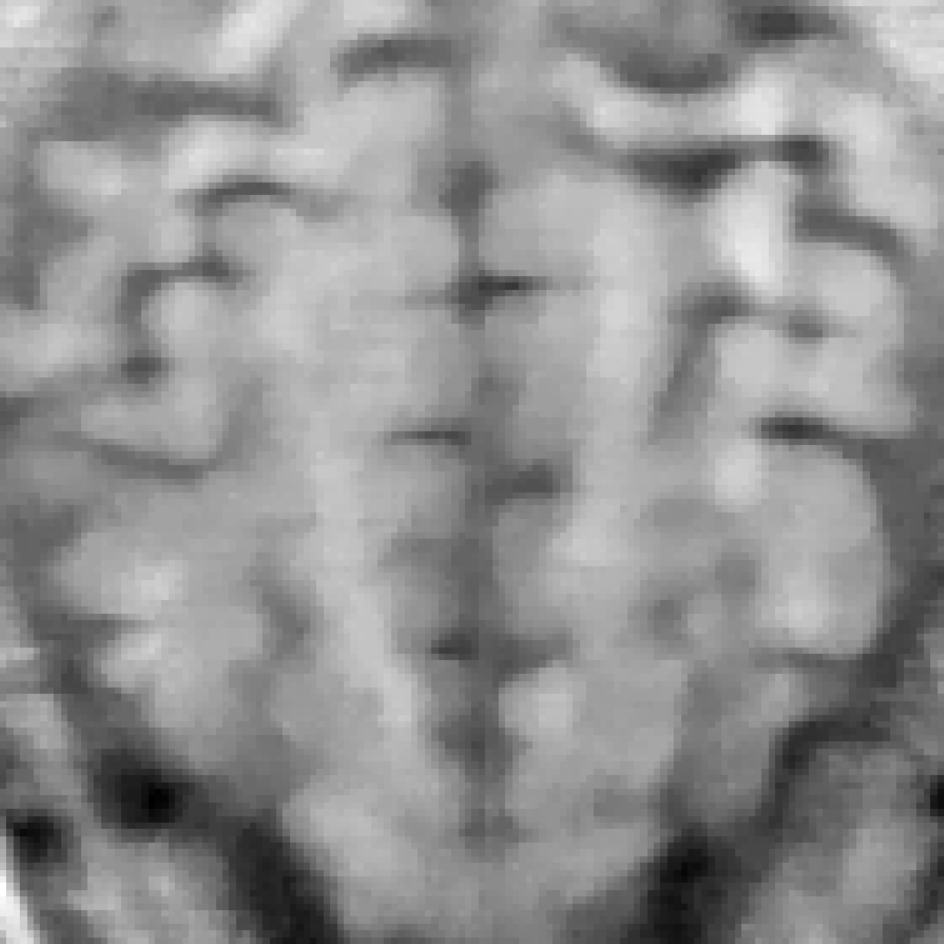}
  \end{subfigure}
  \begin{subfigure}[t]{0.135\textwidth}
  \centering\includegraphics[scale=0.3]{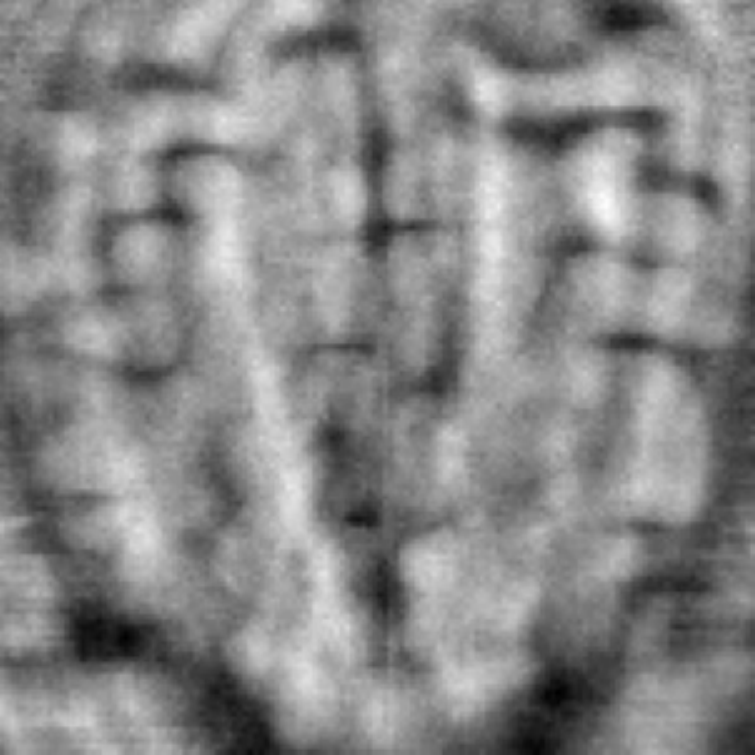}
  \end{subfigure}
  \begin{subfigure}[t]{0.135\textwidth}
  \centering\includegraphics[scale=0.243]{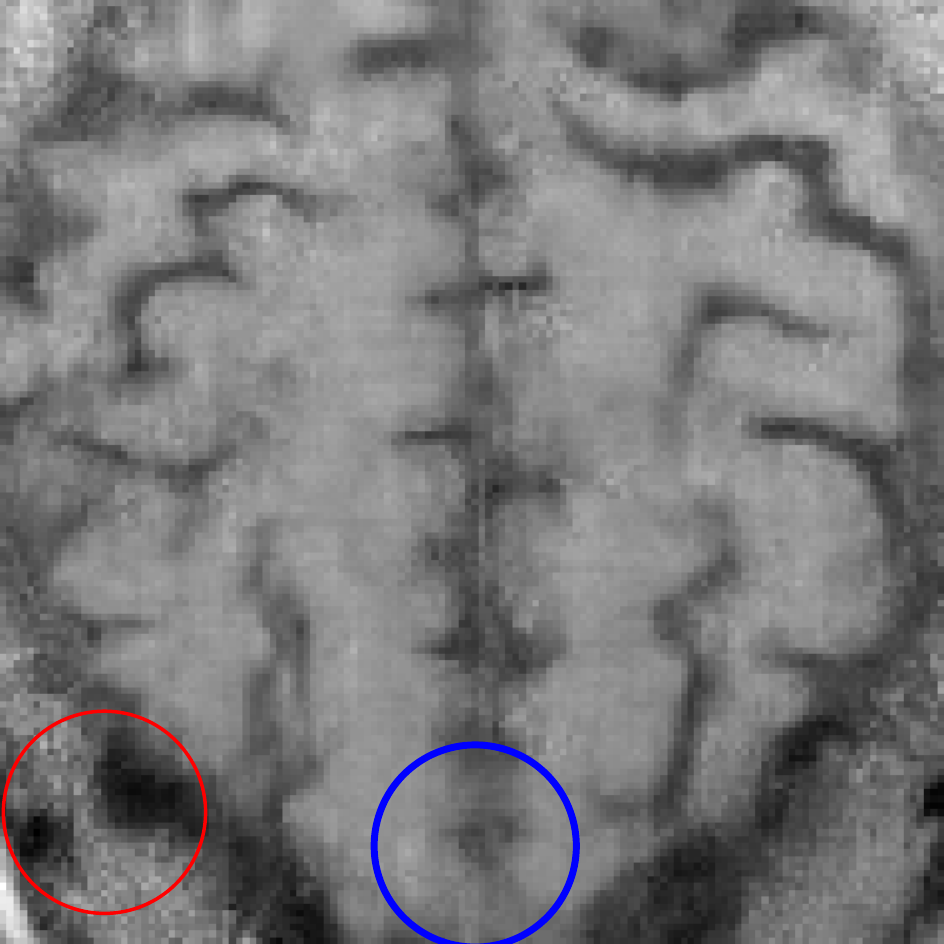}
  \end{subfigure}
  \begin{subfigure}[t]{0.135\textwidth}
  \centering\includegraphics[scale=0.243]{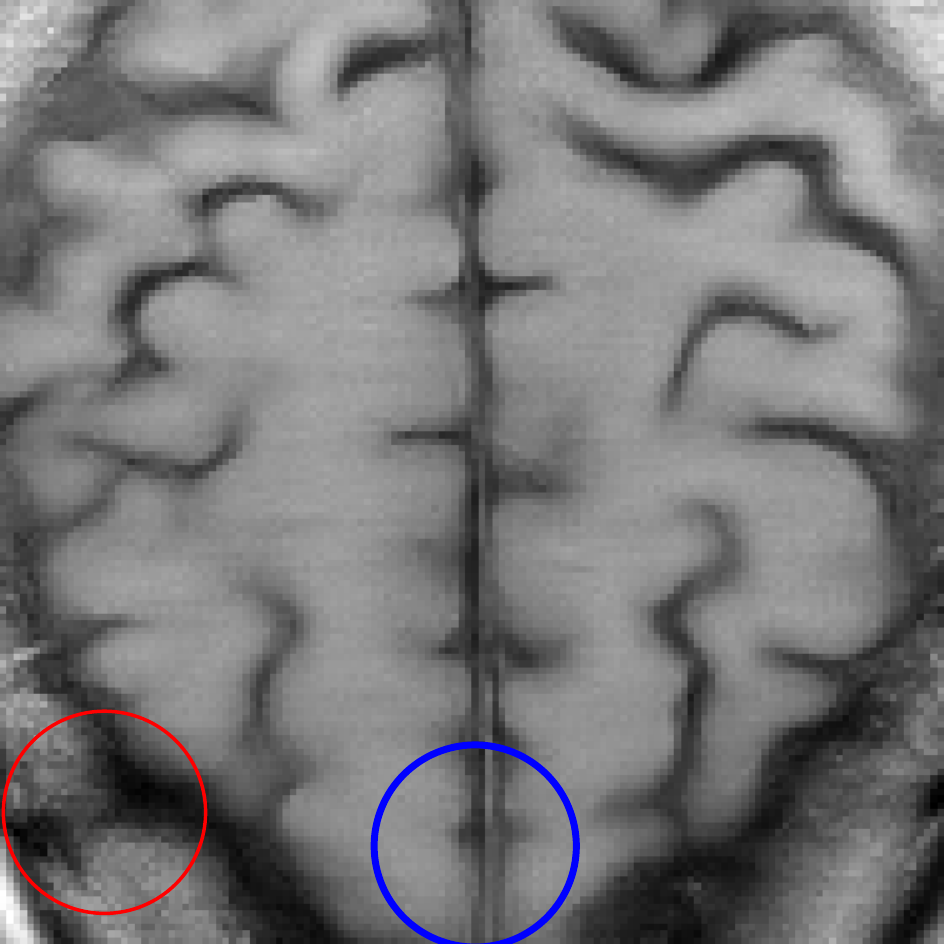}
  \end{subfigure}
  \begin{subfigure}[t]{0.135\textwidth}
  \centering\includegraphics[scale=0.243]{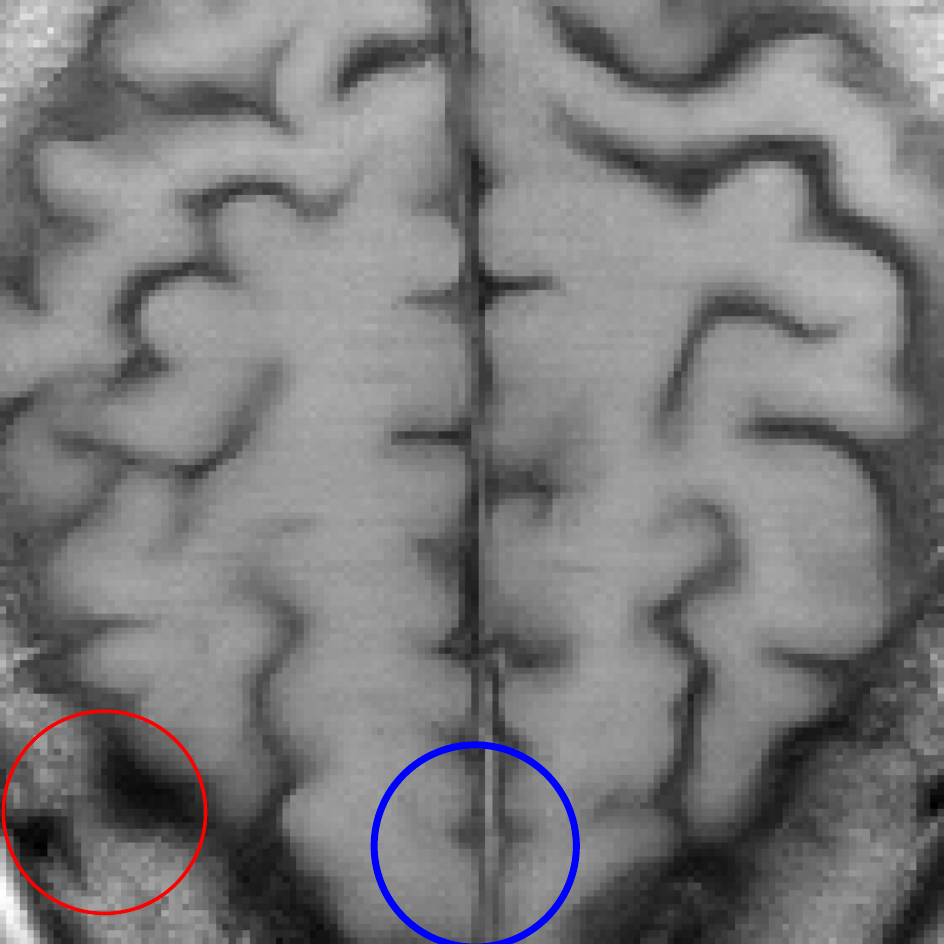}
  \end{subfigure}
  \begin{subfigure}[t]{0.135\textwidth}
  \centering\includegraphics[scale=0.243]{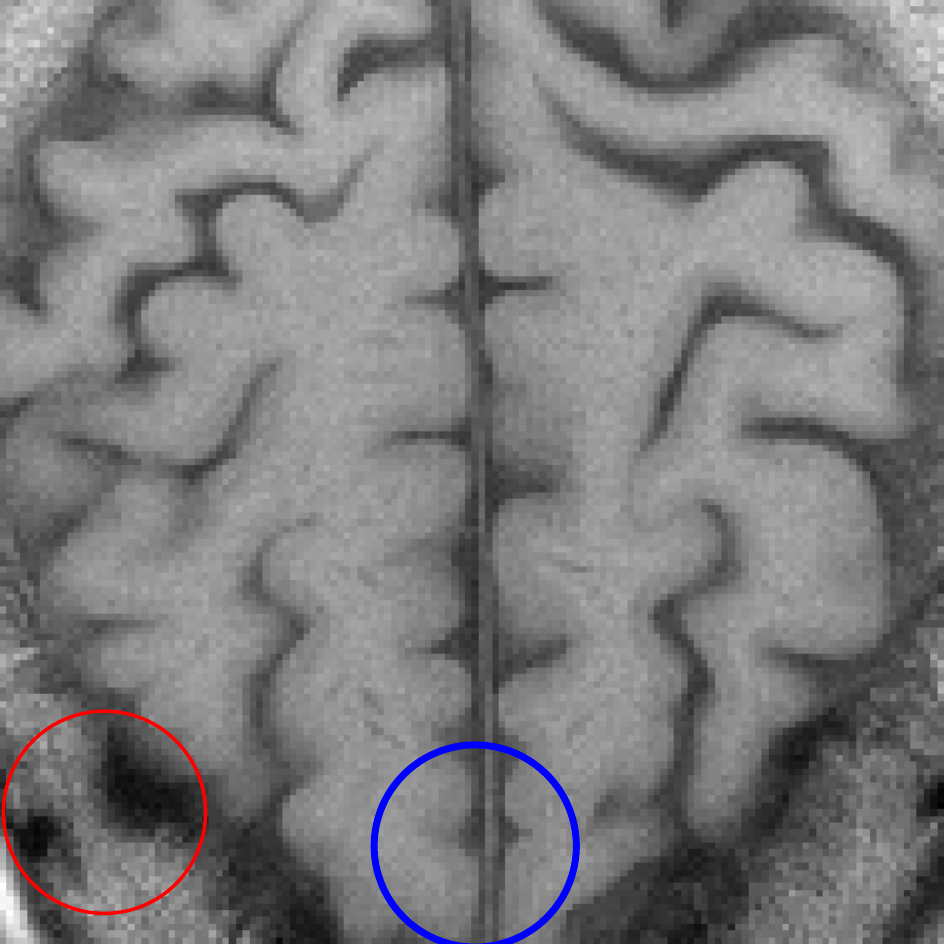}
  \end{subfigure}\par\medskip  
\caption{ConvDecoder-SM, without any training data, yields noticeable improvement over ConvDecoder for brain MRI images, significantly outperforms TV (a traditional compressed sensing method) as well as ENLIVE (a recently-introduced compressed sensing method), and performs slightly worse than U-net (a baseline trained neural network) as well as VarNet (a state-of-the-art trained neural network). Images in the bottom row are zoomed-in versions of the top row.
}
\label{fig:brain}
\end{figure*}

Figure~\ref{fig:brain} depicts sample reconstructions for ConvDecoder-SM (Convdecoder + sensitivity maps), ConvDecoder, U-net, the end-to-end variational network, TV, and ENLIVE for a validation image. Note the significant improvement in the reconstruction quality as a result of using estimated sensitivity maps. Since ConvDecoder-SM, U-net, and the end-to-end variational network (VarNet) are performing best in reconstructing the shown sample, we also annotated two specific parts on their reconstructions to point out their differences. The blue circle denotes a region where U-net and VarNet are giving a sharp reconstruction, while ConvDecoder-SM yields a smooth reconstruction. The red circle on the other hand, denotes a region fully recovered by ConvDecoder-SM, whereas U-net has merged the two black points in the region. 
We observe that for the region denoted by the red circle, VarNet also merges the mentioned black points to some extent, yet this effect is less severe compared to the Unet.

\section{Better performance at the cost of more computations}

In Section~\ref{sec:better-performance}, we showed a better reconstruction accuracy can be achieved via ConvDecoder at the cost of more computation. Specifically, we introduced an ensemble trick to enhance the performance by averaging the outputs of multiple decoders. We further demonstrated monotonic improvement of PSNR as a function of the number of decoders. The outcome of such analysis was getting closer (within 2 dB) to the performance of the state-of-the-art VarNet.

\begin{figure}[th]
\centering
  \begin{subfigure}[t]{0.25\textwidth}
  \caption*{ConvDecoder-A}
  \centering\includegraphics[scale=0.5]{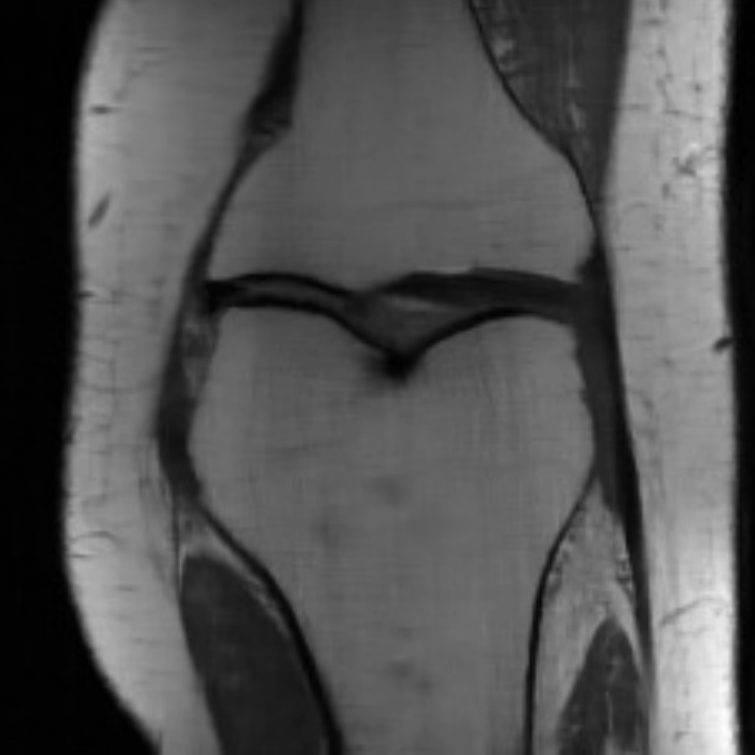}
  \end{subfigure}
  \begin{subfigure}[t]{0.25\textwidth}
  \caption*{ConvDecoder}
  \centering\includegraphics[scale=0.5]{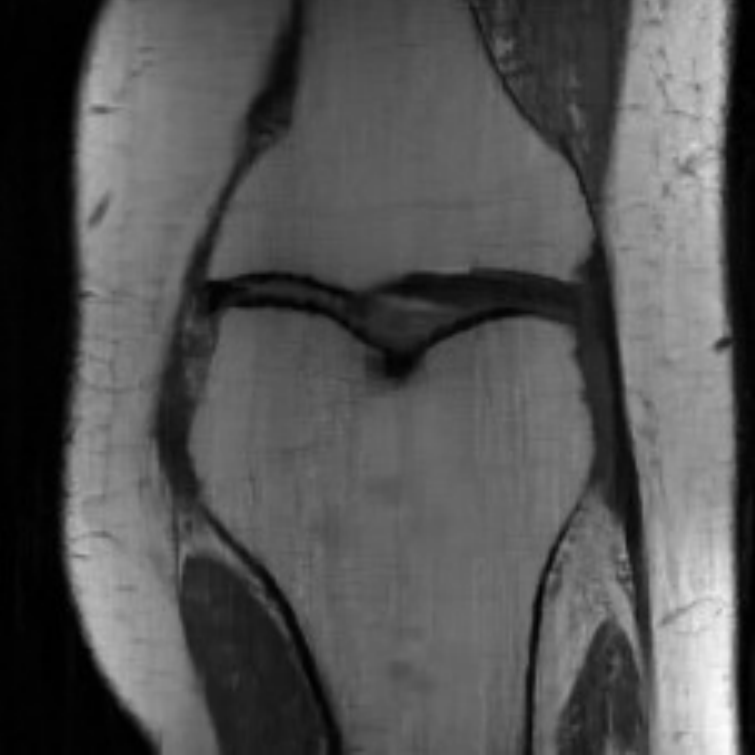}
  \end{subfigure}
  \begin{subfigure}[t]{0.25\textwidth}
  \caption*{ground truth}
  \centering\includegraphics[scale=0.5]{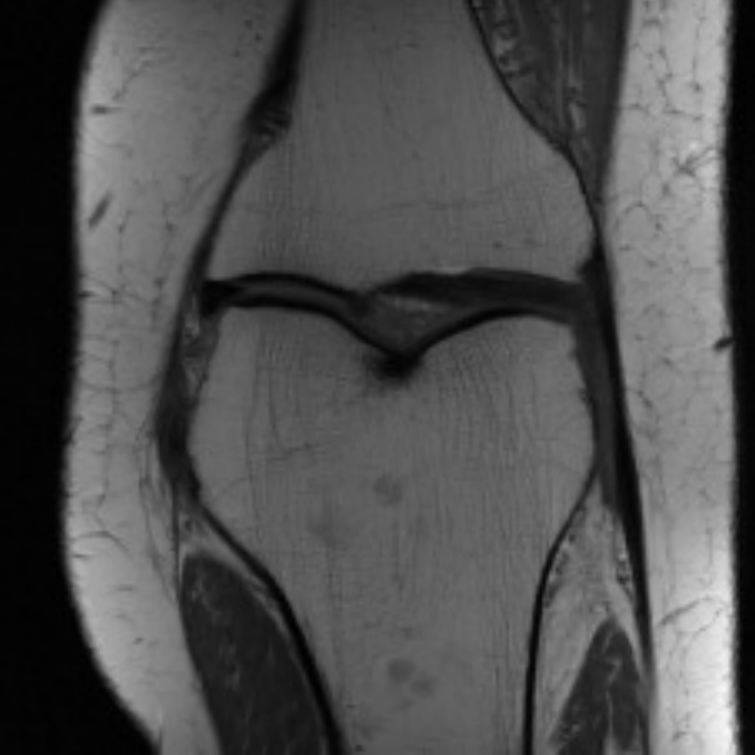}
  \end{subfigure}
\caption{Sample reconstructions for ConvDecoder and ConvDecoder-A (with the averaging technique) for a validation image from multi-coil knee measurements (4x accelerated).}
\label{fig:ensemble}
\end{figure}

Figure~\ref{fig:ensemble} shows sample reconstructions for ConvDecoder with and without our averaging technique (the average image is obtained using ten decoders). As shown, the averaging technique results in slightly smoother regions in the middle parts of the image, yet removes small reconstruction artifacts, and overall yields a higher score (e.g., 0.59 dB higher PSNR as shown in Section~\ref{sec:better-performance}).

\end{document}